%% file: preprint.tex
\documentclass[10pt,letterpaper]{article}
\usepackage[margin=1in]{geometry}
\usepackage{times}
\usepackage[scaled=0.95]{helvet}
\usepackage{titlesec}
\titleformat{\section}{\large\sffamily\bfseries}{\thesection}{0.8em}{}
\titleformat{\subsection}{\normalsize\sffamily\bfseries}{\thesubsection}{0.8em}{}
\titleformat{\subsubsection}{\normalsize\sffamily\bfseries}{\thesubsubsection}{0.8em}{}
\titleformat{\paragraph}[runin]{\normalsize\sffamily\bfseries}{}{0pt}{}
\titlespacing*{\section}{0pt}{2.2ex plus 1ex minus .2ex}{1.2ex}
\usepackage[authoryear,round]{natbib}
\setcitestyle{citesep={;},aysep={,},yysep={;}}
\usepackage[T1]{fontenc}
\usepackage{microtype}
\usepackage{fancyhdr}
\input{math_commands.tex}

\usepackage{algorithm}
\usepackage{algorithmic}

\usepackage{hyperref}
\hypersetup{
  colorlinks=true, linkcolor=linkblue, citecolor=linkblue, urlcolor=linkblue,
  pdftitle={Arbitrage-Free Implied Volatility Surface Reconstruction via Additive Martingales},
  pdfauthor={Haochen Luo, Sifan Lin, Xiaolong Luo, An Binh Minh, Yuan Zhang, Chen Liu}
}
\usepackage{url}
\usepackage{amsmath,amssymb,amsthm}
\usepackage{graphicx}
\usepackage{booktabs}
\usepackage{multirow}

\usepackage{longtable}

\usepackage[table]{xcolor}
\definecolor{linkblue}{HTML}{1463FF}
\definecolor{titlebackground}{HTML}{F2F3F5}
\definecolor{abstractink}{HTML}{25333D}
\usepackage{tcolorbox}
\usepackage{tikz}
\usepackage{wrapfig}
\usepackage{adjustbox}
\usepackage{enumitem}
\usetikzlibrary{arrows.meta,positioning,shapes.geometric,fit,backgrounds}

\theoremstyle{plain}
\newtheorem{theorem}{Theorem}
\newtheorem{proposition}[theorem]{Proposition}
\newtheorem{lemma}[theorem]{Lemma}

\theoremstyle{remark}

\providecommand{\E}{\mathbb{E}}
\renewcommand{\E}{\mathbb{E}}
\providecommand{\R}{\mathbb{R}}
\renewcommand{\R}{\mathbb{R}}

\providecommand{\addmart}{}
\renewcommand{\addmart}{\textsc{AddMart}}

\definecolor{amred}{RGB}{200,30,30}
\definecolor{amblue}{RGB}{30,90,170}
\definecolor{amgreen}{RGB}{20,130,40}

\begin{document}

\begin{tcolorbox}[
  colback=titlebackground, colframe=titlebackground,
  boxrule=0pt, arc=12pt, boxsep=0pt,
  left=17pt, right=17pt, top=17pt, bottom=17pt,
  before skip=0pt, after skip=17pt
]
{\sffamily\bfseries\fontsize{20}{22}\selectfont
Arbitrage-Free Implied Volatility Surface\\
Reconstruction via Additive Martingales\par}
\vspace{8pt}
{\sffamily\bfseries\fontsize{9.2}{12}\selectfont
Haochen Luo\textsuperscript{1}, Sifan Lin\textsuperscript{1},
Xiaolong Luo\textsuperscript{1}, An Binh Minh\textsuperscript{1},
Yuan Zhang\textsuperscript{2}, Chen Liu\textsuperscript{1}\par}
\vspace{4pt}
{\fontsize{10}{12}\selectfont
\textsuperscript{1}City University of Hong Kong\\
\textsuperscript{2}Shanghai University of Finance and Economics\par}
\vspace{10pt}
\begingroup
\color{abstractink}
\setlength{\parskip}{0pt}
This work focuses on reconstructing a continuous implied volatility (IV) surface from a small number of option quotes, a fundamental task for option pricing and risk management. 
We propose \textbf{Additive Martingale} (\addmart{}), a few-shot IV surface reconstruction framework that satisfies static no-arbitrage conditions \textit{by construction}. 
Rather than predicting IVs directly, \addmart{} maps sparse quotes to non-negative increment densities of an exponential additive martingale, from which valid risk-neutral marginals and IVs are recovered. 
By incorporating the no-arbitrage structure directly into its architecture, rather than enforcing it through regularization in the learning objective, \addmart{} produces IV surfaces that satisfy static no-arbitrage conditions throughout the domain. 
Extensive experiments on ten real-option datasets demonstrate that \addmart{} achieves competitive reconstruction accuracy with no material static arbitrage violations, while directly enabling the estimation of VaR, CVaR, and crash probabilities. 

\par\endgroup
\vspace{10pt}
{\small\textsf{\textbf{Correspondence:}}\quad
\href{mailto:chester.hc.luo@my.cityu.edu.hk}{\texttt{chester.hc.luo@my.cityu.edu.hk}},
\href{mailto:chen.liu@cityu.edu.hk}{\texttt{chen.liu@cityu.edu.hk}}\par}
\end{tcolorbox}

\input{sections/intro}

\input{sections/background}

\input{sections/method}

\input{sections/exps}

\vspace{-1em}
\section{Conclusion}

In conclusion, \addmart{} reframes few-shot implied volatility surface reconstruction by learning the increment laws of an exponential additive martingale rather than directly predicting an unconstrained surface. This design enforces static no-arbitrage by construction and directly yields valid risk-neutral marginal distributions without additional arbitrage penalties. Across ten real-market datasets, \addmart{} achieves competitive reconstruction accuracy among amortized methods while enabling downstream distributional risk analysis, demonstrating the benefit of learning structured probability objects rather than only fitting surface values.

The current design nevertheless has several limitations. The additive increment structure induces long-horizon skew flattening, which can reduce accuracy on LEAPS, and the model is less effective for highly asymmetric or heavy-tailed underlyings such as VIX. In addition, serial convolution and implied-volatility inversion limit its suitability for ultra-high-frequency inference, although the resulting latency remains practical at the minute scale. 
Future work could address these limitations through richer state-dependent or nonstationary martingale constructions that better preserve persistent skew and heavy tails, more efficient convolution and implied-volatility inversion for lower-latency inference, and extensions such as uncertainty quantification, adaptive reference selection, and cross-asset pretraining to improve robustness under sparse, noisy, or out-of-distribution market conditions. Detailed analyses of these failure modes, together with latency and memory measurements, are provided in Appendix~\ref{appendix:limitations}.

\section*{AI use statement}

In this work, we used generative AI tools to assist with experimental planning and organization. Additionally, we used generative AI tools for scientific visualization, table formatting, and language editing to improve the readability and presentation of the manuscript. All AI-assisted content, including experimental suggestions, figures, tables, and edited text, was carefully reviewed and verified by the authors for correctness and consistency with the underlying results. Generative AI tools were not used to independently interpret results or draw scientific conclusions. We take responsibility for the final content of this work, including text, claims, or artifacts produced with the aid of generative AI.

\bibliography{references}
\bibliographystyle{plainnat}

\clearpage

\appendix

\input{appendix/background}

\input{appendix/proof}

\input{appendix/data}

\input{appendix/exps}

\input{appendix/analysis}

\end{document}

%% file: math_commands.tex
\usepackage{amsmath,amsfonts,bm}

\def\eqref#1{equation~\ref{#1}}
\def\1{\bm{1}}

\DeclareMathAlphabet{\mathsfit}{\encodingdefault}{\sfdefault}{m}{sl}
\SetMathAlphabet{\mathsfit}{bold}{\encodingdefault}{\sfdefault}{bx}{n}

\def\gL{{\mathcal{L}}}

\def\gN{{\mathcal{N}}}

\def\gQ{{\mathcal{Q}}}

\def\gT{{\mathcal{T}}}

\def\sQ{{\mathbb{Q}}}

\newcommand{\E}{\mathbb{E}}

\newcommand{\R}{\mathbb{R}}

%% file: sections/intro.tex
\section{Introduction}

Options are among the most actively traded derivatives in modern financial markets, indispensable for risk transfer, hedging, and price discovery. 
Within the Black-Scholes-Merton (BSM) framework~\citep{black1973pricing}, market participants analyze and quote an option not through its dollar price but through its \emph{implied volatility} (IV), the volatility input that reconciles the model price with the market price and forms a forward-looking summary of the market's expectation of future fluctuations. 
Aggregating across contracts yields the \emph{implied-volatility surface} $\mathrm{IV}(k,\tau)$ over log-moneyness $k=\log(K/F)$ and time-to-maturity $\tau$, which is the central pricing and risk-management object on a derivatives desk \citep{gatheral2011volatility}.
In addition, the IV surface gives access to the risk-neutral density (RND): via Breeden-Litzenberger \citep{breeden1978prices}, its convexity in strike recovers the full underlying distribution on which downstream tail-risk and exotic pricing depend.
% This surface further serves as the gateway to the risk-neutral density (RND): its convexity in strike recovers, via Breeden-Litzenberger \citep{breeden1978prices}, the full distribution of the underlying on which downstream tail-risk and exotic pricing depend. 
Yet at any instant the market quotes only a sparse, discrete set of strikes and maturities, whereas desks need the \emph{entire continuous} surface to price illiquid strikes, hedge complex portfolios, and compute risk measures consistently.

Reconstructing the entire continuous IV surface from sparse option quotes is a long-standing problem.
Classical approaches include parametric models such as SVI~\citep{gatheral2011volatility}, SSVI~\citep{gatheral2014arbitrage} and SABR~\cite{hagan2002managing}, non-parametric arbitrage-aware interpolation and smoothing \citep{kahale2004arbitrage,fengler2009arbitrage}, and model-based calibration through local or stochastic volatility dynamics \citep{dupire1997pricing,heston1993closed}. 
More recently, learning-based methods have demonstrated flexibility of neural networks for surface smoothing~\cite{ackerer2020deep,chataigner2020deep}, generative modeling~\citep{ning2023arbitrage,vuletic2024volgan} and amortized reconstruction across historical surfaces~\citep{zhang2023two,zhuang2025meta,zhang2025risk}.
In particular, HyperIV~\citep{hyperiv2025} formulates the problem in the practically important \emph{few-shot, real-time, arbitrage-free} regime: given only $9$ reference quotes, it reconstructs the entire IV surface using no-arbitrage regularization in \emph{a single forward pass}.
By mapping quotes to a small pointwise network, HyperIV is fast and has competitive cross-asset performance.

Despite advancements, existing learning-based methods for IV surface construction have two major limitations. First, HyperIV and deep-smoothing approaches \citep{ackerer2020deep} enforce static no-arbitrage conditions, such as butterfly and calendar constraints, through penalties evaluated on a finite grid. 
Such penalties introduce an accuracy-regularization trade-off. In addition, they only provide soft and grid-dependent control to enforce static no-arbitrage conditions.
% Such penalties provide only soft, grid-dependent control and introduce an additional accuracy--regularization trade-off. 
Second, direct IV prediction does not ensure a valid risk-neutral density (RND): applying the Breeden-Litzenberger derivative~\citep{breeden1978prices} can produce negative densities and consequently invalid tail-risk measures. 
We address both issues by changing the model output. 

\begin{figure*}[h]
  \centering
  \newcommand{\addmartLW}{1.3}%      <-- stroke weight, tune here
  \begin{adjustbox}{width=\linewidth}
    \includegraphics[width=\linewidth]{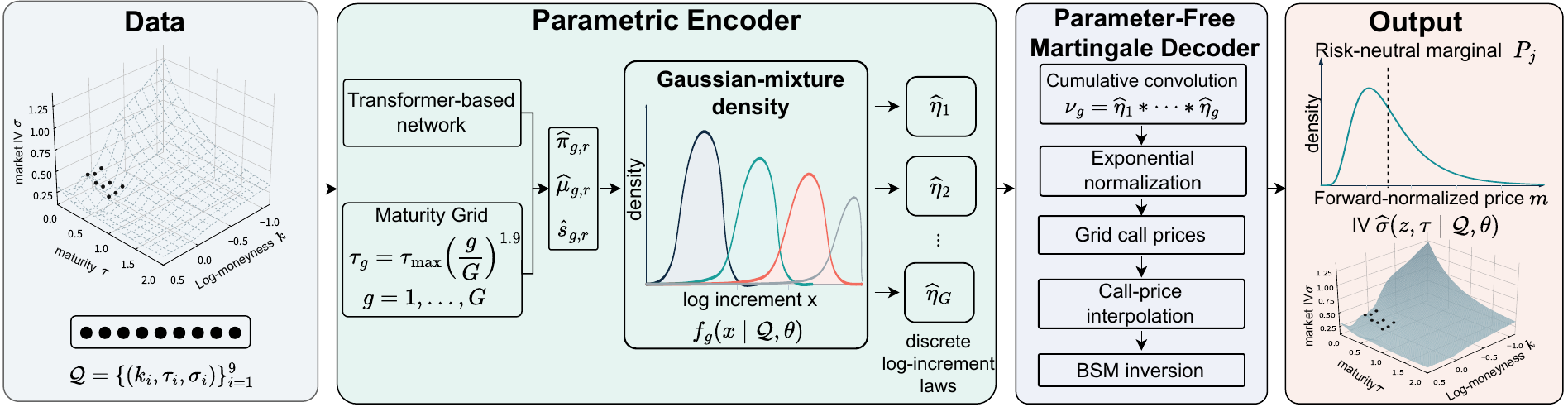}
  \end{adjustbox}
  \caption{\textbf{The ADDMART pipeline.} Reference quotes are encoded into a latent representation for maturity-wise increment prediction, followed by parameter-free deterministic pricing and inversion.}
  \label{fig:addmart-arch}
\end{figure*}

To address the issues above, we propose \textbf{Additive Martingale} (\addmart{}), a learning-based method to reconstruct IV surface satisfying static no-arbitrage conditions \textit{by construction} through a martingale process. 
As demonstrated in Figure~\ref{fig:addmart-arch}, \addmart{} uses a permutation-invariant parameterized set encoder \citep{zaheer2017deep} to map a few reference quotes to non-negative increment densities of an exponential additive martingale. Full-support linear convolution then produces valid martingale marginals, from which call prices and IVs are recovered through a non-parametric decoder. 
Our method connects amortized few-shot reconstruction with the classical density and convex-order literature \citep{bahra1997implied,brigo2002lognormal,dugas2000incorporating,hirsch2011peacocks}. 
The properties of the martingales generated by the encoder guarantee that the IV surface satisfies the static no-arbitrage conditions globally~\citep{kellerer1972markov,hirsch2011peacocks}, which enables us to train the models without any additional constraints.
% Valid marginals guarantee butterfly-arbitrage-freeness, while their convex-order growth guarantees calendar-arbitrage-freeness . 

Our contributions are threefold. (1) We introduce AddMart, an amortized framework that maps sparse option quotes to flexible increment distributions and reconstructs IV surfaces through an exponential additive martingale, producing valid risk-neutral marginals and enforcing static no-arbitrage by construction. (2) We establish no-arbitrage guarantees for the full-support decoder and call-price interpolation, and bound the calendar residual induced by support truncation. The resulting marginals directly support option-implied VaR, CVaR, and crash-probability calculations without numerical differentiation. (3) Across ten real-option datasets, AddMart achieves competitive reconstruction accuracy while exhibiting no material butterfly or calendar violations across five seeds and 2,500 audited model-date surfaces over the 16-month test period, without per-surface parameter optimization at inference.
%Our contributions are: \textbf{(1)} To our knowledge, \addmart{} is the first IV surface reconstruction method based on amortized network architecture that directly produces valid RNDs while enforcing static no-arbitrage constraints by construction. \textbf{(2)} We establish the no-arbitrage guarantees of the full-support decoder, bound the calendar residual induced by grid truncation, and characterize the long-maturity skew flattening of additive increments. These theoretical properties further enable valid-by-construction VaR, CVaR, and crash probabilities to be computed directly from the per-maturity probability masses, without numerically differentiating the IV surface. \textbf{(3)} Across ten real-option datasets (seven underlyings at the daily scale and three broad-market ETFs at the minute scale), \addmart{} achieves competitive reconstruction accuracy in both overall and reliable IV-MAE while exhibiting no material butterfly or calendar violations across five seeds and 2500 audited model-date surfaces over the 16-month test period.

\vspace{-1em}

%% file: sections/background.tex
\section{Related Work}
\label{sec:related}

\textbf{Arbitrage-Free Surface and Density Models.} Risk neutral densities have long been fitted with lognormal mixtures
\citep{bahra1997implied,brigo2002lognormal}, and convexity architectures make butterfly freeness
structural \citep{dugas2000incorporating,amos2017input}. The most accurate models in practice remain
per surface calibrations such as SSVI \citep{gatheral2014arbitrage} and Deep Smoothing
\citep{ackerer2020deep}, but both require expensive optimization per surface and meet no arbitrage only
softly, as a numerical penalty rather than a guarantee. A complementary route specifies an arbitrage free
valuation first and recovers the supporting martingale \citep{carr2005note,madan2002making}; modeling the
log price by an \emph{additive} process makes this constructive, since the marginals of a genuine
martingale increase in convex order, giving calendar freeness, while each marginal is a valid density,
giving butterfly freeness \citep{carr2021additive,azzone2025explicit}. Neural variants exist:
\citet{ning2023arbitrage} train a variational autoencoder over the parameters of additive and Levy models to
generate arbitrage free surfaces, and \citet{lin2024neural} fit an additive logistic martingale whose
deterministic term structure is a feed forward network. \addmart{} shares this additive martingale
backbone but is quote conditioned and amortized few shot via a permutation invariant set encoder
\citep{zaheer2017deep}, places a \emph{nonparametric} mixture density on each increment rather than a fixed
parametric family, which we show is decisive for accuracy, and makes
calendar freeness hard by construction through linear convolution rather than a soft penalty.

\textbf{Few-Shot Amortization.} %To remove the per surface bottleneck, HyperIV \citep{hyperiv2025} maps the nine quotes to a compact
%implied volatility network with a hypernetwork, VolNP \citep{zhuang2025meta} casts reconstruction as a meta
%learning neural process, and operator deep smoothing \citep{wiedemann2025operator} uses a single graph
%neural operator instance. All return a surface, or at best a signed Breeden-Litzenberger density obtained
%by finite differencing it, and all enforce no arbitrage as a soft penalty. However, it provides no hard guarantee, and its residual butterfly violation persists at $0.6\%$
%to $0.9\%$ out of sample. 
HyperIV \citep{hyperiv2025}, VolNP \citep{zhuang2025meta}, and Operator Deep Smoothing \citep{wiedemann2025operator} reconstruct IV surfaces without per-surface optimization. HyperIV and Operator Deep Smoothing use no-arbitrage penalties, while VolNP combines SABR-derived pretraining with likelihood-based fitting. Neither mechanism guarantees global static no-arbitrage or non-negative implied risk-neutral densities. Other structured neural models target different pricing settings.
Concurrent work emits neural Gaussian mixture
densities with closed form prices \citep{vandenberg2026mixtures}, but in a model surrogate setting rather
than from sparse market quotes and without a cross maturity martingale. \addmart{} focuses on reconstruction from sparse market quotes, combining amortized inference with a common martingale construction that produces valid marginals and enforces consistency across maturities.

\vspace{-1em}
 \section{Preliminaries}
\label{sec:prelim}

We summarize the frequently used notation in this work in Table~\ref{tab:notation} of Appendix~\ref{app:notation}. In this work, although listed equity and ETF options are American-style, we use liquid
OTM quotes, for which immediate exercise has no value, and express them as
European-equivalent Black--76 IVs using parity-implied forwards that account
for prevailing carry. This convention limits early-exercise and dividend
effects in our contemporaneous reconstruction setting, so the formal guarantee
applies to the resulting European-equivalent surface.
The Black-Scholes-Merton (BSM) framework provides a closed-form pricing formula for European call options. Given the underlying asset's current price $S_0$, the strike price $K$, the risk-free interest rate $r$, the time-to-maturity $\tau$, and the volatility $\sigma$, the call price is given by $C = S_0 \gN(d_1) - K e^{-r \tau} \gN(d_2)$, where $d_1 = \frac{\ln (S_0 / K) + (r + \sigma^2 / 2)\tau}{\sigma \sqrt{\tau}}$, $d_2 = d_1 - \sigma \sqrt{\tau}$, and $\gN$ denotes the cumulative distribution function of the standard normal distribution. Because all parameters except the volatility $\sigma$ are directly observable in the market, equating the BSM model price to the observed market option price allows us to back out the unique \textit{implied volatility (IV)}. 

% \begin{wrapfigure}{r}{0.5\textwidth}
%     \centering
%     \vspace{-0.8\baselineskip}
%     \includegraphics[width=\linewidth]{figures/no_arbitrage_intuition.pdf}
%     \caption{
%         Butterfly and calendar restrictions implied by static no-arbitrage.
%     }
%     \label{fig:no-arbitrage-intuition}
%     \vspace{-0.6\baselineskip}
% \end{wrapfigure}

\vspace{-0.5em}
\begin{figure}[h]
    \centering
    \includegraphics[width=0.9\linewidth]
        {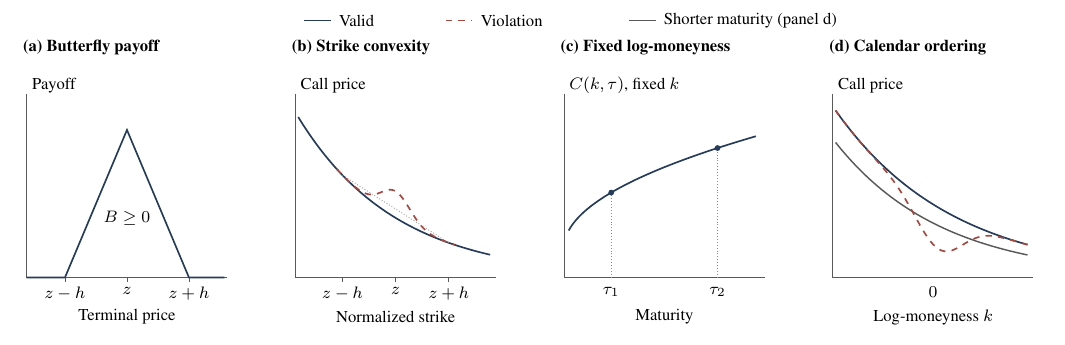}
    \vspace{-1em}
    \caption{Butterfly and calendar restrictions implied by static no-arbitrage.}
    \label{fig:no-arbitrage-intuition}
    \vspace{-1em}
\end{figure}

To facilitate our further analysis, it is convenient to work with forward-normalized quantities. Let $k = \log (K / F(\tau))$ and $M_\tau = S_\tau / F(\tau)$ where $F(\tau)$ is the parity-implied forward price and $S_\tau$ is the asset price at maturity. Under the pricing measure $\sQ$, $\E_\sQ [M_\tau] = 1$. We define the forward-normalized, undiscounted call price as $C(k, \tau) = \E_\sQ[(M_\tau - e^k)^+]$ or equivalently $c(z, \tau) = E_\sQ[(M_\tau - z)^+]$ where $z = e^k = K / F(\tau)$ is the normalized strike. Furthermore, we have $\forall \tau$, $c(0, \tau) = 1$ and $\lim_{z \to +\infty} c(z, \tau) = 0$. In this work, we aim to learn a surface mapping from $k$ and $\tau$ to implied volatility $\text{IV}(k, \tau)$ based on observable quotes.

\textbf{Static No-Arbitrage Conditions.} A valid option IV surface must satisfy both strike-wise and cross-maturity no-arbitrage conditions~\citep{carr2005note,gatheral2014arbitrage}, corresponding to the absence of butterfly and calendar arbitrage, respectively.
The butterfly constraint requires the call price $c(\cdot, \tau)$ to be non-increasing and globally convex in normalized strike $z$, i.e., $\partial c(z, \tau) / \partial z \leq 0$ and $\partial^2 c(z, \tau) / \partial z^2 \geq 0$. By the Breeden--Litzenberger theorem \citep{breeden1978prices}, the distributional second derivative $\partial^2c/\partial z^2$ is the risk-neutral marginal measure of $M_\tau$ and is therefore non-negative.
The calendar constraint requires the call price $c(z, \cdot)$ to be non-decreasing in maturity, i.e., $\forall 0 \leq \tau_1 \leq \tau_2$, $c(z, \tau_1) \leq c(z, \tau_2)$. Figure~\ref{fig:no-arbitrage-intuition} illustrates the butterfly and
calendar restrictions implied by static no-arbitrage.

\textbf{Problem Setting.} Given $N$ option quotes for an underlying asset $\mathcal{Q}=\{(k_n,\tau_n,\sigma_n)\}_{n=1}^{N}$ where $\sigma_n$ is implied volatility indicated by the market price, our goal is to learn a model $\widehat{\sigma} (z, \tau|\gQ, \theta)$ parameterized by $\theta$ that reconstruct the full implied-volatility surface subject to the static no-arbitrage conditions discussed above. Unlike per-surface calibration, the learned model should generate the predicted IV in a single forward pass. 
At a high level, the learning problem is:
\begin{equation} \label{eq:generic_problem}
\begin{aligned}
    &\min_\theta \E_{(k,\tau, \sigma ) \sim \gT} ~l(\widehat{\sigma}(z, \tau|\gQ, \theta) - \sigma)
    \\ s.t.~~\forall z, \tau, \frac{\partial \widehat{c}(z, \tau)}{\partial z} \leq 0, &\frac{\partial^2 \widehat{c}(z, \tau)}{\partial z^2} \geq 0, \quad \forall z, \tau_1, \tau_2, [\widehat{c}(z, \tau_1) - \widehat{c}(z, \tau_2)](\tau_1 - \tau_2) \geq 0.
\end{aligned}
\end{equation}
where $\gT$ denote the training set consisting of sampled quotes to optimize $\theta$, $\widehat{c}(z, \tau)$ is the estimated forward-normalized, undiscounted call price based on $\widehat{\sigma}(z, \tau|\gQ, \theta)$ and $l$ is the loss objective function.
In general, the learning problem aims to minimize the discrepancy between the market IV and the estimated IV subject to static no-arbitrage conditions.

%% file: sections/method.tex
\section{Methodology: \addmart{}}
\label{sec:method}

% We adopt the forward-normalized pricing notation, few-shot reconstruction
% setting, and static-arbitrage conditions introduced in
% Section~\ref{sec:prelim}. Rather than predicting implied volatility directly,
% \addmart{} parameterizes an exponential martingale and recovers option prices,
% implied volatilities, and risk-neutral marginals from its increment laws. The
% method separates a learned amortization network from a deterministic pricing
% decoder: the network selects increment distributions that match the sparse
% quotes, while the decoder supplies the probabilistic and no-arbitrage
% structure.

We introduce \addmart{} in this section, including the exponential additive-martingale construction (Section~\ref{sec:method-construct}), the resulting by-construction static no-arbitrage guarantees (Section~\ref{sec:method-arb}), the amortized architecture that generates estimated IV from sparse quotes (Section~\ref{sec:method-network}), and the end-to-end training and inference pipeline (Section~\ref{sec:method-training}).

\subsection{Martingale Construction for the Forward-Normalized Price}
\label{sec:method-construct}

On a maturity grid $0 = \tau_0 < \tau_1 < \ldots < \tau_G$ and a filtered probability space $(\Omega,\mathcal F,\{\mathcal F_{\tau_j}\}_{j=0}^{G},\mathbb Q)$, we consider $G$ independent random variables $\{L_g\}_{g=1}^{G}$ with probability laws $\{\eta_g\}_{g=1}^{G}$, each satisfying $\mathbb E_{\mathbb Q}[e^{L_g}]<\infty$.
We define the additive state $A_{\tau_j}:=\sum_{g\leq j}L_g$ and its normalized exponential as the normalized terminal price $M_{\tau_j}$. Based on the independent variable, we have:
% Let $0=\tau_0<\tau_1<\cdots<\tau_G$ be a maturity grid. We consider
% independent log-increments $\{L_g\}_{g=1}^{G}$ with probability laws
% $\{q_g\}_{g=1}^{G}$ satisfying
% $\mathbb E_{\mathbb Q}[e^{L_g}]<\infty$ on a filtered
% probability space
% $(\Omega,\mathcal F,\{\mathcal F_{\tau_j}\}_{j=0}^{G},\mathbb Q)$.

\begin{equation}
M_{\tau_j}
:=
\frac{e^{A_{\tau_j}}}
     {\mathbb E_{\mathbb Q}[e^{A_{\tau_j}}]}
=
\prod_{g\leq j}
\frac{e^{L_g}}
     {\mathbb E_{\mathbb Q}[e^{L_g}]} .
\label{eq:mart}
\end{equation}
Each multiplicative $\frac{e^{L_g}}{\mathbb E_{\mathbb Q}[e^{L_g}]}$ in (\ref{eq:mart}) is non-negative and has unit expectation under $\mathbb{Q}$. In addition, since $L_{j + 1}$ is independent of $\mathcal F_{\tau_j}$, we have $\mathbb E_{\mathbb Q}[M_{\tau_{j + 1}}\mid\mathcal F_{\tau_j}] = M_{\tau_j}$. Therefore, $\{M_{\tau_j}\}_{j=0}^{G}$ is a non-negative martingale with $\forall j, \mathbb E_{\mathbb Q}[M_{\tau_j}]=1$.

% Each multiplicative
% factor is non-negative and has expectation one. Since $L_{j+1}$ is independent
% of $\mathcal F_{\tau_j}$, we have
% $\mathbb E[M_{\tau_{j+1}}\mid\mathcal F_{\tau_j}]=M_{\tau_j}$. Therefore,
% $\{M_{\tau_j}\}_{j=0}^{G}$ is a non-negative martingale with
% $\mathbb E[M_{\tau_j}]=1$.

Let $\nu_j:=\eta_1*\cdots*\eta_j$ denote the probability law of $A_{\tau_j}$ and define $Z_j := \mathbb E_{\mathbb{Q}}[e^{A_{\tau_j}}] = \int_{\mathbb R}e^x\nu_j(dx)$.
Therefore, we can calculate the forward-normalized, undiscounted call price $c(z, \tau_j)$ as follows:
\begin{equation} \label{eq:calc-c}
    c(z, \tau_j) = \mathbb E_{\mathbb Q}[(M_{\tau_j} - z)^+] = \int_{\mathbb{R}} \left(\frac{e^x}{Z_j} - z\right)^+ \nu_j(dx)
\end{equation}

% Let $\nu_j:=q_1*\cdots*q_j$ denote the law of $A_{\tau_j}$ and define
% $D_j:=\int_{\mathbb R}e^x\nu_j(dx)$. For normalized strike
% $z:=K/F(\tau_j)=e^k$, the forward-normalized call price is
% \begin{equation}
% c(z,\tau_j)
% =
% \mathbb E[(M_{\tau_j}-z)^+]
% =
% \int_{\mathbb R}
% \left(\frac{e^x}{D_j}-z\right)^+
% \nu_j(dx).
% \label{eq:calc-c}
% \end{equation}
Equivalently, the risk-neutral marginal $\mathsf P_j$ of $M_{\tau_j}$ is the pushforward of $\nu_j$ through $x\mapsto e^x/Z_j$.

Equation (\ref{eq:calc-c}) calculates the forward-normalized, undiscounted call price at the maturity grid. For more general maturity $\tau\in(\tau_j,\tau_{j+1})$, we let $\lambda=(\tau-\tau_j)/(\tau_{j+1}-\tau_j)$ and interpolate the prices at the maturity grid by $c(z,\tau)=(1-\lambda)c(z,\tau_j)+\lambda c(z,\tau_{j+1})$.
This is equivalent to using the mixture marginal $\mathsf P_\tau=(1-\lambda)\mathsf P_j+\lambda\mathsf P_{j + 1}$.
With $c(z, \tau)$, we can calculate the corresponding IV$(k, \tau)$ by BSM framework introduced in Section~\ref{sec:prelim}.

\subsection{Arbitrage-Free by Construction}
\label{sec:method-arb}

Our main theorem indicates the forward-normalized price $c(z, \tau)$ constructed by the martingale above is free of static arbitrage, so we do not need additional constraints to enforce this requirement.

\begin{theorem}[Static no-arbitrage on the modeled maturity interval]
\label{thm:static-arbitrage}
Let $L_1,\ldots,L_G$ be independent real-valued random variables satisfying
$\mathbb{E}_{\mathbb{Q}}[e^{L_g}]<\infty$. Construct the normalized
exponential martingale and its call prices as in Section~4.1, with
$M_0=1$, and interpolate the grid call prices linearly at fixed normalized
strike. For every $\tau\in[0,\tau_G]$, the function
$c(z,\tau)$ is non-increasing and convex in $z>0$, with
$(1-z)^+\le c(z,\tau)\le1$.
For every normalized fixed strike $z>0$,
$c(z,\tau)$ is non-decreasing in $\tau$. Thus the surface satisfies
both strike-wise and cross-maturity no-arbitrage conditions under
the stated pricing assumptions.
\end{theorem}

To prove the theorem above, we notice that each marginal $\mathsf P_j$ of
$M_{\tau_j}$ is a valid probability law and that these marginals increase
in convex order. Full-support convolution in the probability law of
$A_{\tau_j}$ preserves non-negativity and unit mass, while the martingale
property in Equation~(\ref{eq:mart}) implies convex-order monotonicity
\citep{kellerer1972markov}. Both properties hold for every admissible value
of the increment laws $\{\eta_g\}_{g=1}^G$. The complete proof is provided
in Appendix~\ref{app:proof-arb}.

As a direct consequence, each grid-maturity marginal emitted by
\addmart{} is a valid probability measure. Distributional quantities such
as the CDF, survival function, generalized quantiles, VaR, CVaR, and crash
probabilities can therefore be computed directly from the predicted
marginal. The corresponding formulas are provided in
Appendix~\ref{app:proof-risk}.
Theorem~\ref{thm:static-arbitrage} shows that the static no-arbitrage
conditions are enforced by the construction in
Section~\ref{sec:method-construct}, rather than by the training objective.
Therefore, \addmart{} requires no butterfly or calendar arbitrage penalty
during learning.

\subsection{Amortized Network Architecture}
\label{sec:method-network}

\addmart{} employs a model parameterized by $\theta$ to follow the construction process introduced in Section~\ref{sec:method-construct} so that the generated surface is the arbitrage-free by construction.
Figure~\ref{fig:addmart-arch} summarizes the architecture. It consists of
a learned set encoder and increment-density head followed by a deterministic
martingale decoder.

\paragraph{Parametric Encoder.} The encoder maps the input quote reference set $\mathcal Q=\{(k_i, \tau_i, \sigma_i)\}_{i=1}^{N}$ to an estimation of the probability laws $\{\eta_g\}_{g = 1}^G$ of log-increments $\{L_g\}_{g = 1}^G$ with a model parameterized by $\theta$.
Specifically, each quote $(k_i,\tau_i,\sigma_i)$ is mapped to a 128-dimensional feature by a linear transformation, and the resulting sequence is processed by two Transformer encoder layers with two attention heads, followed by mean pooling over the quote dimension and a linear projection to obtain the 64-dimensional representation $h\in\R^{64}$.
We use no positional encoding and mean pooling over the quote dimension to make the representation $h$ invariant to the ordering of input quotes.
Self-attention allows each quote to be interpreted relative to the remaining maturities and moneyness levels, while the pooled representation summarizes the sparse surface as a whole.
We then generate probability laws $\{\eta_g\}_{g = 1}^G$ based on learning-based representation $h$.
In particular, we choose $G = 30$ maturity buckets up to $\tau_{\max} = 2$ years, with an exponential interpolation $\tau_g = \tau_{\max} (g / G)^{1.9}$ so that more buckets are allocated near expiry.
For each bucket, we concatenate $h$ with time features $(\tau_g, \sqrt{\tau_g})$ and employ a two-layer MLP with width 128 and GELU~\citet{hendrycks2016gaussian} activations to map the time-augmented feature $\widehat{h}_g = (h, \tau_g, \sqrt{\tau_g})$ to the 24-dimensional output, which is then organized tuples $\{\widehat{\pi}_{g, r}, \widehat{\mu}_{g, r}, \widehat{s}_{g, r}\}_{r = 1}^8$.
We then model the corresponding probability law $\eta_g$ as a Gaussian mixture $f_g(x|\gQ, \theta) = \sum_{r = 1}^R \pi_{g, r} \gN(x|\mu_{g, r}, s^2_{g, r})$ where $\gN(x|\mu_{g, r}, s^2_{g, r})$ is the probability density of Gaussian distribution with its weight $\pi_{g, r}$, mean $\mu_{g, r}$ and variance $s^2_{g, r}$ calculated as:
\begin{equation}
    \pi_{g, r} = \frac{e^{\widehat{\pi}_{g, r}}}{\sum_{r = 1}^R e^{\widehat{\pi}_{g, r}}},\quad \mu_{g, r} = 0.3 \sqrt{\Delta \tau_g} \widehat{\mu}_{g, r},\quad s_{g, r} = \sqrt{\Delta \tau_g}\left[0.02 + \text{softplus}(\widehat{s}_{g, r} - 1)\right].
\end{equation}
where $\Delta \tau_g = \tau_g - \tau_{g - 1}$.
% We scale each Gaussian component by $\sqrt{\Delta \tau_g}$ to impose a natural diffusive time scale.
% Through the mixture of $8$ Gaussian distributions, $\eta_g$ is flexible enough to represent asymmetric and multimodal increment laws.
%             $\widehat{\eta}_g$ as follows:
% \begin{equation} \label{eq:estimate_eta}
%     \widehat{\eta}_g (x|\gQ, \theta) = \sum_{i = 1}^{2048} \Big[f_g(x_i|\gQ, \theta) \Big/ \sum_{j = 1}^{2048} f_g(x_j|\gQ, \theta) \Big] \delta_a (x_i).
% \end{equation}
% where $\delta_a$ is Dirac measure. Compared with the Gaussian mixture law $f_g(x|\gQ, \theta)$, the discrete estimation law $\widehat{\eta}_g (x|\gQ, \theta)$ is an approximation that facilitate the calculations of subsequent operations in Section~\ref{sec:method-construct}. In addition, $\widehat{\eta}_g (x|\gQ, \theta)$ is still differentiable w.r.t. $\theta$ and facilitate model training.
We scale the mean and standard deviation of each Gaussian component by
$\sqrt{\Delta\tau_g}$ to impose a natural diffusive time scale.
The mixture of eight Gaussian components allows $\eta_g$ to represent
asymmetric and multimodal increment laws.
We discretize this law on a fixed, equally spaced grid
$\{x_i\}_{i=1}^{X}\subset[-5,5]$, with $X=2048$, by evaluating
and normalizing the mixture density:
\begin{equation}
\label{eq:estimate_eta}
\widehat{\eta}_g
=
\sum_{i=1}^{X} m_{g,i}\,\delta_{x_i},
\qquad
m_{g,i}
=
\frac{f_g(x_i\mid\gQ,\theta)}
{\sum_{\ell=1}^{X} f_g(x_\ell\mid\gQ,\theta)}.
\end{equation}
Here, $\delta_{x_i}$ denotes the Dirac probability measure at $x_i$.
The dependence of $\widehat{\eta}_g$ and $m_{g,i}$ on $\gQ$ and
$\theta$ is omitted for brevity.
The resulting discrete law approximates the continuous increment law
$\eta_g$ and enables the convolution and pricing operations in
Section~\ref{sec:method-construct}.
Its probability masses are non-negative, sum to one, and are
differentiable with respect to $\theta$, enabling end-to-end training.

\paragraph{Parameter-free Decoder.} Encoder's output $\widehat{\eta}_g (x|\gQ, \theta)$ are passed to the decoder for construction in Section~\ref{sec:method-construct}.
Specifically, we estimate the probability law of $A_{\tau_j}$ by convolution, and then the forward-normalized, undiscounted call price $\widehat{c}(z, \tau|\gQ, \theta)$ by Equation~(\ref{eq:calc-c}) and interpolation.
The decoder contains no trainable parameters and supplies the martingale structure of Theorem~\ref{thm:static-arbitrage}.
We apply BSM framework on the log-moneyness $k = \log z$, time-to-maturity $\tau$ and estimated call price $\widehat{c}(z, \tau|\gQ, \theta)$ to obtain the estimated implied volatility, which is a function of $z$, $\tau$ conditioned on $\gQ$, $\theta$ and denoted as $\widehat{\sigma}(z, \tau|\gQ, \theta)$.

\subsection{Training and Inference}
\label{sec:method-training}

\paragraph{Training Phase.} We follow~\citet{hyperiv2025} to include $9$ reference quotes as the conditioning set $\gQ$ and another $256$ quotes sampled with replacement as the training set $\gT$ to train the model.
Each quote $(k_n, \tau_n, \sigma_n)$ include its log-moneyness $k_n = \log z_n$, time-to-maturity $\tau_n$ and market IV $\sigma_n$.
For each quote in the training set, we apply parametric encoder and non-parametric decoder discussed in Section~\ref{sec:method-network} on $k_n$ and $\tau_n$ to obtain the estimated IV $\widehat{\sigma}(e^{k_n}, \tau_n | \gQ, \theta)$.
We minimize the estimated discrepancy between the estimated IV and the market IV as follows:
\begin{equation} \label{eq:objective}
    \min_\theta \gL(\theta) = \frac{1}{|\gT|} \sum_{n = 1}^{|\gT|} l_\beta \big(\widehat{\sigma}(e^{k_n}, \tau_n | \gQ, \theta) - \sigma_n\big)
\end{equation}
where $l_\beta$ is a Huber loss~\citet{huber1992robust} with $\beta = 0.05$. The quadratic part of $l_\beta$ supplies smooth gradients for accurately reconstructed contracts, while its linear tail limits the influence of noisy or stale quotes.
Compared with the generic form in (\ref{eq:generic_problem}), our objective in (\ref{eq:objective}) has no penalty term, regularization term or any constraints, because static no-arbitrage is supplied by the decoder and is therefore independent of the attained loss value.
We defer more technical details of the training phase in Appendix~\ref{app:implicit-iv}.

\paragraph{Inference Phase.} In the inference phase, the trained encoder processes the sparse reference quotes $\gQ$ once and emits all $G$ increment distributions in a single forward pass. The decoder then compute the estimated IV based on convolution, interpolation and inverse BSM. It returns both the queried IV surface and the underlying risk-neutral marginals. The implementation is detailed in Appendix~\ref{app:exact-inference}.

% \paragraph{Inference.}
% At inference phase, the trained network processes the sparse reference quotes $\gQ$ once and emits all $G$ increment distributions in a single forward pass. The
% full-support decoder computes the cumulative base and exponentially
% tilted laws using zero-padded FFT convolution. Once these laws have been
% constructed, prices at arbitrary strikes are obtained from tail probabilities
% and indexed lookup, while off-grid maturities are evaluated by linear
% interpolation in call-price space. A robust BSM bisection converts the
% resulting prices to implied volatilities. Inference uses fixed trained weights and requires no per-surface fitting,
% optimization, or gradient computation. It returns both the queried IV surface
% and the underlying risk-neutral marginals. The full-support inference path is
% the predictor to which Theorem~\ref{thm:static-arbitrage} applies; its
% implementation is detailed in Appendix~\ref{app:exact-inference}. At test
% time, the nine conditioning contracts are excluded from the reconstruction
% targets.

%% file: sections/exps.tex
\vspace{-1em}
\section{Experiments}
\label{sec:exp}

\subsection{Experiment settings}

\textbf{Data and protocol.}
We use OPRA top-of-book option quotes obtained from a commercial data
provider,\footnote{\url{https://massive.com}} and price each contract at the
bid--ask midpoint. We infer the forward from put--call parity and invert
implied volatility under Black--76 \citep{black1976pricing}. Each surface
contains OTM options and is conditioned on $9$ reference quotes: the contracts
nearest to three target maturities (7, 30, and 90 days) and three deltas
($0.50$, $0.25$, and $-0.25$). All remaining contracts are reconstruction
targets.
We evaluate $7$ daily books: SPY, QQQ, IWM, AAPL, GOOGL, META,
NVDA, and $3$ intraday-minute books: SPY, QQQ, IWM.\footnote{SPY,
QQQ, and IWM are liquid ETFs tracking the S\&P 500, Nasdaq-100, and Russell
2000, respectively.} All datasets use the same strict out-of-time split:
surfaces before February 1, 2025 are used for training and validation, and
surfaces from that date onward for testing. The resulting ten datasets contain
approximately 40.1 million option quotes across 70,027 surfaces. Full
preprocessing and dataset statistics are provided in
Appendix~\ref{appendix:data}.

\textbf{Evaluation metrics.}
We report IV-MAE (\ref{app:metrics}) over all reconstruction targets (\emph{overall}) and over the
liquid region $\{|k|\le0.1,\ \mathrm{ttm}\ge14\text{d}\}$
(\emph{reliable}), with mean and standard deviation over five seeds for
amortized methods. Static arbitrage is evaluated for every method using the same
finite-difference grid. We report violations of calendar monotonicity
($\partial_\tau w\ge0$) and butterfly positivity ($g(k,\tau)\ge0$), where
$w=\sigma^2\tau$. Metric definitions are given in
Appendix~\ref{app:metrics}.

\textbf{Baselines.}
We compare \addmart{} with the amortized few-shot models, including HyperIV
\citep{hyperiv2025}, GNO \citep{li2020neural}, and VAE
\citep{kingma2014auto}, and with the per-surface methods, including SSVI
\citep{gatheral2014arbitrage} and Deep Smoothing
\citep{ackerer2020deep}. All amortized methods use the same $9$
reference quotes, out-of-time split, and five random seeds. Implementation and
training details are reported in Appendix~\ref{app:baselines}.

\subsection{Reconstruction Performance}

Tables~\ref{tab:mae} and~\ref{tab:arb} summarize reconstruction accuracy and static-arbitrage violations under the unified evaluation protocol. All amortized methods are evaluated over five seeds, whereas SSVI and Deep Smoothing are fitted independently to each test surface. On the daily data, \addmart{} achieves the lowest overall IV-MAE on six of the seven assets, with HyperIV performing slightly better only on NVDA. The reliable-region comparison is closer: \addmart{} leads on SPY, QQQ, and IWM, while HyperIV is marginally better on the four single names. Consequently, the asset-equal daily averages favor \addmart{} for overall reconstruction ($1.65$ versus $1.72$ for HyperIV), while HyperIV has a small advantage in the reliable region ($1.27$ versus $1.29$). At the minute scale, \addmart{} is best on both metrics for SPY and IWM, whereas GNO performs best on QQQ. Averaged across the three minute-scale assets, \addmart{} again gives the lowest reliable and overall errors, at $0.61$ and $1.08$, respectively. The seed-level results also show that \addmart{} is substantially more stable than GNO, whose large standard deviations on SPY and IWM indicate frequent optimization instability.

% \vspace{-1em}

The arbitrage audit provides a sharper distinction between the methods. On a common $801\times30$ full-support price grid, \addmart{} produces no calendar or butterfly violations beyond numerical precision across any asset or scale. SSVI also attains zero violations, but requires a separate optimization for every surface. Deep Smoothing remains close to arbitrage-free in pairwise terms, although small butterfly violations persist. In contrast, the amortized neural baselines exhibit material violations: on the daily assets, HyperIV violates $0.60$--$2.62\%$ of calendar pairs and $1.63$--$8.17\%$ of butterfly pairs, while GNO reaches calendar-violation rates as high as $22.28\%$. The violations become more pronounced at the minute scale, particularly for GNO and VAE. Thus, among methods that reconstruct a surface in a single amortized forward pass, \addmart{} provides the strongest overall accuracy while uniquely preserving the no-arbitrage conditions to numerical precision.

\vspace{-1em}
\input{tables/tab_acc}

\vspace{-1em}
\input{tables/tab_arb}

\subsection{Downstream Distribution Validity.}
\begin{figure}[h]
    \centering
    \includegraphics[width=\linewidth]{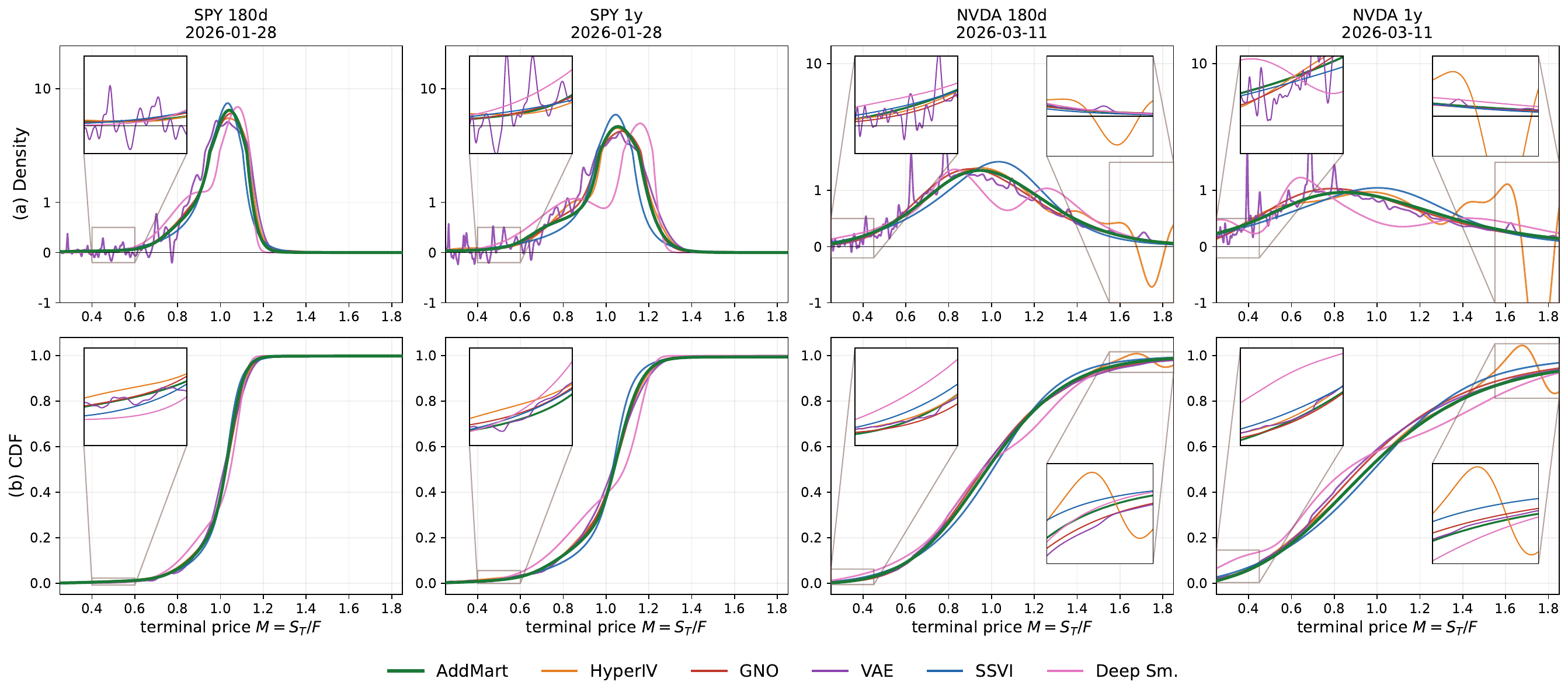}
    \caption{\textbf{Risk-neutral densities and CDFs.} AddMart guarantees valid densities and monotone CDFs, while IV prediction-based baselines may violate these properties, particularly for one-year NVDA. Insets highlight the tails.}
    \label{fig:density}
    \vspace{-1em}
\end{figure}

Figure~\ref{fig:density} examines whether the reconstructed surfaces imply
coherent risk-neutral distributions. The upper row shows the
Breeden--Litzenberger density and the lower row its CDF, with insets
highlighting small violations that are obscured at the full-distribution
scale. Across the representative SPY and NVDA surfaces, \addmart{} produces
non-negative densities and monotone CDFs by construction. For the baselines that directly predict or fit the IV surface, second differentiation amplifies local curvature errors in the
reconstructed surface, producing negative density and locally decreasing
CDFs. The effect is particularly pronounced for the one-year NVDA example,
where HyperIV has $21.8\%$ integrated negative mass despite competitive
IV reconstruction accuracy.

Distributional validity also affects downstream risk estimates.
Appendix~\ref{app:tail_risk} evaluates option-implied downside CVaR against a
full-information reference fitted using all available quotes. \addmart{}
achieves the lowest CVaR error in five of the six asset--horizon settings,
while its predicted marginals remain valid in every evaluated case. The
baseline distributions frequently require ex-post clipping and
renormalization before CVaR can be computed, making the resulting tail
estimates dependent on the repair procedure. These quantities describe
risk-neutral tail exposure implied by option prices and should not be
interpreted as forecasts of realized losses.

\vspace{-1em}

\vspace{-1em}

\begin{table}[h]
\caption{Key ablations on SPY and NVDA daily
(mean$\pm$std over five seeds). IV-MAE is reported in
$\times10^{-2}$. All mixture and training-loss variants use the exact
linear decoder and have zero calendar and butterfly violations.}
\label{tab:ablation}
\centering
\footnotesize

\begin{minipage}[t]{0.39\linewidth}
\vspace{0pt}
\centering
\textbf{(a) Convolution}
\vspace{2pt}

\setlength{\tabcolsep}{3.5pt}
\begin{tabular}{@{}l cc@{}}
\toprule
Variant & SPY & NVDA \\
\midrule
\multicolumn{3}{l}{\emph{Reliable IV-MAE}} \\
Linear exact & $0.56{\pm}0.04$ & $1.84{\pm}0.06$ \\
Linear crop  & $0.56{\pm}0.02$ & $1.82{\pm}0.06$ \\
Circular     & $\mathbf{0.55{\pm}0.04}$ &
               $\mathbf{1.76{\pm}0.06}$ \\
\addlinespace[2pt]
\midrule
\multicolumn{3}{l}{\emph{Calendar violation (\%)}} \\
Linear exact & $\mathbf{0.000}$ & $\mathbf{0.000}$ \\
Linear crop  & $3.855$ & $0.023$ \\
Circular     & $3.652$ & $0.070$ \\
\bottomrule
\end{tabular}
\end{minipage}
\hfill
\begin{minipage}[t]{0.58\linewidth}
\vspace{0pt}
\centering
\textbf{(b) Architecture and training}
\vspace{2pt}

\setlength{\tabcolsep}{4pt}
\begin{tabular}{@{}ll cc@{}}
\toprule
Group & Variant & SPY & NVDA \\
\midrule
\multirow{5}{*}{Mixture $n_{\mathrm{inc}}$}
 & $1$ & $2.99{\pm}0.15$ & $2.11{\pm}0.13$ \\
 & $2$ & $0.67{\pm}0.05$ & $\mathbf{1.74{\pm}0.07}$ \\
 & $4$ & $\mathbf{0.55{\pm}0.05}$ & $1.75{\pm}0.11$ \\
 & $8$ (default) & $0.56{\pm}0.04$ & $1.84{\pm}0.06$ \\
 & $16$ & $0.56{\pm}0.03$ & $1.75{\pm}0.09$ \\
\midrule
\multirow{3}{*}{Training loss}
 & Implicit IV (default)
 & $0.56{\pm}0.04$ & $\mathbf{1.84{\pm}0.06}$ \\
 & Newton IV
 & $\mathbf{0.53{\pm}0.02}$ & $4.91{\pm}0.35$ \\
 & Price
 & $0.56{\pm}0.02$ & $1.88{\pm}0.09$ \\
\bottomrule
\end{tabular}
\end{minipage}
\end{table}

\subsection{Further Analysis}

\textbf{Ablations.}
Table~\ref{tab:ablation} examines the main architectural and training choices on the SPY and NVDA daily datasets. The convolution scheme primarily affects the structural guarantee rather than reconstruction accuracy. The exact linear decoder achieves zero calendar violations on both assets, while cropped and circular convolution~\citep{pfister2017discrete} introduce measurable violations despite similar reliable-region IV-MAE. In particular, circular convolution attains slightly lower IV-MAE than the exact decoder, but produces calendar-violation rates of $3.652\%$ on SPY and $0.070\%$ on NVDA. This comparison confirms that the no-arbitrage property follows from the full-support linear construction rather than from reconstruction accuracy alone.
The mixture ablation shows that a single component is insufficient, especially on SPY, whereas using two or more components substantially improves accuracy. Increasing the mixture size beyond this point yields only modest and asset-dependent changes. The default $n_{\mathrm{inc}}{=}8$ therefore provides a practical capacity choice without relying on fine-grained tuning. The training-loss comparison shows a similar pattern. Direct price supervision performs comparably to the implicit-IV objective, while Newton-based IV inversion improves SPY accuracy but becomes unstable on NVDA. We therefore use the implicit-IV objective as the default because it provides consistently strong performance across both assets.
% \vspace{-1em}

\begin{wrapfigure}{r}{0.4\textwidth}
\vspace{-2em}
\centering
\includegraphics[width=\linewidth]{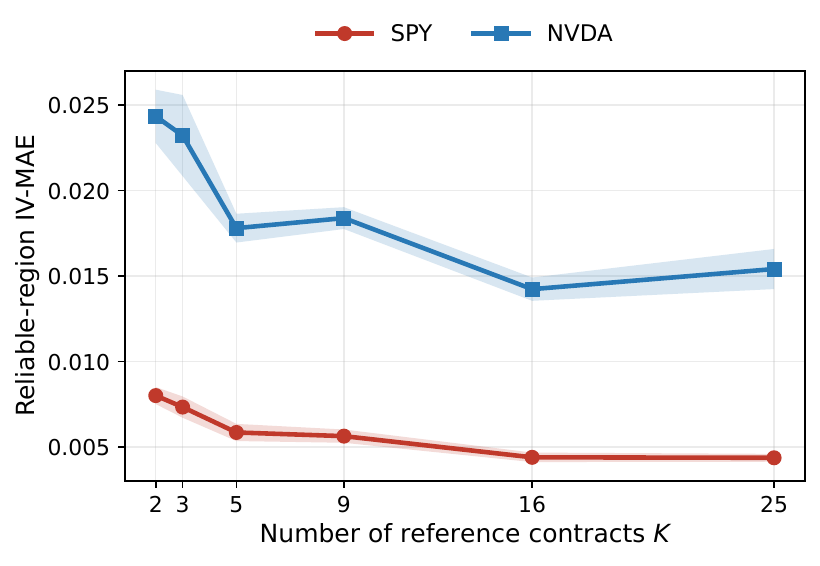}
\caption{Effect of reference-set size on \addmart{} for SPY and NVDA on daily data. Lines and shaded regions report the mean and standard deviation of reliable-region IV-MAE over five seeds.}
\label{fig:kref}
\vspace{-1em}
\end{wrapfigure}

\textbf{Analysis of Reference Budget.}
We examine how \addmart{} uses increasingly informative reference sets by
varying the number of observed contracts
$K\in\{2,3,5,9,16,25\}$. Each setting uses the same architecture, is trained
over five seeds, and is evaluated on an identical set of non-reference
targets. As shown in Figure~\ref{fig:kref}, additional references generally
improve reconstruction accuracy, although the gains are asset-dependent and
not strictly monotone: SPY approaches saturation between $K=16$ and $K=25$,
whereas NVDA performs best at $K=16$. This suggests that both reference-set
size and quote placement affect performance. We use $K=9$ in the main
experiments as a practical balance between accuracy and sparse-market
applicability, with a structured $3$-maturity $\times$ $3$-delta design.

% \textbf{Temporal Robustness.} A model trained once is deployed for months; both accuracy and safety should not decay. We train once
% on SPX 2022-03\,--\,2025-02 and evaluate per month over the following 16 consecutive months
% (Figure~\ref{fig:decay}). Accuracy barely decays for \addmart{} (slope $+0.0003$/month, flat)
% and is the lowest reliable-MAE in almost every month; aggregated over the 16 months,
% \addmart{} $0.0091\ll$ GNO $0.0147\ll$ HyperIV $0.0276$, the latter about $2\times$ its one-month-ahead
% error. The April-2025 regime shock spiked HyperIV to $0.085$ and GNO to $0.033$ but \addmart{} only to
% $0.010$, consistent with the constraint regularizing against drift. Safety does not decay: \addmart{}'s
% butterfly-arbitrage stays $\approx0$ (mean $0.002\%$) across all 16 months, while HyperIV's is
% persistently $0.6$--$0.9\%$ ($\sim$300$\times$ higher) and does not diminish as the surface drifts,
% since its soft penalty only held near the training distribution. Because \addmart{}'s no-arbitrage is by
% construction, it is invariant to deployment time: the model is trained once and remains arbitrage-free
% over the full 16-month period.

\textbf{Cross-Asset Performance.}
We evaluate zero-shot transfer by applying each source-asset checkpoint to
the other daily assets, using only the target surface's nine reference
quotes and no retraining. Across the 42 directed source--target pairs,
\addmart{} attains the lowest overall mean MAE ($5.51\times10^{-2}$) and
outperforms HyperIV on $76.2\%$ of the pairs, while HyperIV is stronger in
the reliable region ($4.81$ versus $5.33\times10^{-2}$). Transfer is most
effective between related broad-market ETFs and deteriorates, often
asymmetrically, when volatility levels and smile shapes differ substantially.
These results indicate that \addmart{} learns transferable surface structure,
although its representation is not fully asset-invariant. The complete
protocol, summary statistics, and directed transfer matrices are reported in
Appendix~\ref{sec:cross_asset}.

%% file: tables/tab_acc.tex
\begin{table}[h]
\caption{Reliable and overall IV-MAE ($\times10^{-2}$, mean$\pm$std over 5 seeds). Lowest per metric within each asset row in bold. $^\ddagger$SSVI and Deep Smoothing are per-surface fits (one optimization per surface, not amortized over a training seed), reported as a single value.}
\label{tab:mae}
\centering\small\setlength{\tabcolsep}{3pt}
\resizebox{\linewidth}{!}{%
\begin{tabular}{@{}ll cc cc cc cc cc cc @{}}
\toprule
& & \multicolumn{2}{c}{\addmart{}} & \multicolumn{2}{c}{HyperIV} & \multicolumn{2}{c}{GNO} & \multicolumn{2}{c}{VAE} & \multicolumn{2}{c}{SSVI$^\ddagger$} & \multicolumn{2}{c}{Deep Sm.$^\ddagger$} \\
\cmidrule(lr){3-4}\cmidrule(lr){5-6}\cmidrule(lr){7-8}
\cmidrule(lr){9-10}\cmidrule(lr){11-12}\cmidrule(lr){13-14}
Scale & Asset
& Reliable & Overall
& Reliable & Overall
& Reliable & Overall
& Reliable & Overall
& Reliable & Overall
& Reliable & Overall \\
\midrule
\multirow{7}{*}{Daily}
& SPY
 & $\mathbf{0.56}{\scriptscriptstyle\,\pm0.02}$
 & $\mathbf{0.89}{\scriptscriptstyle\,\pm0.05}$
 & $0.57{\scriptscriptstyle\,\pm0.02}$
 & $0.99{\scriptscriptstyle\,\pm0.03}$
 & $0.62{\scriptscriptstyle\,\pm0.05}$
 & $0.95{\scriptscriptstyle\,\pm0.07}$
 & $1.33{\scriptscriptstyle\,\pm0.15}$
 & $1.75{\scriptscriptstyle\,\pm0.11}$
 & $1.29$ & $2.24$ & $1.07$ & $2.23$ \\

 & QQQ
 & $\mathbf{0.68}{\scriptscriptstyle\,\pm0.03}$
 & $\mathbf{1.25}{\scriptscriptstyle\,\pm0.05}$
 & $0.74{\scriptscriptstyle\,\pm0.03}$
 & $1.31{\scriptscriptstyle\,\pm0.04}$
 & $0.90{\scriptscriptstyle\,\pm0.10}$
 & $1.45{\scriptscriptstyle\,\pm0.13}$
 & $1.30{\scriptscriptstyle\,\pm0.15}$
 & $1.98{\scriptscriptstyle\,\pm0.07}$
 & $1.20$ & $2.19$ & $1.11$ & $2.57$ \\

 & IWM
 & $\mathbf{0.77}{\scriptscriptstyle\,\pm0.03}$
 & $\mathbf{1.17}{\scriptscriptstyle\,\pm0.03}$
 & $0.78{\scriptscriptstyle\,\pm0.08}$
 & $1.27{\scriptscriptstyle\,\pm0.09}$
 & $1.09{\scriptscriptstyle\,\pm0.53}$
 & $1.77{\scriptscriptstyle\,\pm0.94}$
 & $1.19{\scriptscriptstyle\,\pm0.12}$
 & $1.93{\scriptscriptstyle\,\pm0.14}$
 & $1.16$ & $1.95$ & $1.20$ & $2.57$ \\

 & AAPL
 & $\mathbf{1.18}{\scriptscriptstyle\,\pm0.07}$
 & $\mathbf{1.40}{\scriptscriptstyle\,\pm0.05}$
 & $\mathbf{1.18}{\scriptscriptstyle\,\pm0.07}$
 & $1.55{\scriptscriptstyle\,\pm0.08}$
 & $1.52{\scriptscriptstyle\,\pm0.25}$
 & $1.77{\scriptscriptstyle\,\pm0.31}$
 & $1.76{\scriptscriptstyle\,\pm0.04}$
 & $2.37{\scriptscriptstyle\,\pm0.10}$
 & $2.08$ & $3.14$ & $2.90$ & $5.11$ \\

 & GOOGL
 & $2.22{\scriptscriptstyle\,\pm0.14}$
 & $\mathbf{2.37}{\scriptscriptstyle\,\pm0.11}$
 & $\mathbf{2.07}{\scriptscriptstyle\,\pm0.25}$
 & $2.43{\scriptscriptstyle\,\pm0.24}$
 & $2.59{\scriptscriptstyle\,\pm0.32}$
 & $2.76{\scriptscriptstyle\,\pm0.24}$
 & $3.12{\scriptscriptstyle\,\pm0.09}$
 & $3.42{\scriptscriptstyle\,\pm0.10}$
 & $3.21$ & $4.01$ & $5.87$ & $8.10$ \\

 & META
 & $1.82{\scriptscriptstyle\,\pm0.16}$
 & $\mathbf{1.95}{\scriptscriptstyle\,\pm0.12}$
 & $\mathbf{1.79}{\scriptscriptstyle\,\pm0.07}$
 & $2.03{\scriptscriptstyle\,\pm0.08}$
 & $2.05{\scriptscriptstyle\,\pm0.22}$
 & $2.30{\scriptscriptstyle\,\pm0.26}$
 & $2.29{\scriptscriptstyle\,\pm0.04}$
 & $2.73{\scriptscriptstyle\,\pm0.10}$
 & $3.69$ & $4.46$ & $6.80$ & $10.40$ \\

 & NVDA
 & $\mathbf{1.80}{\scriptscriptstyle\,\pm0.11}$
 & $2.54{\scriptscriptstyle\,\pm0.07}$
 & $\mathbf{1.80}{\scriptscriptstyle\,\pm0.11}$
 & $\mathbf{2.46}{\scriptscriptstyle\,\pm0.12}$
 & $2.66{\scriptscriptstyle\,\pm0.65}$
 & $3.08{\scriptscriptstyle\,\pm0.55}$
 & $3.31{\scriptscriptstyle\,\pm0.29}$
 & $3.50{\scriptscriptstyle\,\pm0.27}$
 & $3.83$ & $4.97$ & $5.19$ & $7.72$ \\

\midrule
\multirow{3}{*}{Minute}
 & SPY
 & $0.56{\scriptscriptstyle\,\pm0.03}$
 & $\mathbf{0.92}{\scriptscriptstyle\,\pm0.04}$
 & $0.57{\scriptscriptstyle\,\pm0.03}$
 & $0.96{\scriptscriptstyle\,\pm0.04}$
 & $\mathbf{0.55}{\scriptscriptstyle\,\pm0.08}$
 & $0.95{\scriptscriptstyle\,\pm0.16}$
 & $1.13{\scriptscriptstyle\,\pm0.01}$
 & $1.46{\scriptscriptstyle\,\pm0.03}$
 & $1.25$ & $2.10$ & $1.07$ & $2.07$ \\

 & QQQ
 & $0.65{\scriptscriptstyle\,\pm0.07}$
 & $1.23{\scriptscriptstyle\,\pm0.07}$
 & $0.66{\scriptscriptstyle\,\pm0.04}$
 & $1.20{\scriptscriptstyle\,\pm0.06}$
 & $\mathbf{0.59}{\scriptscriptstyle\,\pm0.05}$
 & $\mathbf{1.01}{\scriptscriptstyle\,\pm0.06}$
 & $1.13{\scriptscriptstyle\,\pm0.01}$
 & $1.69{\scriptscriptstyle\,\pm0.03}$
 & $1.20$ & $2.29$ & $1.06$ & $2.44$ \\

 & IWM
 & $\mathbf{0.61}{\scriptscriptstyle\,\pm0.04}$
 & $\mathbf{1.09}{\scriptscriptstyle\,\pm0.05}$
 & $0.63{\scriptscriptstyle\,\pm0.01}$
 & $1.11{\scriptscriptstyle\,\pm0.03}$
 & $0.77{\scriptscriptstyle\,\pm0.15}$
 & $1.31{\scriptscriptstyle\,\pm0.28}$
 & $1.00{\scriptscriptstyle\,\pm0.01}$
 & $1.51{\scriptscriptstyle\,\pm0.00}$
 & $1.19$ & $2.02$ & $1.16$ & $2.33$ \\
\bottomrule
\end{tabular}}
\end{table}

%% file: tables/tab_arb.tex
\begin{table}[h]
\caption{Calendar and butterfly arbitrage-violation pair rates (\%) on the common $801\times30$ full-support price grid, using a numerical tolerance of $10^{-10}$. Amortized methods are pooled over 5 seeds; $^\ddagger$SSVI and Deep Smoothing are per-surface fits and are reported as a single value. Lowest per metric within each asset row in bold.}
\label{tab:arb}
\centering\small\setlength{\tabcolsep}{2pt}
\resizebox{\linewidth}{!}{%
\begin{tabular}{@{}ll cc cc cc cc cc cc@{}}
\toprule
& & \multicolumn{2}{c}{\addmart{}}
& \multicolumn{2}{c}{HyperIV}
& \multicolumn{2}{c}{GNO}
& \multicolumn{2}{c}{VAE}
& \multicolumn{2}{c}{SSVI$^\ddagger$}
& \multicolumn{2}{c}{Deep Sm.$^\ddagger$} \\
\cmidrule(lr){3-4}\cmidrule(lr){5-6}\cmidrule(lr){7-8}
\cmidrule(lr){9-10}\cmidrule(lr){11-12}\cmidrule(lr){13-14}
Scale & Asset
& Calendar & Butterfly
& Calendar & Butterfly
& Calendar & Butterfly
& Calendar & Butterfly
& Calendar & Butterfly
& Calendar & Butterfly \\
\midrule
\multirow{7}{*}{Daily}
 & SPY
 & $\mathbf{0}$ & $\mathbf{0}$
 & $0.7310$ & $1.6295$
 & $15.6910$ & $4.4026$
 & $0.0546$ & $8.4843$
 & $\mathbf{0}$ & $\mathbf{0}$
 & $\mathbf{0}$ & $0.0113$ \\

 & QQQ
 & $\mathbf{0}$ & $\mathbf{0}$
 & $1.0047$ & $2.3451$
 & $22.2821$ & $5.2844$
 & $0.2145$ & $7.8386$
 & $\mathbf{0}$ & $\mathbf{0}$
 & $\mathbf{0}$ & $0.0222$ \\

 & IWM
 & $\mathbf{0}$ & $\mathbf{0}$
 & $0.6008$ & $4.8701$
 & $18.8903$ & $7.6170$
 & $0.3043$ & $6.8536$
 & $\mathbf{0}$ & $\mathbf{0}$
 & $\mathbf{0}$ & $0.0642$ \\

 & AAPL
 & $\mathbf{0}$ & $\mathbf{0}$
 & $1.0690$ & $6.6294$
 & $20.6356$ & $5.0270$
 & $0.0785$ & $5.0351$
 & $\mathbf{0}$ & $\mathbf{0}$
 & $\mathbf{0}$ & $0.0160$ \\

 & GOOGL
 & $\mathbf{0}$ & $\mathbf{0}$
 & $0.9958$ & $7.9404$
 & $14.6969$ & $3.4538$
 & $0.2408$ & $5.5603$
 & $\mathbf{0}$ & $\mathbf{0}$
 & $0.0051$ & $0.0195$ \\

 & META
 & $\mathbf{0}$ & $\mathbf{0}$
 & $2.6205$ & $8.1659$
 & $11.6960$ & $3.5165$
 & $0.0906$ & $4.4891$
 & $\mathbf{0}$ & $\mathbf{0}$
 & $\mathbf{0}$ & $0.0108$ \\

 & NVDA
 & $\mathbf{0}$ & $\mathbf{0}$
 & $1.6370$ & $7.8105$
 & $7.2435$ & $2.3606$
 & $0.3796$ & $2.9705$
 & $\mathbf{0}$ & $\mathbf{0}$
 & $\mathbf{0}$ & $\mathbf{0}$ \\

\midrule
\multirow{3}{*}{Minute}
 & SPY
 & $\mathbf{0}$ & $\mathbf{0}$
 & $0.3779$ & $1.9884$
 & $22.4997$ & $8.7293$
 & $21.0067$ & $18.5972$
 & $\mathbf{0}$ & $\mathbf{0}$
 & $0.0053$ & $0.0336$ \\

 & QQQ
 & $\mathbf{0}$ & $\mathbf{0}$
 & $1.5006$ & $6.1951$
 & $25.4820$ & $8.7914$
 & $20.2924$ & $13.6071$
 & $\mathbf{0}$ & $\mathbf{0}$
 & $0.0162$ & $0.0274$ \\

 & IWM
 & $\mathbf{0}$ & $\mathbf{0}$
 & $0.7306$ & $4.2616$
 & $23.2808$ & $8.3368$
 & $18.2714$ & $13.4086$
 & $\mathbf{0}$ & $\mathbf{0}$
 & $\mathbf{0}$ & $0.0745$ \\
\bottomrule
\end{tabular}}
\end{table}

%% file: appendix/background.tex
\section*{Ethics Statement}
This study uses option-market quotes rather than human-subject or individual-level data. The proposed model produces implied-volatility surfaces and risk-neutral distributions; these are model-based quantities, not forecasts of realized returns. Although the construction enforces static no-arbitrage under its stated assumptions, this property does not eliminate market-data errors, model risk, or uncertainty in downstream decisions. The outputs should therefore not be treated as standalone investment or risk-management advice.

\section*{Reproducibility Statement}
The paper specifies the model architecture, increment-law parameterization, martingale decoder, training objective, data preprocessing, reference-quote selection, chronological data split, baselines, and evaluation metrics in the main text and appendices. Amortized methods are evaluated with five independent random seeds. The experiments use OPRA option quotes obtained through a commercial data provider; access to the underlying data is subject to the provider's licensing terms. These details are intended to support replication by researchers with access to appropriately licensed data.

\section{Notation}
\label{app:notation}

Table~\ref{tab:notation} summarizes the notation used throughout the paper.
We use lowercase letters for scalar quantities, bold lowercase letters for
vectors, calligraphic letters for sets, and uppercase blackboard-bold letters
for probability measures.

\input{tables/notations}

\section{Extended Background on Options, Volatility, and Arbitrage}
\label{sec:appendix_arbitrage}

\subsection{Options and the Black Scholes Merton Model}
\label{subsec:app_options_bsm}

A European call option is a foundational financial contract. It provides the buyer with the right, but absolutely no obligation, to purchase a specific underlying asset (such as a stock or an index) at a predetermined price called the strike price $K$, on a specific future date called the maturity date $T$. Let $t$ denote the current time. The duration remaining until the option expires is known as the time to maturity, denoted by $\tau = T - t$. Because this contract limits downside risk while preserving upside potential, it carries a premium (the option price) that must be paid upfront.

Under the standard option pricing framework \citep{black1973pricing,merton1971theory}, the theoretical price of a European call option can be computed if we assume geometric Brownian motion for the underlying asset. Assuming a constant risk free interest rate $r$, a continuous dividend yield $q$, and a constant volatility $\sigma$ representing the standard deviation of the asset returns, the price of a call option is given by the Black Scholes Merton formula:
$$C_{\mathrm{BSM}}(S_t, K, \tau, r, q, \sigma) = S_t e^{-q\tau} \Phi(d_1) - K e^{-r\tau} \Phi(d_2)$$
where $\Phi(\cdot)$ is the cumulative distribution function of the standard normal distribution. The terms $d_1$ and $d_2$ are explicitly defined as:
$$d_1 = \frac{\ln(S_t/K) + (r - q + \frac{1}{2}\sigma^2)\tau}{\sigma\sqrt{\tau}}$$
$$d_2 = d_1 - \sigma\sqrt{\tau}$$

While the Black Scholes Merton mathematical model assumes that volatility $\sigma$ is a constant parameter, real financial markets behave differently. Empirical market prices systematically deviate from this constant volatility assumption, demonstrating that the market prices different strike prices and different maturity dates with different effective volatilities.

\subsection{Implied Volatility and the Volatility Surface}
\label{subsec:app_implied_vol_surface}

Because all inputs to the Black Scholes Merton formula---the current stock price $S_t$, the strike price $K$, the time to maturity $\tau$, the risk free rate $r$, and the dividend yield $q$---are directly observable in the market, practitioners use the formula in reverse. They treat the model as a universal mapping function rather than an absolute pricing truth.

Given an observable market price of a call option $C_{\mathrm{mkt}}$, the \emph{implied volatility}, denoted as $\sigma_{\mathrm{imp}}$, is defined as the single unique scalar value that forces the theoretical model price to exactly equal the observed market price:
$$C_{\mathrm{mkt}} = C_{\mathrm{BSM}}(S_t, K, \tau, r, q, \sigma_{\mathrm{imp}})$$
Because the option price strictly increases as volatility increases (a property known as positive vega, mathematically $\partial C_{\mathrm{BSM}} / \partial \sigma > 0$), this mapping from price to implied volatility is unique and universally invertible. Quoting implied volatility instead of raw dollar prices allows traders to compare options across different underlying assets and different price levels using a standardized metric.

To further standardize the representation of these options, it is standard market practice to map the absolute strike price $K$ into a relative measure called \emph{log moneyness}, denoted by $k$. Let $F(\tau) = S_t e^{(r-q)\tau}$ denote the forward price of the underlying asset at maturity $\tau$. The log moneyness is defined as:
$$k = \ln\left(\frac{K}{F(\tau)}\right)$$
When $k=0$, the option is termed "at the money". When we compute the implied volatility across a continuous range of log moneyness $k$ and maturities $\tau$, we construct a two dimensional mathematical function $\mathrm{IV}(k, \tau)$. This continuous geometric map is widely known as the \emph{Implied Volatility Surface}. It acts as the central pricing tool used by financial institutions to price exotic derivatives, hedge complex portfolios, and calculate future market risk scenarios.

\subsection{Static Arbitrage and the Risk Neutral Density}
\label{subsec:app_static_no_arbitrage}

For an implied volatility surface $\mathrm{IV}(k, \tau)$ to be theoretically valid and economically tradable, it must be strictly free of static arbitrage. In finance, an arbitrage is a theoretical mechanism that guarantees a riskless profit with zero net investment. If a model generates a surface containing arbitrage opportunities, it produces mathematically absurd outputs, such as negative probability values. 

According to the celebrated Breeden Litzenberger theorem \citep{breeden1978prices}, the second derivative of the undiscounted call price with respect to the strike price directly yields the market probability mass function, commonly referred to as the risk neutral density. 

Complete absence of static arbitrage requires the surface to rigorously satisfy two primary geometric conditions:

\begin{enumerate}[leftmargin=12pt]
    \item \textbf{Butterfly No Arbitrage:} A butterfly spread is an option strategy involving three different strike prices. For this strategy to never yield a guaranteed riskless payout, the implied risk neutral density must be nonnegative across all possible strikes. Mathematically, this demands that the call price must be globally convex with respect to the strike price $K$:
    $$\frac{\partial^2 C}{\partial K^2} \geq 0$$
    If this second derivative were negative in any region, it would imply that the market assigns a negative probability to the asset reaching that specific price range, which is fundamentally impossible.

    \item \textbf{Calendar No Arbitrage:} A calendar spread involves options with the same strike price but different maturity dates. Because an option with a longer time to maturity provides the holder with more time for the asset to move favorably, it must be at least as valuable as an option with a shorter maturity. Therefore, the forward option prices must not decrease as the time to maturity increases:
    $$\frac{\partial C}{\partial \tau} \geq 0$$
    A violation here would imply that time carries negative value.
\end{enumerate}

When translating these concepts into forward normalized spaces, preserving calendar no arbitrage is exactly equivalent to requiring that the sequence of asset distributions over time increases in \emph{convex order}. In probability theory, a sequence of distributions increasing in convex order forms a martingale structure known as a \emph{peacock} (from the French \emph{processus croissant pour l'ordre convexe}). Therefore, explicitly modeling the implied volatility surface through the lens of additive martingales strictly guarantees that both the butterfly and calendar no arbitrage conditions are satisfied by design.

%% file: tables/notations.tex
\renewcommand{\arraystretch}{1.15}
\begin{longtable}{
@{}
>{\raggedright\arraybackslash}p{0.21\linewidth}
>{\raggedright\arraybackslash}p{0.18\linewidth}
>{\raggedright\arraybackslash}p{0.53\linewidth}
@{}
}
\caption{Notation used throughout the paper.}
\label{tab:notation}\\
\toprule
Symbol & Domain or type & Description \\
\midrule
\endfirsthead

\multicolumn{3}{c}{\tablename\ \thetable\ -- continued} \\
\toprule
Symbol & Domain or type & Description \\
\midrule
\endhead

\midrule
\multicolumn{3}{r}{Continued on next page} \\
\endfoot

\bottomrule
\endlastfoot

\multicolumn{3}{l}{\textit{Option market and pricing}} \\

$B(k,s_{\mathrm{tot}})$ & $[0,1]$ &
Forward-normalized Black call price. \\

$C(k,\tau)$ & $[0,1]$ &
Call price in log-moneyness coordinates,
$C(k,\tau)=c(e^k,\tau)$. \\

$c(z,\tau)$ & $[0,1]$ &
Forward-normalized undiscounted call price,
$c(z,\tau)=\mathbb{E}_{\mathbb Q}[(M_\tau-z)^+]$. \\

$d_{+},d_{-}$ & $\mathbb{R}$ &
Black variables,
$d_{+}=-k/s_{\mathrm{tot}}+s_{\mathrm{tot}}/2$ and
$d_{-}=-k/s_{\mathrm{tot}}-s_{\mathrm{tot}}/2$. \\

$\mathrm{DF}(\tau)$ & $(0,1]$ &
Discount factor for maturity $\tau$, defined as
$\mathrm{DF}(\tau)=\exp[-r(\tau)\tau]$. \\

$\mathbb{E}_{\mathbb Q},
 \mathbb{P}_{\mathbb Q}$ & -- &
Expectation and probability under the pricing measure $\mathbb Q$. \\

$F(\tau)$ & $\mathbb{R}_{+}$ &
Parity-implied forward price of the underlying for maturity $\tau$. \\

$k$ & $\mathbb{R}$ &
Log-moneyness, defined as $k=\log(K/F(\tau))$. \\

$K$ & $\mathbb{R}_{+}$ &
Option strike price. \\

$M_\tau$ & $\mathbb{R}_{+}$ &
Forward-normalized terminal price,
$M_\tau=S_\tau/F(\tau)$, with
$\mathbb{E}_{\mathbb Q}[M_\tau]=1$. \\

$\Phi,\phi$ & -- &
CDF and density of the standard normal distribution. \\

$q$ & $\mathbb{R}$ &
Continuous dividend yield or implied carry. \\

$\mathbb Q$ & -- &
Pricing measure used for normalized option valuation. \\

$r(\tau)$ & $\mathbb{R}$ &
Continuously compounded risk-free rate. \\

$S_t$ & $\mathbb{R}_{+}$ &
Underlying asset price at observation time $t$. \\

$S_\tau$ & $\mathbb{R}_{+}$ &
Underlying asset price at maturity. \\

$s_{\mathrm{tot}}$ & $\mathbb{R}_{+}$ &
Black total volatility,
$s_{\mathrm{tot}}=\sigma\sqrt{\tau}$. \\

$\sigma(k,\tau)$ & $\mathbb{R}_{+}$ &
Black implied volatility. \\

$t$ & $\mathbb{R}_{+}$ &
Current observation time. \\

$T$ & $\mathbb{R}_{+}$ &
Option expiration time. \\

$\tau=T-t$ & $\mathbb{R}_{+}$ &
Time to maturity, measured in years. \\

$w(k,\tau)$ & $\mathbb{R}_{+}$ &
Total implied variance,
$w(k,\tau)=\sigma(k,\tau)^2\tau$. \\

$(x)^+$ & $\mathbb{R}_{+}$ &
Positive part, defined as $(x)^+=\max(x,0)$. \\

$z=e^k$ & $\mathbb{R}_{+}$ &
Normalized strike, equal to $K/F(\tau)$. \\
% have checked?

\addlinespace
\multicolumn{3}{l}{\textit{Additive martingale construction}} \\

$A_{\tau_j}$ & $\mathbb{R}$ &
Cumulative additive state,
$A_{\tau_j}=\sum_{g\leq j}L_g$. \\

$\Delta\tau_g$ & $\mathbb{R}_{+}$ &
Width of maturity bucket $g$,
$\Delta\tau_g=\tau_g-\tau_{g-1}$. \\

$\eta_g$ & $\mathcal{P}(\mathbb{R})$ &
Probability law of the additive increment $L_g$. \\

$G$ & $\mathbb{N}$ &
Number of maturity buckets. \\

$L_g$ & $\mathbb{R}$ &
Independent additive log-increment in bucket $g$. \\

$M_{\tau_j}$ & $\mathbb{R}_{+}$ &
Normalized exponential martingale,
\[
M_{\tau_j}
=
\frac{e^{A_{\tau_j}}}{Z_j}
=
\prod_{g\leq j}Y_g.
\] \\

$\nu_j$ & $\mathcal{P}(\mathbb{R})$ &
Law of $A_{\tau_j}$, given by
$\nu_j=\eta_1*\cdots*\eta_j$. \\

$\lambda(\tau)$ & $[0,1]$ &
Interpolation weight for
$\tau\in(\tau_j,\tau_{j+1})$,
\[
\lambda(\tau)
=
\frac{\tau-\tau_j}{\tau_{j+1}-\tau_j}.
\] \\

$\mathsf{P}_j$ & $\mathcal{P}(\mathbb{R}_{+})$ &
Risk-neutral marginal law of $M_{\tau_j}$, obtained by pushing
$\nu_j$ forward through $x\mapsto e^x/Z_j$. \\

$\mathsf{P}_\tau$ & $\mathcal{P}(\mathbb{R}_{+})$ &
Interpolated marginal,
\[
\mathsf{P}_\tau
=
(1-\lambda)\mathsf{P}_j
+
\lambda\mathsf{P}_{j+1}.
\] \\

$\preceq_{\mathrm{cx}}$ & -- &
Convex order between probability distributions or random variables. \\

$\tau_j$ & $\mathbb{R}_{+}$ &
The $j$-th maturity-grid point, with
$0=\tau_0<\tau_1<\cdots<\tau_G$. \\

$Y_g$ & $\mathbb{R}_{+}$ &
Unit-mean multiplicative increment,
\[
Y_g
=
\frac{e^{L_g}}
{\mathbb{E}_{\mathbb Q}[e^{L_g}]}.
\] \\

$Z_j$ & $\mathbb{R}_{+}$ &
Exponential normalizer,
\[
Z_j
=
\mathbb{E}_{\mathbb Q}[e^{A_{\tau_j}}]
=
\int_{\mathbb R}e^x\nu_j(dx).
\] \\

$(\Omega,\mathcal F,
\{\mathcal F_{\tau_j}\}_{j=0}^{G},
\mathbb Q)$ & -- &
Filtered probability space supporting the additive process. \\

\addlinespace
\multicolumn{3}{l}{\textit{Learned parameterization and grid implementation}} \\

$\mathcal Q$ & finite set &
Sparse reference-quote set used to condition the network. \\

$\theta$ & -- &
Trainable parameters of the set encoder and increment head. \\

$f_{\theta,g}(x\mid\mathcal Q)$ & $\mathbb{R}_{+}$ &
Continuous Gaussian-mixture density predicted for increment $g$. \\

$x_i$ & $\mathbb{R}$ &
The $i$-th point of the base log-return grid. \\

$m_{\theta,g,i}$ & $[0,1]$ &
Normalized probability mass assigned to $x_i$ for increment $g$, with
$\sum_i m_{\theta,g,i}=1$. \\

$\widehat{\eta}_{\theta,g}$ & $\mathcal{P}(\mathbb{R})$ &
Discrete increment law used by the decoder,
\[
\widehat{\eta}_{\theta,g}
=
\sum_i m_{\theta,g,i}\delta_{x_i}.
\] \\

$v_j,\ v_j^\star$ & probability vectors &
Cumulative base and Esscher-tilted log-state mass vectors at
maturity $\tau_j$. \\

$w_{j,i}$ & $[0,1]$ &
Probability mass of the emitted discrete marginal
$\widehat{\mathsf P}_j$ at atom $M_{j,i}$. \\

$R$ & $\mathbb N$ &
Number of Gaussian components in each increment density. \\

$X$ & $\mathbb N$ &
Number of points in the base log-return grid. \\

\end{longtable}

%% file: appendix/proof.tex
\section{Proofs and Additional Theory}
\label{app:theory}

Throughout this section, $\eta_1,\dots,\eta_G$ are probability laws on
$\mathbb R$, $L_g\sim\eta_g$ are independent under $\mathbb Q$ with
$\mathbb E_{\mathbb Q}[e^{L_g}]<\infty$, and
\[
Y_g
:=
\frac{e^{L_g}}{\mathbb E_{\mathbb Q}[e^{L_g}]},
\qquad
M_{\tau_j}
=
\prod_{g\leq j}Y_g,
\qquad
\mathcal F_{\tau_j}
=
\sigma(L_1,\dots,L_j).
\]

\subsection{Proof of Theorem~\ref{thm:static-arbitrage}}
\label{app:proof-arb}

\begin{proof}
Each $Y_g$ is non-negative and satisfies $\mathbb E_{\mathbb Q}[Y_g]=1$. Since
$Y_{j+1}$ is independent of $\mathcal F_{\tau_j}$ and $M_{\tau_j}$ is
$\mathcal F_{\tau_j}$-measurable,
\begin{equation}
\mathbb E_{\mathbb Q}
\left[
M_{\tau_{j+1}}
\mid
\mathcal F_{\tau_j}
\right]
=
M_{\tau_j}\mathbb E_{\mathbb Q}[Y_{j+1}]
=
M_{\tau_j}.
\label{eq:proof-martingale}
\end{equation}
Hence $\{M_{\tau_j}\}$ is a non-negative martingale with unit expectation.

For any convex function $\varphi$ for which the expectations exist,
conditional Jensen's inequality and
Equation~\ref{eq:proof-martingale} give
\[
\varphi(M_{\tau_j})
=
\varphi\!\left(
\mathbb E_{\mathbb Q}[
M_{\tau_{j+1}}\mid\mathcal F_{\tau_j}]
\right)
\leq
\mathbb E_{\mathbb Q}[
\varphi(M_{\tau_{j+1}})
\mid\mathcal F_{\tau_j}].
\]
Taking expectations yields
\begin{equation}
\mathbb E_{\mathbb Q}[\varphi(M_{\tau_j})]
\leq
\mathbb E_{\mathbb Q}[\varphi(M_{\tau_{j+1}})].
\label{eq:proof-convex-order}
\end{equation}
Thus,
$M_{\tau_j}\preceq_{\mathrm{cx}}M_{\tau_{j+1}}$, and the grid-maturity
marginals form a peacock
\citep{kellerer1972markov,hirsch2011peacocks}.

For fixed normalized strike $z>0$, the call payoff
$\varphi_z(m)=(m-z)^+$ is convex in $m$. Applying
Equation~\ref{eq:proof-convex-order} gives
$c(z,\tau_j)\leq c(z,\tau_{j+1})$, which is the calendar condition at grid
maturities.

For fixed $m\geq0$, the map $z\mapsto(m-z)^+$ is non-increasing and convex.
Integration against the marginal law of $M_{\tau_j}$ therefore shows that
$z\mapsto c(z,\tau_j)$ is non-increasing and convex. Its distributional
second derivative is the non-negative marginal measure:
\[
\partial_{zz}^{\,2}c(\cdot,\tau_j)=\mathsf P_j\geq0.
\]
Moreover,
\[
c(0,\tau_j)=\mathbb E_{\mathbb Q}[M_{\tau_j}]=1,
\qquad
\lim_{z\rightarrow\infty}c(z,\tau_j)=0,
\qquad
c(z,\tau_j)\geq(1-z)^+.
\]
The lower bound follows from Jensen's inequality, and the limit follows from
dominated convergence because $(M_{\tau_j}-z)^+\leq M_{\tau_j}$.

For $\tau\in(\tau_j,\tau_{j+1})$, the interpolation defined in
Section~\ref{sec:method-construct} is a convex combination of the endpoint
call curves and therefore remains non-increasing and convex in $z$. Since
$c(z,\tau_{j+1})-c(z,\tau_j)\geq0$, it is also non-decreasing in $\tau$
within each maturity interval. Equivalently, the interpolated marginal
$\mathsf P_\tau=(1-\lambda)\mathsf P_j+\lambda\mathsf P_{j+1}$ is a probability measure with unit
mean.

It remains to verify that the arbitrage-free price admits a BSM IV
representation. For fixed $k$ and positive maturity, write the normalized
Black call price as $B(k,s_{\mathrm{tot}})$, where
$s_{\mathrm{tot}}=\sigma\sqrt{\tau}$ is total volatility. The function
$s_{\mathrm{tot}}\mapsto B(k,s_{\mathrm{tot}})$ is continuous and strictly
increasing because $\partial B/\partial s_{\mathrm{tot}}=\phi(d_+)>0$.
Its limiting values are $B(k,0^+)=(1-e^k)^+$ and $B(k,\infty)=1$. Therefore every interior price
$c\in((1-e^k)^+,1)$ has a unique non-negative implied volatility. Combining
these properties proves butterfly- and calendar-arbitrage-freeness at every
strike and maturity.
\end{proof}

% \subsection{Proof of Corollary~\ref{cor:risk-functionals}}
\subsection{Distributional Risk Functionals}
\label{app:proof-risk}

\begin{proof}
At a grid maturity, let
\[
\widehat{\mathsf P}_j
=
\sum_i w_{j,i}\delta_{M_{j,i}},
\qquad
w_{j,i}\geq0,
\qquad
\sum_i w_{j,i}=1.
\]
Then $\widehat{\mathsf P}_j$ is a finitely supported probability measure. For every
measurable function $\varphi$ that is integrable under $\widehat{\mathsf P}_j$,
\begin{equation}
\mathbb E_{\widehat{\mathsf P}_j}
[\varphi(M_{\tau_j})]
=
\sum_i w_{j,i}\varphi(M_{j,i})
=
\int\varphi\,d\widehat{\mathsf P}_j .
\label{eq:proof-expectation}
\end{equation}

Its CDF and survival function are
\[
F_j(a)
=
\sum_i w_{j,i}\mathbf 1\{M_{j,i}\leq a\},
\qquad
\overline F_j(a)
=
\sum_i w_{j,i}\mathbf 1\{M_{j,i}>a\}.
\]
They are monotone and take values in $[0,1]$. The generalized quantile
\[
F_j^{-1}(\alpha)
=
\inf\{a:F_j(a)\geq\alpha\}
\]
is therefore well defined. Quantile-based VaR and crash probabilities such
$F_j(a)=\sum_iw_{j,i}\mathbf 1\{M_{j,i}\leq a\}$ and
$\overline F_j(a)=\sum_iw_{j,i}\mathbf 1\{M_{j,i}>a\}$, respectively.
They are monotone and take values in $[0,1]$, so the generalized quantile
$F_j^{-1}(\alpha)=\inf\{a:F_j(a)\geq\alpha\}$ is well defined.

For any loss variable $\mathsf L=\ell(M_{\tau_j})$ that is finite on the emitted
atoms, its CVaR has the finite convex representation
\begin{equation}
\operatorname{CVaR}_{p}(L)
=
\min_{c\in\mathbb{R}}
\left\{
c+\frac{1}{1-p}
\sum_i w_{j,i}(L_i-c)^+
\right\},
\qquad
L_i=\ell(M_{\tau_j}).
\label{eq:proof-cvar}
\end{equation}
following \citet{rockafellar2000optimization}. The objective is convex in
$c$ because it is a non-negative weighted sum of convex hinge functions.
Every quantity above is evaluated under one valid probability measure and
is therefore well defined.
\end{proof}

\subsection{Approximation Error of a Cropped Decoder}
\label{app:proof-crop}

The formal \addmart{} decoder uses full-support convolution. A fixed-width
training decoder may instead crop and renormalize intermediate convolutions
to reduce memory use. A bound based only on discarded probability mass is
insufficient because a small amount of far-right mass can have a large
effect after exponentiation. The relevant quantity is an exponentially
weighted measure discrepancy.

\begin{lemma}[Calendar residual under an approximate marginal]
\label{lem:crop-error}
Let $\nu_j$ be the exact law of $A_{\tau_j}$ and let
$\widetilde\nu_j$ be any approximate probability law with finite exponential
moment. Define
\[
Z_j=\int e^x\nu_j(dx),
\qquad
\widetilde Z_j=\int e^x\widetilde\nu_j(dx),
\qquad
\delta_j
=
\frac{1}{Z_j}
\int e^x
\left|
\nu_j-\widetilde\nu_j
\right|(dx),
\]
where $|\nu_j-\widetilde\nu_j|$ denotes the total-variation measure of the
signed difference. Let $c_j$ and $\widetilde c_j$ be the corresponding
forward-normalized call prices, each using its own exponential normalizer.
Then, uniformly over normalized strike $z$,
\begin{equation}
\left|
c_j(z)-\widetilde c_j(z)
\right|
\leq
2\delta_j.
\label{eq:crop-price-error}
\end{equation}
Consequently,
\[
\widetilde c_{j+1}(z)-\widetilde c_j(z)
\geq
-2(\delta_j+\delta_{j+1}).
\]
\end{lemma}

\begin{proof}
Define $g_z(m)=(m-z)^+$,
$\psi_Z(x)=g_z(e^x/Z_j)$, and
$\psi_{\widetilde Z}(x)=g_z(e^x/\widetilde Z_j)$. Since
$0\leq \psi_Z(x)\leq e^x/Z_j$,
\[
\left|
\int \psi_Z\,d\nu_j
-
\int \psi_Z\,d\widetilde\nu_j
\right|
\leq
\frac{1}{Z_j}
\int e^x
\left|
\nu_j-\widetilde\nu_j
\right|(dx)
=
\delta_j.
\]
The hinge function is one-Lipschitz, so
\[
\left|
\psi_Z(x)-\psi_{\widetilde Z}(x)
\right|
\leq
e^x
\left|
\frac{1}{Z_j}
-
\frac{1}{\widetilde Z_j}
\right|.
\]
Integrating under $\widetilde\nu_j$ gives
\[
\int
\left|
\psi_Z-\psi_{\widetilde Z}
\right|
d\widetilde\nu_j
\leq
\frac{|Z_j-\widetilde Z_j|}{Z_j}
\leq
\delta_j.
\]
The triangle inequality proves
Equation~\ref{eq:crop-price-error}. Since the exact prices satisfy
$c_{j+1}(z)\geq c_j(z)$ by
Theorem~\ref{thm:static-arbitrage},
\[
\begin{aligned}
\widetilde c_{j+1}(z)-\widetilde c_j(z)
&=
\bigl[\widetilde c_{j+1}(z)-c_{j+1}(z)\bigr]
+\bigl[c_{j+1}(z)-c_j(z)\bigr]
+\bigl[c_j(z)-\widetilde c_j(z)\bigr] \\
&\geq
-2\delta_{j+1}-2\delta_j.\end{aligned}
\]
\end{proof}

The bound converges to zero whenever the approximate marginals converge to
the exact marginals in exponentially weighted total variation. The
full-support inference decoder has no structural cropping error; only
floating-point roundoff remains.

\subsection{Long-Horizon Behavior of the Additive Class}
\label{app:proof-clt}

\begin{proposition}[Standardized log-skew decay]
\label{prop:clt}
Suppose
$L_g=\mu_g+s\sqrt{\Delta\tau}\,\xi_g$, where $s>0$ and
$\{\xi_g\}$ are i.i.d.\ random variables with zero mean, unit variance, and
$\rho:=\mathbb E[|\xi_1|^3]<\infty$. Then
\begin{equation}
\operatorname{Skew}(A_{\tau_j})
=
\frac{\mathbb E[\xi_1^3]}{\sqrt j}
=
\mathcal O(j^{-1/2}).
\label{eq:proof-skew}
\end{equation}
Moreover, the centered and standardized additive state converges in
distribution to a standard Gaussian as $j\rightarrow\infty$.
\end{proposition}

\begin{proof}
Independence gives
\[
\kappa_2(A_{\tau_j})
=
j s^2\Delta\tau,
\qquad
\kappa_3(A_{\tau_j})
=
j s^3(\Delta\tau)^{3/2}\mathbb E[\xi_1^3].
\]
Therefore,
\[
\operatorname{Skew}(A_{\tau_j})
=
\frac{\kappa_3(A_{\tau_j})}
     {\kappa_2(A_{\tau_j})^{3/2}}
=
\frac{
j s^3(\Delta\tau)^{3/2}\mathbb E[\xi_1^3]
}{
(j s^2\Delta\tau)^{3/2}
}
=
\frac{\mathbb E[\xi_1^3]}{\sqrt j}.
\]
The Lyapunov ratio with exponent one is
\[
\frac{
\sum_{g\leq j}\mathbb E[|L_g-\mu_g|^3]
}{
\left(
\sum_{g\leq j}\operatorname{Var}(L_g)
\right)^{3/2}
}
=
\frac{\rho}{\sqrt j}
\longrightarrow0.
\]
The Lyapunov central limit theorem therefore yields the stated Gaussian
limit.
\end{proof}

Proposition~\ref{prop:clt} establishes a structural tendency of the additive
log-state toward Gaussianity under repeated aggregation. This is consistent
with the long-horizon smile flattening discussed in conclusion part; the proposition concerns standardized log-return
skewness and does not assume a pointwise convergence rate for the IV skew.

\section{Architecture, Algorithms, and Implementation Details}
\label{app:implementation}

\subsection{Network Architecture and Increment Parameterization}
\label{app:architecture}

Each reference quote is represented by
$u_i=(k_i,\tau_i,\sigma_i)\in\mathbb R^3$. A linear layer first embeds each
quote into $\mathbb R^{128}$, after which two Transformer encoder layers
process the resulting set. Each layer uses two attention heads, a
feed-forward width of 128, ReLU activation, zero dropout, residual
connections, and layer normalization. We use no positional encoding.
Consequently, the Transformer is permutation equivariant over the input
quotes, and the subsequent mean pooling makes the surface representation
permutation invariant. A final linear map produces
$h\in\mathbb R^{64}$.

We use a non-uniform maturity grid with $G=30$ points,
\[
\tau_g=\tau_{\max}\left(\frac{g}{G}\right)^\gamma,
\qquad
\tau_{\max}=2,
\qquad
\gamma=1.9,
\]
and define $\tau_0=0$ and
$\Delta\tau_g=\tau_g-\tau_{g-1}$. The power grid allocates more buckets to
short maturities, where option prices and implied volatilities vary most
rapidly.

For each bucket, the density head receives
$\bar h_g=[h,\tau_g,\sqrt{\tau_g}]\in\mathbb R^{66}$. It consists of two
hidden layers of width 128 with GELU activations, followed by an output layer
of width $3R$, where $R=8$. The output is divided into mixture logits
$a_{g,r}$, raw locations $\widetilde\mu_{g,r}$, and raw scales
$\ell_{g,r}$. These are transformed as
\[
\pi_{g,r}
=
\frac{\exp(a_{g,r})}
{\sum_{r'=1}^{R}\exp(a_{g,r'})},
\qquad
\mu_{g,r}
=
0.3\sqrt{\Delta\tau_g}\,\widetilde\mu_{g,r},
\]
and
\[
s_{g,r}
=
\sqrt{\Delta\tau_g}
\left[
0.02+\operatorname{softplus}(\ell_{g,r}-1)
\right].
\]
The factor $\sqrt{\Delta\tau_g}$ provides a diffusive time scale, while the
mixture components allow the increment law to be asymmetric and multimodal.

Let $\{x_i\}_{i=1}^{X}$ be an equally spaced grid with $X=2048$ and
$x_i\in[-5,5]$. Conditional on the reference set $\mathcal Q$, the density
head defines
\[
f_{\theta,g}(x\mid\mathcal Q)
=
\sum_{r=1}^{R}
\pi_{g,r}
\frac{1}{\sqrt{2\pi}\,s_{g,r}}
\exp\left(
-\frac{(x-\mu_{g,r})^2}{2s_{g,r}^{2}}
\right).
\]
Evaluating this density on the grid and normalizing gives
\[
\widetilde m_{\theta,g,i}
=
f_{\theta,g}(x_i\mid\mathcal Q),
\qquad
m_{\theta,g,i}
=
\frac{\widetilde m_{\theta,g,i}}
{\sum_{\ell=1}^{X}\widetilde m_{\theta,g,\ell}}.
\]
Thus, $m_{\theta,g,i}\geq0$ and
$\sum_i m_{\theta,g,i}=1$. The resulting discrete increment law is
\[
\widehat\eta_{\theta,g}
=
\sum_{i=1}^{X}m_{\theta,g,i}\delta_{x_i}.
\]
The complete model contains 236,120 trainable parameters.
Table~\ref{tab:addmart-architecture} summarizes the architecture.

\begin{table}[t]
\caption{AddMart architecture used in the main experiments. The number of
reference quotes $K$ may vary without changing the network.}
\label{tab:addmart-architecture}
\centering
\small
\setlength{\tabcolsep}{5pt}
\begin{tabular}{@{}lll@{}}
\toprule
Component & Tensor mapping & Configuration \\
\midrule
Quote embedding
& $\mathbb R^{K\times3}\rightarrow\mathbb R^{K\times128}$
& Linear projection \\
Set encoder
& $\mathbb R^{K\times128}\rightarrow\mathbb R^{K\times128}$
& 2 layers, 2 heads \\
Surface pooling
& $\mathbb R^{K\times128}\rightarrow\mathbb R^{64}$
& Mean pooling, linear map \\
Bucket conditioning
& $\mathbb R^{64}\rightarrow\mathbb R^{66}$
& Append $(\tau_g,\sqrt{\tau_g})$ \\
Density head
& $\mathbb R^{66}\rightarrow\mathbb R^{3R}$
& MLP $(66,128,128,3R)$ \\
Increment law
& $\mathbb R^{3R}\rightarrow\mathbb R^{X}$
& $R=8$, $X=2048$ \\
Martingale decoder
& $\{m_g\}_{g=1}^{G}\rightarrow\{c(k,\tau)\}$
& $G=30$, linear convolution \\
IV inversion
& $c(k,\tau)\rightarrow\widehat\sigma(k,\tau)$
& Black bisection \\
\midrule
\multicolumn{2}{l}{Total trainable parameters} & 236,120 \\
\bottomrule
\end{tabular}
\end{table}

\subsection{Full-Support Convolution and Exact Inference}
\label{app:exact-inference}

For readability, write $m_{g,i}:=m_{\theta,g,i}$. For each increment law,
define its exponential normalizer and Esscher-tilted mass by
\[
d_g=\sum_i m_{g,i}e^{x_i},
\qquad
m_{g,i}^{\star}
=
\frac{m_{g,i}e^{x_i}}{d_g}.
\]
At maturity $\tau_j$, the cumulative base and tilted laws are represented by
\[
v_j=m_1*_{\mathrm{lin}}\cdots*_{\mathrm{lin}}m_j,
\qquad
v_j^\star
=
m_1^\star*_{\mathrm{lin}}\cdots*_{\mathrm{lin}}m_j^\star,
\]
with exponential normalizer $Z_j=\prod_{g\leq j}d_g$.

If each increment law contains $X$ atoms, the cumulative law at bucket $j$
contains $j(X-1)+1$ atoms. The largest required support therefore has
\[
L=G(X-1)+1
\]
atoms. The implementation zero-pads every FFT to at least this length, using
the next power of two for computational efficiency. Spectral multiplication
then implements full linear convolution without circular wraparound or
intermediate support truncation.

The exact decoder performs these operations in float64. Inverse FFTs may
produce negative coefficients at the level of floating-point roundoff. We
clip such coefficients to zero and renormalize each mass vector. This
correction changes option prices only at numerical precision while preventing
remote-tail errors from being amplified by the exponential transformation.

Directly forming the atoms $e^{A_{\tau_j}}/Z_j$ is numerically unstable in
the far-right tail. The tilted representation permits call prices to be
computed from tail probabilities instead. For normalized strike $z=e^k$,
let $a_j(k)=k+\log Z_j$. Then
\begin{align}
c(z,\tau_j)
&=
\mathbb E_{\mathbb Q}
\left[
\left(
\frac{e^{A_{\tau_j}}}{Z_j}-z
\right)^+
\right] \nonumber\\
&=
\frac{1}{Z_j}
\mathbb E_{\mathbb Q}
\left[
e^{A_{\tau_j}}
\mathbf 1\{A_{\tau_j}\geq a_j(k)\}
\right]
-
z\,\mathbb P_{\mathbb Q}
\left(A_{\tau_j}\geq a_j(k)\right) \nonumber\\
&=
\nu_j^\star([a_j(k),\infty))
-
z\,\nu_j([a_j(k),\infty)),
\label{eq:appendix-tail-price}
\end{align}
where
\[
d\nu_j^\star(x)
=
\frac{e^x}{Z_j}\,d\nu_j(x).
\]
After computing suffix sums of $v_j$ and $v_j^\star$, pricing a strike
requires one threshold lookup and two tail probabilities. For maturities
between grid points, we use the call-price interpolation defined in
Section~\ref{sec:method-construct}. The interpolated price is then converted
to IV using Black inversion.
Algorithm~\ref{alg:addmart-inference} gives the complete inference procedure.

\begin{algorithm}[h]
\caption{Exact AddMart inference}
\label{alg:addmart-inference}
\begin{algorithmic}[1]
\REQUIRE Reference set
$\mathcal Q=\{(k_i,\tau_i,\sigma_i)\}_{i=1}^{K}$;
queries $\mathcal T=\{(k_n,\tau_n)\}_{n=1}^{N}$
\STATE $h\leftarrow\operatorname{SetEncoder}_{\theta}(\mathcal Q)$
\FOR{$g=1,\ldots,G$}
    \STATE
    $(\pi_g,\mu_g,s_g)
    \leftarrow
    \operatorname{DensityHead}_{\theta}
    (h,\tau_g,\sqrt{\tau_g})$
    \STATE Evaluate the Gaussian mixture on $\{x_i\}_{i=1}^{X}$ and
    normalize it to obtain $m_g$
    \STATE
    $d_g\leftarrow\sum_i m_{g,i}e^{x_i}$
    \STATE
    $m_{g,i}^{\star}\leftarrow m_{g,i}e^{x_i}/d_g$
\ENDFOR
\STATE $v_0\leftarrow\delta_0$,
$v_0^\star\leftarrow\delta_0$, and $Z_0\leftarrow1$
\FOR{$j=1,\ldots,G$}
    \STATE
    $v_j\leftarrow v_{j-1}*_{\mathrm{lin}}m_j$
    \STATE
    $v_j^\star\leftarrow
    v_{j-1}^\star*_{\mathrm{lin}}m_j^\star$
    \STATE
    $Z_j\leftarrow Z_{j-1}d_j$
    \STATE Compute suffix sums of $v_j$ and $v_j^\star$
\ENDFOR
\FOR{each query $(k_n,\tau_n)\in\mathcal T$}
    \STATE Locate the maturity buckets bracketing $\tau_n$
    \STATE At each bracketing bucket $j$, set
    $a_j\leftarrow k_n+\log Z_j$
    \STATE Compute
    $c_{n,j}\leftarrow
    \nu_j^\star([a_j,\infty))
    -e^{k_n}\nu_j([a_j,\infty))$
    \STATE Interpolate the bracketing call prices to obtain $c_n$
    \STATE
    $\widehat\sigma_n
    \leftarrow
    \operatorname{BlackInverse}(k_n,\tau_n,c_n)$
\ENDFOR
\RETURN Reconstructed IVs
$\{\widehat\sigma_n\}_{n=1}^{N}$ and marginal laws
$\{\widehat{\mathsf P}_j\}_{j=1}^{G}$
\end{algorithmic}
\end{algorithm}

The exact decoder also exposes the complete discrete marginal laws
$\{\widehat{\mathsf P}_j\}_{j=1}^{G}$ and their maturity-dependent atom
locations. We use these objects to evaluate distributional risk functionals
and to audit probability mass, exponential normalization, butterfly
convexity, and calendar monotonicity directly in price space.

\subsection{Training Protocol}
\label{app:optimization}

At each timestamp, the main experiments use a fixed $3\times3$
maturity--delta reference set as input. We independently draw $N=256$ target
contracts with replacement from the complete surface. Reference contracts
remain in this sampling pool during training. For validation and testing,
however, the nine conditioning contracts are removed from the target set, so
the reported metrics measure reconstruction at contracts that were not
provided to the network.

Training uses batches of 16 surfaces and the objective in
Equation~\ref{eq:appendix-huber}. All target contracts receive unit weight;
the optional reliable-region weight is set to one. The reliable evaluation
region, defined by $|k|\leq0.1$ and $\tau\geq14$ days, therefore does not
receive additional weight during training.

We optimize the model with Adam, using an initial learning rate of
$10^{-3}$ and cosine annealing to $10^{-5}$. Gradients are clipped to unit
norm, and an update is skipped if either the loss or the gradient norm is
non-finite. Daily models are trained for at most 400 epochs, evaluated every
20 epochs, and stopped after eight validation evaluations without
improvement. Minute models are trained for at most 80 epochs, evaluated every
five epochs, and stopped after six validation evaluations without
improvement. Checkpoints are selected using validation overall IV-MAE. We
train each asset and sampling frequency independently over five random seeds.

\paragraph{Training and inference decoders.}
The main five-seed checkpoints are optimized with a memory-efficient,
fixed-width differentiable convolution. During training, this approximation
centrally crops and renormalizes each intermediate convolution. At evaluation,
the learned encoder and density head are kept fixed, while all reported
predictions and arbitrage audits are recomputed with the full-support decoder
of Appendix~\ref{app:exact-inference}. The structural guarantee therefore
applies to the exact inference map used to generate the reported surfaces.
Lemma~\ref{lem:crop-error} bounds the approximation introduced during
optimization. We additionally evaluate end-to-end training with the exact
decoder in the corresponding ablation.

\subsection{Implicit Implied-Volatility Layer}
\label{app:implicit-iv}

The martingale decoder outputs forward-normalized call prices, whereas the
training targets and reconstruction metric are expressed in implied
volatility. Differentiating through an unrolled Newton solver is unstable in
the deep wings, where Vega is small and inaccurate iterates can create large
gradients or become trapped at a clipping boundary. We instead solve the
Black inverse accurately by bisection in the forward pass and specify its
gradient through implicit differentiation.

\paragraph{Forward Black inversion.}
Let $s_{\mathrm{tot}}=\sigma\sqrt{\tau}$ denote total volatility. The
forward-normalized Black call price is
\begin{equation}
B(k,s_{\mathrm{tot}})
=
\Phi(d_+)-e^k\Phi(d_-),
\qquad
d_+
=
-\frac{k}{s_{\mathrm{tot}}}
+\frac{s_{\mathrm{tot}}}{2},
\qquad
d_-
=
-\frac{k}{s_{\mathrm{tot}}}
-\frac{s_{\mathrm{tot}}}{2}.
\label{eq:appendix-black-map}
\end{equation}
Given a decoder price $c$, implied total volatility solves
$B(k,s_{\mathrm{tot}})=c$. Before inversion, we project the price into the
interior of its no-arbitrage interval:
\begin{equation}
\widetilde c
=
\min
\left\{
1-\epsilon,\,
\max
\left[
c,\,
(1-e^k)^++\epsilon
\right]
\right\},
\qquad
\epsilon=10^{-7}.
\label{eq:appendix-price-projection}
\end{equation}
This projection is inactive for a valid interior price and acts only as a
numerical safeguard near intrinsic value and the unit upper bound.

During training, we solve
$B(k,s_{\mathrm{tot}})=\widetilde c$ using 60 bisection iterations over
$s_{\mathrm{tot}}\in[10^{-6},6]$. The returned implied volatility is
\[
\widehat\sigma
=
\frac{s_{\mathrm{tot}}}
{\sqrt{\max(\tau,10^{-12})}}.
\]
Bisection uses the strict monotonicity of $B$ in total volatility and does
not require an initial volatility estimate.

\paragraph{Implicit derivative.}
Define
\[
\Psi(\sigma,c;k,\tau)
=
B(k,\sigma\sqrt{\tau})-c.
\]
At the solution,
$\Psi(\widehat\sigma,c;k,\tau)=0$. The implicit-function theorem gives
\begin{equation}
\frac{\partial\widehat\sigma}{\partial c}
=
-\frac{\partial\Psi/\partial c}
{\partial\Psi/\partial\sigma}
=
\frac{1}{V},
\qquad
V:=\sqrt{\tau}\,\phi(d_+),
\label{eq:appendix-implicit-gradient}
\end{equation}
where $V$ is the Vega of the forward-normalized Black price. Hence, for a
model parameter $\theta$,
\[
\frac{\partial\widehat\sigma}{\partial\theta}
=
\frac{1}{V}
\frac{\partial c_\theta}{\partial\theta}.
\]

The inverse $1/V$ becomes unstable as $V\rightarrow0$. We therefore use the
Tikhonov-regularized scalar pseudoinverse
\begin{equation}
\frac{\partial\widehat\sigma}{\partial c}
\approx
\frac{V}{V^2+\lambda_{\mathrm V}^2},
\qquad
\lambda_{\mathrm V}=0.01.
\label{eq:appendix-damped-gradient}
\end{equation}
For $V\gg\lambda_{\mathrm V}$, this approaches the exact derivative $1/V$.
As $V\rightarrow0$, it approaches zero rather than diverging. Its maximum
magnitude is $1/(2\lambda_{\mathrm V})=50$, which explicitly bounds the
amplification of price gradients in low-Vega regions. The damping affects
only the backward pass; the forward value remains the bisection solution.

\paragraph{Huber residual.}
For residual
$r_n=\widehat\sigma_{\theta,n}-\sigma_n$, the elementwise Huber loss is
\begin{equation}
\rho_\beta(r)
=
\begin{cases}
r^2/(2\beta), & |r|\leq\beta,\\
|r|-\beta/2, & |r|>\beta,
\end{cases}
\qquad
\beta=0.05.
\label{eq:appendix-huber}
\end{equation}
Its derivative is
\[
\rho_\beta'(r)
=
\begin{cases}
r/\beta, & |r|\leq\beta,\\
\operatorname{sign}(r), & |r|>\beta.
\end{cases}
\]
Combining this derivative with the damped implicit inverse yields
\begin{equation}
\frac{\partial\mathcal L_{\mathrm{IV}}}{\partial\theta}
=
\frac{1}{|\mathcal T|}
\sum_{n\in\mathcal T}
\rho_\beta'(r_n)
\frac{V_n}{V_n^2+\lambda_{\mathrm V}^2}
\frac{\partial c_{\theta,n}}{\partial\theta}.
\label{eq:appendix-complete-gradient}
\end{equation}

The objective is expressed in IV units and directly matches the reported
reconstruction metric. The Huber influence function limits the contribution
of large residuals, while the damped inverse controls gradient amplification
from low-Vega contracts. No gradient is propagated through the bisection
iterations; the custom backward pass evaluates
Equation~\ref{eq:appendix-damped-gradient} once at the converged solution.

\paragraph{Inference-time inversion.}
Evaluation does not require gradients. We project prices into the same valid
numerical interval and apply 50 bisection iterations over
$s_{\mathrm{tot}}\in[10^{-3},5]$. At this precision, the inversion error is
negligible relative to the reported IV-MAE. Training and evaluation therefore
use the same normalized Black pricing map, while only the training layer
defines a custom implicit backward pass.

%% file: appendix/data.tex
\section{Data Preparation}
\label{appendix:data}

In this section, we provide detail data processing pipeline with used data scale
\subsection{Data Preparation}
\label{app:data}

\paragraph{Asset universe and snapshot construction.}
We evaluate the actual option tickers rather than relabelling ETF options as index books. At the
daily scale, the universe consists of three broad-market ETFs (SPY, QQQ, and IWM) and four liquid
single names (AAPL, GOOGL, META, and NVDA). At the intraday scale, we use SPY, QQQ, and IWM.
A daily surface is sampled at 15:59 ET, the final minute of continuous trading before the closing
auction. Intraday surfaces are sampled every 15 minutes from 09:45 to 15:45 ET, giving at most
25 snapshots per trading day. 
For each target timestamp \(T\), we retrieve the last OPRA NBBO observed at or before \(T\) for every
listed contract. The resulting cross-section represents the prevailing option book at \(T\); the
distance between the quote timestamp and \(T\) measures quote staleness rather than a synchronization
offset. We retain quotes updated within 500\,ms of \(T\). The underlying price is the contemporaneous
midpoint of the Nasdaq BBO from Databento XNAS.ITCH. Thus, the underlying quote is time-aligned with
the option snapshot, although it represents the Nasdaq order book rather than the consolidated SIP
NBBO.

\paragraph{Quote filtering and price construction.}
We retain contracts with positive bid and ask prices, positive displayed sizes, and
\(\mathrm{ask}\geq\mathrm{bid}\). The option price is the arithmetic NBBO midpoint,
\[
    P_{\mathrm{mid}}
    = \frac{P_{\mathrm{bid}}+P_{\mathrm{ask}}}{2}.
\]
We further require a relative spread
\((P_{\mathrm{ask}}-P_{\mathrm{bid}})/P_{\mathrm{mid}}\leq 0.35\), an option price of at least
\(0.1\), and a time to expiry in \([0.5/365,2]\) years.

The risk-free curve is obtained by linearly interpolating the most recent available Treasury yields
across maturities and setting \(D(\tau)=\exp(-r(\tau)\tau)\). For each timestamp--expiry slice, the
forward is estimated from put--call parity,
\[
    F_K = K + \frac{C(K)-P(K)}{D(\tau)}.
\]
We first take the median across strikes with both call and put quotes, and then recompute the median
using strike pairs within \(5\%\) of this initial estimate. At least three paired strikes are required.
When parity cannot be estimated reliably, we use
\(F=S\exp((r-q)\tau)\), where \(S\) is the aligned Nasdaq-BBO midpoint and \(q\) is the median carry
inferred from the parity-resolved expiries of the same surface.

Given \(F\), we define log-moneyness as \(k=\log(K/F)\) and retain the out-of-the-money side of each
strike: puts for \(K<F\) and calls for \(K\geq F\). Implied volatility is obtained by a vectorized
damped-Newton inversion of the Black--76 price, initialized at \(50\%\) volatility and constrained to
\([10^{-4},5]\). Contracts whose reconstructed Black price does not match the observed midpoint within
the numerical acceptance tolerance are removed. We use forward delta,
\(\Delta_C=\Phi(d_+)\) for calls and \(\Delta_P=\Phi(d_+)-1\) for puts.

\paragraph{Surface-level quality control.}
The remaining contracts undergo the same static-arbitrage cleaning for every asset and scale.
Within each maturity, we iteratively remove the interior strike with the largest convexity violation
until the normalized call-price curve is convex. Calendar consistency is then enforced in
16 log-moneyness bins by retaining the longest non-decreasing subsequence of median total variance
\(\sigma^2\tau\) across maturities, followed by a second butterfly-convexity pass. A timestamp is
retained only if at least four distinct expiries and 40 contracts remain.

\paragraph{Reference-set selection.}
Every retained surface provides exactly nine conditioning quotes. We first select three distinct
listed maturities nearest to 7, 30, and 90 calendar days. At each selected maturity, we choose three
distinct contracts nearest to the forward-delta targets
\[
    \Delta\in\{0.50,\;0.25,\;-0.25\},
\]
corresponding to an approximately at-the-money call, a 25-delta call, and a 25-delta put. A surface
that cannot provide all nine contracts is discarded. Direct inspection of the final files confirms
that every retained surface contains exactly nine reference quotes; all remaining contracts form the
reconstruction evaluation set.

\paragraph{Temporal split and dataset scale.}
We use a common chronological cutoff of 1 February 2025. The final 10\% of the pre-cutoff timestamps
form the validation window, while all timestamps on or after the cutoff are reserved for testing.
No surface is shuffled across temporal partitions. The final reported universe contains 70,027
surfaces and 40,129,821 retained contracts. Table~\ref{tab:datastats} reports statistics computed
directly from the final Parquet files. Because Meta Platforms changed its ticker from FB to META in
June 2022, we include only contracts listed under the META ticker and do not retrospectively
relabel FB options. Consequently, the META series begins on June 13, 2022 at the daily scale and
June 10, 2022 at the intraday scale, whereas the other assets begin in March 2022. 

\begin{table}[t]
\caption{Statistics of the final experimental datasets. ``Split'' reports the number of
train/validation/test surfaces. A surface is one retained quote timestamp, and ``Avg.\ contracts''
is the mean number of retained contracts per surface, including the nine reference quotes.
Maturity \(\tau\) is measured in ACT/365 years from the exact snapshot time to expiration at
16:00 ET, and \(k=\log(K/F)\).}
\label{tab:datastats}
\centering\small\setlength{\tabcolsep}{4pt}
\resizebox{\linewidth}{!}{%
\begin{tabular}{@{}ll r r r r cc@{}}
\toprule
Scale & Asset & Surfaces & Split (train/val/test) &
Contracts & Avg.\ contracts & \(\tau\) range & \(k\) range \\
\midrule
\multirow{8}{*}{Daily}
 & SPY   & \(992\) & \(601/67/324\) & \(920{,}977\) & \(928.4\)
 & \([0.0027,1.9726]\) & \([-2.482,\,0.696]\) \\
 & QQQ   & \(958\) & \(584/65/309\) & \(563{,}724\) & \(588.4\)
 & \([0.0027,1.9726]\) & \([-1.355,\,0.680]\) \\
 & IWM   & \(727\) & \(406/45/276\) & \(270{,}787\) & \(372.5\)
 & \([0.0027,1.9726]\) & \([-1.418,\,0.540]\) \\
 & AAPL  & \(465\) & \(198/22/245\) & \(85{,}619\) & \(184.1\)
 & \([0.0027,1.9726]\) & \([-2.454,\,0.798]\) \\
 & GOOGL & \(573\) & \(321/36/216\) & \(134{,}580\) & \(234.9\)
 & \([0.0027,1.9671]\) & \([-1.587,\,0.903]\) \\
 & META  & \(591\) & \(282/31/278\) & \(190{,}279\) & \(322.0\)
 & \([0.0027,1.9726]\) & \([-2.715,\,1.488]\) \\
 & NVDA  & \(797\) & \(442/49/306\) & \(318{,}337\) & \(399.4\)
 & \([0.0027,1.9726]\) & \([-2.916,\,1.332]\) \\
\cmidrule(lr){2-8}
 & \textbf{Total} & \(\mathbf{5{,}103}\) & \(\mathbf{2{,}834/315/1{,}954}\)
 & \(\mathbf{2{,}484{,}303}\) & \(\mathbf{486.8}\)
 & \([0.0027,1.9726]\) & \([-2.916,\,1.488]\) \\
\midrule
\multirow{4}{*}{15-min}
 & SPY & \(23{,}699\) & \(14{,}470/1{,}608/7{,}621\)
 & \(18{,}542{,}864\) & \(782.4\)
 & \([0.0028,1.9733]\) & \([-2.551,\,0.764]\) \\
 & QQQ & \(23{,}610\) & \(14{,}786/1{,}643/7{,}181\)
 & \(13{,}498{,}089\) & \(571.7\)
 & \([0.0028,1.9733]\) & \([-1.393,\,0.790]\) \\
 & IWM & \(17{,}615\) & \(10{,}688/1{,}188/5{,}739\)
 & \(5{,}604{,}565\) & \(318.2\)
 & \([0.0028,1.9732]\) & \([-1.446,\,0.711]\) \\
\cmidrule(lr){2-8}
 & \textbf{Total} & \(\mathbf{64{,}924}\) & \(\mathbf{39{,}944/4{,}439/20{,}541}\)
 & \(\mathbf{37{,}645{,}518}\) & \(\mathbf{579.8}\)
 & \([0.0028,1.9733]\) & \([-2.551,\,0.790]\) \\
\bottomrule
\end{tabular}}
\end{table}

\paragraph{Short-expiry coverage.}
The construction filter removes contracts with less than half a calendar day to expiry. Because the
latest sampled timestamp is before the 16:00 ET expiration time, this eliminates all same-day-expiry
contracts in the final files. The minimum observed maturity is 1.0007 days for the daily data and
1.0104 days for the intraday data, and no retained contract has \(\tau<1\) day. Short-dated contracts
beyond one day remain in the dataset: contracts with \(1\leq\tau<7\) days account for 145,312 quotes
(\(5.85\%\)) at the daily scale and 3,419,407 quotes (\(9.08\%\)) at the intraday scale. The
reliable-region metric excludes contracts with fewer than 14 days to maturity, whereas the overall
metric and the training data retain them.

%% file: appendix/exps.tex
\section{Experimental setup}
\label{app:setup}
\subsection{Evaluation Metrics}
\label{app:metrics}

\paragraph{Accuracy.}
For each held-out surface, every method receives the same nine reference
quotes. Accuracy is evaluated on all remaining observed contracts; the
reference contracts themselves are excluded from the targets. For a target
contract \(q\), with log-forward-moneyness \(k_q\), maturity \(\tau_q\), and
market implied volatility \(\sigma_q\), the absolute reconstruction error is
\[
e_q
=
\left|
\hat{\sigma}(k_q,\tau_q)-\sigma_q
\right|.
\]
For evaluation region \(\mathcal A\), we use the contract-weighted IV mean
absolute error
\[
\operatorname{IV\text{-}MAE}_{\mathcal A}
=
\frac{
\sum_s\sum_{q\in\mathcal A_s} e_q
}{
\sum_s |\mathcal A_s|
},
\]
where \(s\) indexes test surfaces. Results are reported in volatility points
(\(\times10^{-2}\)). For amortized methods, we first compute the metric
separately for each of five independently trained seeds and then report the
mean and standard deviation across seeds. SSVI and Deep Smoothing are
optimized independently on every test surface and therefore have no training
seed.

\paragraph{Reliable and overall regions.}
We report accuracy on two fixed evaluation regions:
\begin{itemize}[leftmargin=1.4em,itemsep=2pt,topsep=2pt]
    \item \textbf{Reliable} contains contracts satisfying
    \[
    |k|\leq0.1,
    \qquad
    \operatorname{TTM}\geq14\text{ days}.
    \]
    Since \(k=\log(K/F)\), the moneyness restriction corresponds approximately
    to \(K/F\in[0.905,1.105]\). It retains the central portion of the smile
    while excluding very short-dated contracts.
    \item \textbf{Overall} contains every observed non-reference target
    contract, including short maturities and deep wings.
\end{itemize}

The reliable region is intended as a high-confidence evaluation subset rather
than as an additional data filter. Near the money, option vega is sufficiently
large that the mapping from price to implied volatility is comparatively
stable. At very short maturities or in the deep wings, vega can be close to
zero, so a small price difference or even one bid--ask tick can translate
into a large IV difference. Quotes in those regions are also less liquid. In
our final dataset, the relative bid--ask spread
\[
\frac{\mathrm{ask}-\mathrm{bid}}
     {(\mathrm{ask}+\mathrm{bid})/2}
\]
has a quote-weighted median of \(1.25\%\) and a 90th percentile of
\(6.90\%\) overall, compared with \(0.81\%\) and \(2.54\%\), respectively,
inside the reliable region. The reliable metric therefore measures accuracy
where the market mid and its implied volatility are most statistically
meaningful. The region is defined once and applied identically to every asset, scale, and
method; it is not selected using model performance. Reporting the overall
metric alongside it prevents a method from appearing strong merely by
performing well near the money. The two metrics consequently answer different
questions: reliable MAE measures reconstruction in the liquid interior,
whereas overall MAE measures whether the fitted surface remains well behaved
under short-end and wing extrapolation.

\paragraph{Static-arbitrage validity.}
A surface can fit observed IV quotes accurately while assigning inconsistent
prices to nearby strikes or maturities. We test two elementary forms of static
arbitrage. A \emph{calendar violation} occurs when a longer-maturity call is
cheaper than a shorter-maturity call at the same forward moneyness. A
\emph{butterfly violation} occurs when call price is locally non-convex in
strike; this would permit a butterfly portfolio with a non-negative terminal
payoff and a negative initial cost, and corresponds to a negative discrete
risk-neutral probability.

For each asset and scale, we select 50 deterministic, time-spanning surfaces
from the out-of-time test period. Every method is evaluated on 801
log-moneyness points \(k_i\in[-1,1]\) and the same 30 maturity nodes
\(\tau_j\) used by AddMart. Let \(M_i=e^{k_i}=K_i/F\) and let
\(c_{j,i}=C(K_i,\tau_j)/(D(\tau_j)F(\tau_j))\) denote normalized forward call
price. HyperIV, GNO, VAE, SSVI, and Deep Smoothing are converted from predicted
IV to \(c_{j,i}\) using Black--76. For \addmart{}, we directly evaluate the
normalized prices produced by its exact decoder. All methods are then tested
on the same grid with the same price-space inequalities.

Calendar validity requires adjacent-maturity prices to be non-decreasing.
Butterfly validity requires adjacent strike slopes to be non-decreasing:
\begin{align}
\Delta^{\mathrm{cal}}_{j,i}
&=c_{j+1,i}-c_{j,i}\geq0,
\\
\Delta^{\mathrm{bf}}_{j,i}
&=
\frac{c_{j,i+2}-c_{j,i+1}}{M_{i+2}-M_{i+1}}
-
\frac{c_{j,i+1}-c_{j,i}}{M_{i+1}-M_i}
\geq0.
\end{align}
Each surface therefore contributes \(29\times801\) calendar pairs and
\(30\times799\) butterfly pairs. For either condition \(a\in
\{\mathrm{cal},\mathrm{bf}\}\), the reported violation rate is
\[
R_a
=
100\,
\frac{
\sum_{r,s,j,i}
\mathbf 1[\Delta^{a,(r,s)}_{j,i}<-\varepsilon]
}{
\sum_{r,s}N_a^{(r,s)}
},
\]
where \(r\) indexes training seeds and \(s\) indexes audited surfaces.
Learned methods use five seeds, whereas per-surface baselines have no seed
dimension. Thus, a reported rate of \(0.5\%\) means that \(0.5\%\) of all
evaluated price pairs violate the corresponding inequality; it does not mean
that only \(0.5\%\) of complete surfaces contain a violation.

\paragraph{Numerical tolerance.}
Finite-precision price evaluation and differencing can produce tiny negative
residuals around an exact zero. The headline pair-rate audit uses
\(\varepsilon_{\mathrm{pair}}=10^{-10}\). We separately define a surface as
having a materially meaningful violation only if its minimum residual is below
\(-10^{-6}\). Consequently, a residual such as \(-10^{-9}\) is visible in the
strict pair diagnostic but is not interpreted as a material arbitrage failure.
The spatial heatmaps and convolution ablations use
\(\varepsilon_{\mathrm{vis}}=10^{-7}\) to suppress isolated numerical noise;
their percentages are therefore kept distinct from the stricter headline
pair rates.

The exact AddMart decoder is audited in both float32 and float64. Both
precisions produce zero calendar and butterfly pair violations in the formal
audit; the smallest residuals are approximately \(10^{-13}\), consistent with
floating-point error. We additionally check for non-finite prices, invalid
price bounds, and failed surface evaluations before aggregation.

\subsection{Baselines and Reproduction Details}
\label{app:baselines}

We compare \addmart{} with three amortized neural baselines and two
per-surface calibration methods. All methods receive information from the
same nine observed reference contracts. Accuracy is evaluated on the
remaining, non-reference contracts of each surface.

\paragraph{Common protocol.}
Each amortized model is trained independently for every asset and sampling
frequency using seeds $\{0,1,2,3,4\}$. We use the chronological split
described in Appendix~\ref{app:data}: the test period begins on
February 1, 2025, while the last 10\% of the pre-test timestamps form the
validation set. The test period is never used for hyperparameter tuning or
checkpoint selection.

During training, each surface is represented by the nine reference tuples
$(k_i,\tau_i,\sigma_i)$. We sample $N=1024$ query contracts from the same
surface with replacement and minimize the corresponding reconstruction
objective. HyperIV uses a batch size of 16 surfaces, while GNO and VAE use
20. All three neural baselines are optimized with Adam at an initial learning
rate of $10^{-3}$ and cosine annealing to $10^{-5}$. The checkpoint with the
lowest validation overall IV-MAE is retained. At test time, the nine
conditioning quotes are excluded from both reliable-region and overall
accuracy metrics. The reliable region is
\[
    \mathcal{R}
    =\{(k,\tau): |k|\le 0.1,\ \tau\ge 14/365\}.
\]
We report the mean and standard deviation across the five independently
trained seeds. SSVI and Deep Smoothing perform a new optimization for every
test surface and are therefore reported as single values.

\begin{table}[t]
\caption{Implementation settings for the amortized baselines. Epoch counts
are maximum values for daily/intraday experiments.}
\label{tab:baseline_impl}
\centering
\small
\setlength{\tabcolsep}{5pt}
\begin{tabular}{@{}lrrrr@{}}
\toprule
Method & Trainable parameters & Batch & Queries/surface
       & Epochs (daily/intraday) \\
\midrule
HyperIV & $243{,}153$ & 16 & 1024 & $1500/150$ \\
GNO     & $102{,}529$ & 20 & 1024 & $300/60$ \\
VAE     & $332{,}015$ & 20 & 1024 & $200/60$ \\
\bottomrule
\end{tabular}
\end{table}

\begin{itemize}[leftmargin=1.4em,itemsep=4pt,topsep=3pt]

\item \textbf{HyperIV} \citep{hyperiv2025}.
We retain the released set-embedding and hypernetwork construction and adapt
its data interface to our nine-reference protocol. The generated
surface-specific IV network has architecture
$2\!\rightarrow\!16\!\rightarrow\!16\!\rightarrow\!1$, with \texttt{tanh}
hidden activations and a \texttt{softplus} output. It contains 337 generated
weights. A permutation-invariant set encoder maps the nine
$(k,\tau,\sigma)$ tuples to these weights and contains $243{,}153$ trainable
parameters.

Our reproduction uses the HyperIV reconstruction loss together with its three
soft regularizers: calendar monotonicity, the Gatheral--Jacquier butterfly
condition, and density normalization. These terms are evaluated by automatic
differentiation on a $31\times9$ auxiliary grid over
$k\in[-1.5,0.5]$ and $\tau\in[0.01,2]$. All four loss components have unit
weight. We replace non-finite gradients by zero and clip the global gradient
norm to one, which prevents occasional spikes produced by the
second-derivative penalties. Validation is performed every 50 epochs for the
daily experiments and every five epochs for the intraday experiments.

\item \textbf{GNO} \citep{li2020neural,wiedemann2025operator}.
Our GNO baseline uses the published Operator Deep Smoothing
\texttt{GNOVolSmoother} architecture, adapted so that the nine reference
contracts constitute the observed nodes and arbitrary held-out contracts
constitute the query nodes. Following the published operator coordinates, we
transform each query to
\[
    (\rho,z)=\left(\sqrt{\tau},\,\frac{k}{\sqrt{\tau}}\right).
\]
The model uses three graph-neural-operator layers, 16 operator channels, a
64-dimensional pointwise network, radius 10, and subsample size 16, for a
total of $102{,}529$ trainable parameters. It is trained only with query-IV
MSE; no calendar or butterfly penalty is added.

The original positive-output layer can enter a numerically saturated region
when fitting low-volatility assets. In our adapter, both reference and target
IVs are divided by the median IV of the training set before being passed to
the network, and predictions are multiplied by the same constant afterward.
The scale is estimated exclusively from the training set and stored in the
checkpoint. This transformation leaves the relative reconstruction problem
unchanged and prevents seed-dependent zero-output collapse without modifying
the released GNO implementation.

\item \textbf{VAE} \citep{kingma2014auto}.
The VAE requires a fixed ordering of its nine-dimensional input. Because the
three selected market maturities vary across timestamps, we first calibrate
the four-parameter SSVI form using only the nine observed references. We then
evaluate that fitted surface at a fixed virtual grid consisting of
7-, 30-, and 90-day maturities crossed with a 25-delta put, ATM, and
25-delta call. This produces a consistently ordered nine-dimensional IV
vector without using any reconstruction target.

The encoder has two width-322 ReLU layers and outputs the mean and log
variance of a 16-dimensional latent variable. The pointwise decoder receives
the latent variable together with $(k,\tau)$ and uses three width-322 ReLU
layers followed by a \texttt{softplus} output. The resulting model contains
$332{,}015$ trainable parameters. We minimize
\[
    \mathcal{L}_{\mathrm{VAE}}
      =\operatorname{MSE}(\widehat{\sigma},\sigma)
       +10^{-3}D_{\mathrm{KL}}
          \bigl(q_\phi(z\mid\sigma_{\mathrm{ref}})
          \,\|\,\mathcal{N}(0,I)\bigr).
\]
Latent samples are used during training, while the posterior mean is used for
validation and testing.

\item \textbf{SSVI}$^\ddagger$ \citep{gatheral2014arbitrage}.
For every test surface, we fit the four-parameter power-law SSVI
parameterization
\[
\theta(\tau)=\sigma_0^2\tau,\qquad
\phi(\theta)=\eta\theta^{-\gamma},
\]
using only the same nine reference quotes. The parameters satisfy
$\sigma_0>0$, $0<\gamma<1/2$, $\eta>0$, and $-1<\rho<1$.
Calibration minimizes IV-MSE with SLSQP and imposes the
Gatheral--Jacquier sufficient butterfly-arbitrage inequalities over the
maturity range of the surface. We use three initializations, at most 1000
SLSQP iterations per initialization, and retain the feasible solution with
the lowest reference-point objective. SSVI has no learned parameters,
training split, validation selection, or seed averaging.

\item \textbf{Deep Smoothing}$^\ddagger$
\citep{ackerer2020deep}.
For each test surface, Deep Smoothing independently fits a network
$2\!\rightarrow\!32\!\rightarrow\!32\!\rightarrow\!1$ with \texttt{tanh}
hidden activations and a \texttt{softplus} IV output. Its objective is
\[
\mathcal{L}_{\mathrm{DS}}
  = \operatorname{MSE}_{\mathrm{ref}}
    + \lambda\left[
       \mathbb{E}\operatorname{ReLU}(-\partial_\tau w)
       +\mathbb{E}\operatorname{ReLU}(-g)
      \right],
\qquad \lambda=1,
\]
where $w(k,\tau)=\sigma(k,\tau)^2\tau$ and $g$ is the
Gatheral--Jacquier butterfly function. The penalties are evaluated on a
$41\times16$ collocation grid whose range is determined from the test
asset's observed $(k,\tau)$ support. Each surface is optimized for 1000 Adam
steps with learning rate $10^{-2}$.

For computational efficiency, the test surfaces of an asset are optimized
simultaneously on a GPU. Every surface retains its own parameters and its own
loss; this batching changes only execution time and does not amortize the
per-surface optimization.

\end{itemize}

%% file: appendix/analysis.tex
\section{Additional Results and Analysis}

\subsection{Downstream Tail-Risk Evaluation}
\label{app:tail_risk}

The discrete marginal distribution produced by \addmart{} can be used
directly for downstream risk calculations. We examine this property through
option-implied downside CVaR at tail probabilities
$\alpha\in\{1\%,5\%,10\%\}$. For each asset, we select 50 dates spanning the
out-of-time test period and evaluate the expiries closest to 180 days and one
year. The amortized models observe only the nine reference contracts used in
the reconstruction task.

\paragraph{Full-information reference.}
Since the physical return distribution is unobserved, we construct an
option-implied reference from all available quotes at the selected expiry.
Let $M_T=S_T/F_T$ and $z=K/F_T$. After converting OTM puts to equivalent calls
using put--call parity, the normalized option price satisfies
\[
    \widetilde C(z)=\mathbb{E}\big[(M_T-z)^+\big].
\]
We fit a nonnegative discrete distribution on
$M_T\in[0.01,4]$ by matching these normalized prices, with quote weights
derived from the bid--ask spread and a second-difference penalty that
suppresses unidentified grid-scale oscillations. Unit mass and the martingale
moment are imposed numerically,
\[
    \sum_i p_i=1,
    \qquad
    \sum_i M_i p_i=1.
\]
The resulting distribution uses approximately 30--36 quotes per expiry and
serves as a full-information option-implied reference. It is not treated as
physical-distribution ground truth.

For a probability mass function $p$ on normalized terminal prices, we define
downside CVaR by
\[
    \operatorname{CVaR}_{\alpha}
    =
    1-
    \frac{1}{\alpha}
    \int_{0}^{\alpha}F_M^{-1}(u)\,du ,
\]
which measures the average normalized loss in the lower
$\alpha$-tail. For \addmart{}, this quantity is computed directly from its
predicted marginal. For IV-only baselines, we recover the density by finite
differences of the predicted call curve. Negative density is clipped and the
remaining mass is renormalized before CVaR is evaluated; we separately record
how often this repair is required.

\begin{table*}[h]

\caption{Option-implied downside-CVaR error on daily test surfaces.
Entries are CVaR-MAE ($\times10^{-2}$), averaged over 50 test dates and
$\alpha\in\{1\%,5\%,10\%\}$, with mean$\pm$std over five seeds.
Lowest error in each column is bold.}
\label{tab:downstream_cvar}

\centering
\small
\setlength{\tabcolsep}{4pt}
\resizebox{0.82\textwidth}{!}{%
\begin{tabular}{@{}l cc cc cc@{}}
\toprule
& \multicolumn{2}{c}{SPY}
& \multicolumn{2}{c}{QQQ}
& \multicolumn{2}{c}{NVDA} \\
\cmidrule(lr){2-3}
\cmidrule(lr){4-5}
\cmidrule(lr){6-7}
Method & 180d & 1y & 180d & 1y & 180d & 1y \\
\midrule
\addmart{}
& $\mathbf{1.87{\pm}.37}$
& $\mathbf{2.27{\pm}.35}$
& $3.05{\pm}.27$
& $\mathbf{3.12{\pm}.19}$
& $\mathbf{3.86{\pm}.20}$
& $\mathbf{2.42{\pm}.21}$ \\
HyperIV
& $1.89{\pm}.18$
& $4.76{\pm}.66$
& $\mathbf{2.88{\pm}.24}$
& $4.65{\pm}.39$
& $4.72{\pm}.93$
& $5.39{\pm}1.91$ \\
GNO
& $3.47{\pm}2.48$
& $4.23{\pm}.97$
& $19.74{\pm}9.79$
& $6.73{\pm}2.57$
& $5.23{\pm}3.74$
& $3.44{\pm}.72$ \\
VAE
& $12.51{\pm}3.78$
& $6.18{\pm}1.44$
& $10.09{\pm}5.42$
& $14.79{\pm}5.32$
& $12.19{\pm}5.00$
& $3.25{\pm}1.22$ \\
\bottomrule
\end{tabular}}
\end{table*}

\begin{figure*}[h]
    \centering
    \includegraphics[width=0.96\textwidth]
    {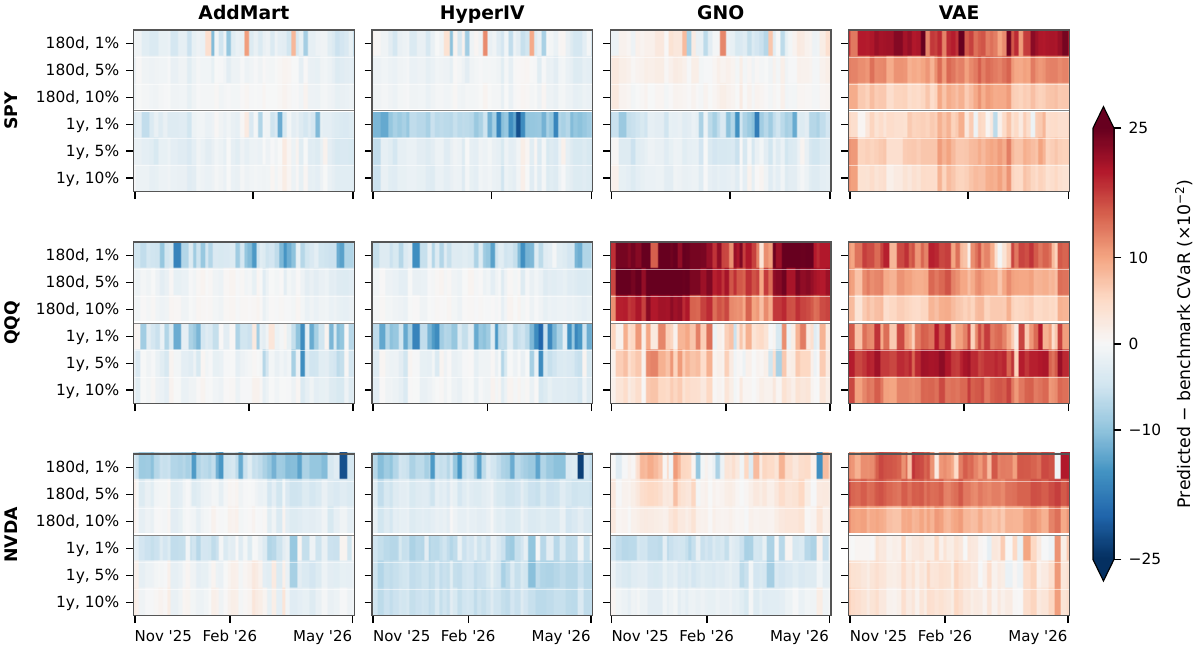}
    \caption{Signed CVaR residuals over the test period. Columns denote
    reconstruction methods and rows denote assets. Within each panel, the
    heatmap rows combine the 180-day and one-year horizons with tail
    probabilities of $1\%$, $5\%$, and $10\%$. Blue indicates that the model
    underestimates the full-information option-implied CVaR, while red
    indicates overestimation. Values are averaged over five seeds.}
    \label{fig:cvar_residuals}
\end{figure*}

Table~\ref{tab:downstream_cvar} shows that \addmart{} obtains the lowest CVaR
error in five of the six asset--horizon settings. On SPY at 180 days it is
nearly tied with HyperIV, while on QQQ at 180 days HyperIV is marginally
better after averaging the three tail levels. The latter advantage is
concentrated at $\alpha=1\%$; \addmart{} is more accurate at the $5\%$ and
$10\%$ tails. At the one-year horizon, \addmart{} is best on all three assets.

Figure~\ref{fig:cvar_residuals} exposes information hidden by the aggregate
MAE. The \addmart{} residuals remain relatively stable over time, with a
moderate tendency to underestimate the most extreme $1\%$ tail. HyperIV
exhibits a stronger negative bias at long horizons, particularly on QQQ and
NVDA. GNO substantially overestimates QQQ tail risk, whereas VAE exhibits
persistent positive bias on SPY and QQQ. These patterns indicate systematic
calibration differences rather than errors driven by a small number of test
dates.

The distributional validity check further separates the methods. The raw
marginals of \addmart{} are valid on every evaluated asset, date, maturity,
and seed. Before post-processing, the invalid-density rate ranges from
$10\%$ to $100\%$ for HyperIV and from $5.2\%$ to $100\%$ for GNO, and is
$100\%$ for VAE. Their CVaR values in Table~\ref{tab:downstream_cvar}
therefore depend on ex-post density repair, while \addmart{} provides the
risk functional directly from its predicted martingale marginal.

\paragraph{Interpretation.}
The experiment evaluates risk-neutral tail exposure implied by option
prices. It does not constitute a forecast of realized portfolio losses.
The $5\%$ and $10\%$ results are better identified by the observed strike
range; the $1\%$ results should be interpreted as an extreme-tail stress
test because they are more sensitive to sparse wing quotes and distributional
extrapolation.

\subsection{Stress Tests}

\textbf{Protocol.}
We evaluate all amortized methods during two pre-specified SPY market-stress
windows: April 1--10, 2025 (8 trading days) and March 23--31, 2026
(7 trading days). To avoid look-ahead bias, we use a separate set of
checkpoints whose out-of-time test period begins on February 1, 2025.
Each model is trained once on the preceding data and then kept frozen
through both events. The reference set, reconstruction targets, and
reliable-region definition are identical to the main experiments.
The two windows contain 10,273 and 14,557 held-out contracts,
respectively.

\begin{figure}[h]
    \centering
    \includegraphics[width=\linewidth]{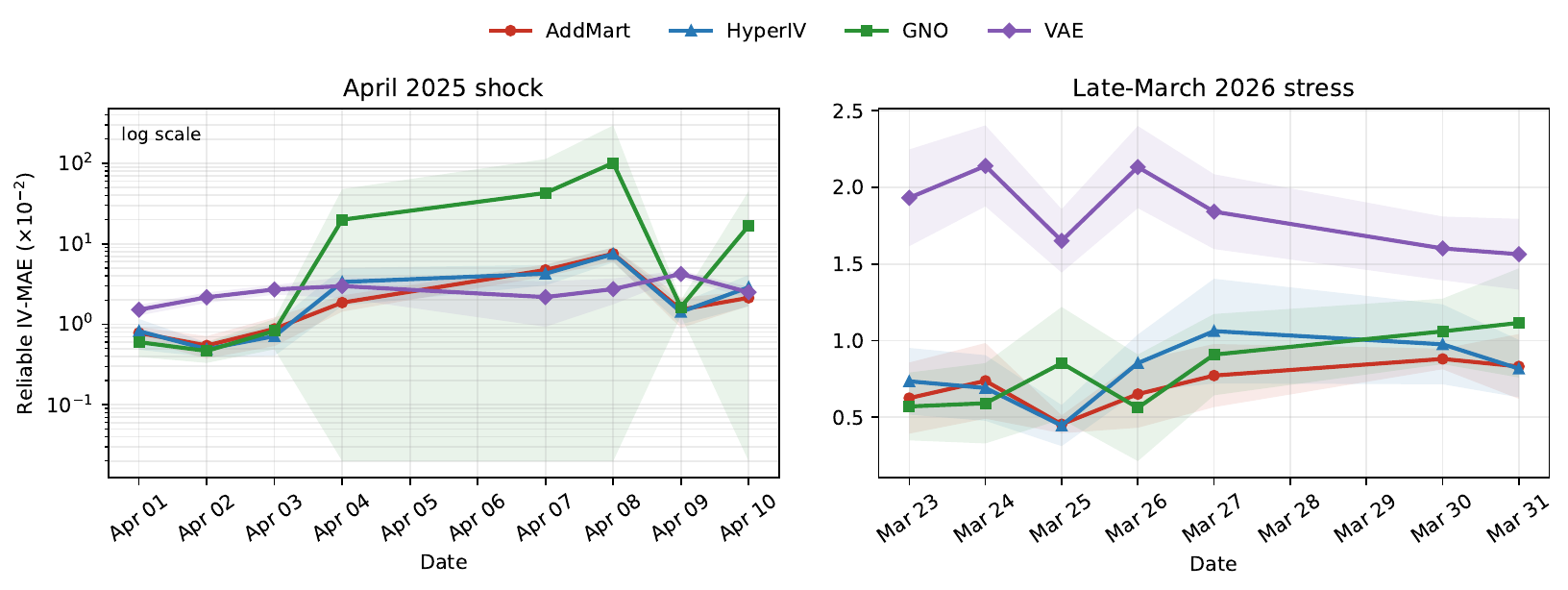}
    \caption{Daily reliable-region IV-MAE on the two SPY stress
    windows. Lines and shaded regions denote the five-seed mean and
    standard deviation. The April panel uses a logarithmic vertical
    scale because one GNO seed diverges during the most volatile days.}
    \label{fig:stress}
\end{figure}

\begin{table}[!htbp]
\caption{SPY performance during two out-of-sample market-stress
windows. IV-MAE is reported in $\times10^{-2}$ as mean$\pm$std over
five seeds. Best results are shown in bold.}
\label{tab:stress}
\centering
\small
\setlength{\tabcolsep}{8pt}
\begin{tabular}{@{}l cc cc@{}}
\toprule
& \multicolumn{2}{c}{Apr.\ 1--10, 2025}
& \multicolumn{2}{c}{Mar.\ 23--31, 2026} \\
\cmidrule(lr){2-3}\cmidrule(lr){4-5}
Method & Reliable & Overall & Reliable & Overall \\
\midrule
\addmart{}
 & $\mathbf{2.03{\pm}.23}$ & $4.92{\pm}.84$
 & $\mathbf{0.72{\pm}.13}$ & $\mathbf{0.98{\pm}.09}$ \\
HyperIV
 & $2.23{\pm}.45$ & $4.99{\pm}.86$
 & $0.80{\pm}.12$ & $1.58{\pm}.19$ \\
GNO
 & $16.74{\pm}28.25$ & $23.85{\pm}37.66$
 & $0.82{\pm}.20$ & $1.19{\pm}.16$ \\
VAE
 & $2.68{\pm}.45$ & $\mathbf{4.76{\pm}.53}$
 & $1.83{\pm}.24$ & $2.06{\pm}.19$ \\
\bottomrule
\end{tabular}
\end{table}

\textbf{Results.}
All methods deteriorate during the April 2025 shock, confirming that the
event induces a substantial distribution shift rather than a
model-specific failure. Relative to March 2025, \addmart{}'s overall and
reliable MAE increase by $5.9\times$ and $4.0\times$, respectively.
Nevertheless, it retains the lowest reliable-region error and remains
essentially tied with HyperIV in overall MAE. VAE obtains a slightly
lower overall error, but remains worse in the liquid reliable region.
GNO is considerably less stable: four seeds produce window-level
overall MAE between $4.5$ and $8.2\times10^{-2}$, while one seed rises
to $91.2\times10^{-2}$, causing the large reported variance.

The late-March 2026 window is less severe but occurs more than one year
after the training cutoff. Here, \addmart{} achieves the lowest error on
both metrics. Its overall MAE is approximately $18\%$ below GNO,
$38\%$ below HyperIV, and $53\%$ below VAE. The daily curves in
Figure~\ref{fig:stress} also show that its seed dispersion remains
controlled throughout the event. Relative to February 2026,
\addmart{}'s overall MAE increases by $35\%$, compared with $71\%$ for
GNO and $86\%$ for HyperIV. VAE exhibits a smaller relative increase,
but from a substantially higher absolute-error level.

These results suggest that the additive construction acts as a useful
structural regularizer under abrupt surface shifts: its accuracy still
degrades when the market leaves the training distribution, but it does
not exhibit the catastrophic seed failure observed for GNO. Moreover,
the exact decoder preserves its calendar and butterfly constraints
independently of the market regime. The stress tests therefore support
robustness in the sense of controlled degradation and retained validity,
rather than invariance of reconstruction error during extreme events.

\subsection{Computational Performance}
\label{sec:computational-performance}

We benchmark the amortized models on a single NVIDIA RTX 5090 D using nine
reference contracts and 300 reconstruction targets per surface. Each
measurement uses 20 warm-up iterations followed by 100 synchronized
iterations. We report the median across seven daily assets and three
minute-scale assets using the seed-0 checkpoints. Preprocessing includes the
construction of model-specific inputs but excludes disk I/O. The batch-one
end-to-end latency is estimated as preprocessing time plus neural forward
time. SSVI and Deep Smoothing are excluded because
they solve a separate optimization problem for every input surface, whereas
the learned models perform amortized inference with fixed trained weights.

\begin{table}[h]
\caption{Single-GPU inference performance. Forward and end-to-end latency are
reported in milliseconds per surface. Throughput measures neural forward
inference at batch size 32. Peak memory is the maximum allocated GPU memory
for batch sizes 1 and 32, respectively.}
\label{tab:computation}
\centering
\small
\setlength{\tabcolsep}{3.5pt}
\resizebox{\linewidth}{!}{%
\begin{tabular}{@{}llrrrrrr@{}}
\toprule
Scale & Method
& \multicolumn{1}{c}{B1 forward}
& \multicolumn{1}{c}{B1 end-to-end}
& \multicolumn{1}{c}{B32 forward}
& \multicolumn{1}{c}{B32 surface/s}
& \multicolumn{1}{c}{Peak MiB B1/B32}
& \multicolumn{1}{c}{Weights MiB} \\
\midrule
\multirow{4}{*}{Daily}
& \addmart{} & 6.98 & 7.60 & 1.040 & 961 & 127.64 / 2939.44 & 0.918 \\
& HyperIV   & 0.58 & \textbf{1.19} & 0.019 & 53,201 & 32.99 / 34.84 & 0.928 \\
& GNO       & 1.99 & 2.60 & 1.874 & 533 & 36.69 / 36.85 & \textbf{0.402} \\
& VAE       & \textbf{0.20} & 29.60 & \textbf{0.009} & \textbf{113,918} & 34.04 / 58.05 & 1.272 \\
\midrule
\multirow{4}{*}{Minute}
& \addmart{} & 6.80 & 7.43 & 1.042 & 960 & 127.64 / 2939.44 & 0.918 \\
& HyperIV   & 0.57 & \textbf{1.21} & 0.019 & 52,994 & 32.99 / 34.84 & 0.928 \\
& GNO       & 1.99 & 2.63 & 1.883 & 531 & 36.69 / 36.85 & \textbf{0.402} \\
& VAE       & \textbf{0.20} & 33.69 & \textbf{0.009} & \textbf{114,110} & 34.04 / 58.05 & 1.272 \\
\bottomrule
\end{tabular}}
\end{table}

For single-surface inference, \addmart{} requires approximately $7$\,ms for the forward pass and $7.4$--$7.6$\,ms including preprocessing. Although slower than HyperIV and GNO due to full-support convolution and implied-volatility inversion, this latency remains below $10$\,ms and is sufficient for minute-scale surface updates and conventional market-making applications. Batch inference further reduces the cost to approximately $1.04$\,ms per surface, or 960 surfaces per second. The model weights occupy less than $1$\,MiB, while batch-one inference uses approximately $128$\,MiB of GPU memory. In return for this moderate computational overhead, \addmart{} achieves the lowest asset-averaged overall IV-MAE at both daily and minute scales while preserving structural no-arbitrage guarantees.

\subsection{Cross-Asset Performance}
\label{sec:cross_asset}

We next examine whether a surface model trained on one underlying can be
applied to another without retraining. For each ordered source--target pair,
we retain the source-asset checkpoint and provide only the target surface's
nine reference quotes at inference time. All remaining target contracts are
used for evaluation under the same out-of-time split beginning on
February 1, 2025. The experiment covers seven daily assets and therefore
contains $7\times7$ directed pairs per method. Learned models are evaluated
over five seeds; SSVI is fitted independently on each target surface and is
included only as a target-calibrated reference.

Table~\ref{tab:cross_asset_summary} shows that transfer is substantially more
difficult than in-domain reconstruction. The overall MAE of \addmart{}
increases from $1.65$ to $5.51$ after excluding diagonal pairs, corresponding
to a $3.34\times$ transfer penalty. Nevertheless, it retains the lowest
cross-asset overall mean and outperforms HyperIV on $76.2\%$ of the 42 directed
pairs when each pair is first averaged over seeds. HyperIV is stronger on the
reliable region after transfer, attaining a mean of $4.81$ versus
\addmart{}'s $5.33$. Thus, the in-domain advantage of \addmart{} does not
uniformly extend to zero-shot accuracy: its benefit is clearer under the
overall metric, while HyperIV transfers better within the liquid interior.

\begin{table*}[h]
\caption{Daily cross-asset transfer over the seven-asset universe.
We report IV-MAE ($\times10^{-2}$). In-domain columns average diagonal
source--target pairs, while cross-asset columns aggregate the 42 directed
off-diagonal pairs. All methods are evaluated over five seeds. The lowest
error in each column is shown in bold.}
\label{tab:cross_asset_summary}
\centering
\setlength{\tabcolsep}{4.5pt}
\begin{adjustbox}{max width=0.75\linewidth,center}
\begin{tabular}{@{}l cc cc cc c@{}}
\toprule
& \multicolumn{2}{c}{In-domain}
& \multicolumn{2}{c}{Cross-asset overall}
& \multicolumn{2}{c}{Cross-asset reliable}
& \multirow{2}{*}{Transfer penalty} \\
\cmidrule(lr){2-3}
\cmidrule(lr){4-5}
\cmidrule(lr){6-7}
Method
& Overall & Reliable
& Mean & Median
& Mean & Median
& \\
\midrule
\addmart{}
& $\mathbf{1.65}$ & $1.29$
& $\mathbf{5.51}$ & $4.67$
& $5.33$ & $4.25$
& $3.34\times$ \\
HyperIV
& $1.72$ & $\mathbf{1.27}$
& $5.53$ & $\mathbf{4.24}$
& $\mathbf{4.81}$ & $\mathbf{3.52}$
& $3.21\times$ \\
GNO
& $2.01$ & $1.63$
& $6.99{\times}10^{3}$ & $5.04$
& $7.62{\times}10^{3}$ & $4.15$
& $2921.6\times$ \\
VAE
& $2.52$ & $2.04$
& $6.03$ & $5.57$
& $5.21$ & $5.04$
& $2.39\times$ \\
\bottomrule
\end{tabular}
\end{adjustbox}
\end{table*}

\begin{table*}[h]
\caption{Daily cross-asset reliable-region IV-MAE
($\times10^{-2}$, five-seed mean). Rows denote training assets and columns
denote evaluation assets. Gray cells indicate in-domain evaluations, and the
lowest value in each target column is shown in bold.}
\label{tab:cross_asset}
\centering
\small
\setlength{\tabcolsep}{3pt}

\begin{minipage}[t]{0.49\textwidth}
\centering
\textbf{(a) AddMart}\\[3pt]
\resizebox{\linewidth}{!}{%
\begin{tabular}{@{}lrrrrrrr@{}}
\toprule
Train $\backslash$ Test
& SPY & QQQ & IWM & AAPL & GOOGL & META & NVDA \\
\midrule
SPY
& \cellcolor{gray!15}\textbf{0.56}
& 0.76 & 0.97 & 2.05 & 6.41 & 7.03 & 15.61 \\
QQQ
& 1.25
& \cellcolor{gray!15}\textbf{0.68}
& 0.79 & 1.55 & 4.03 & 4.84 & 11.14 \\
IWM
& 1.87 & 0.82
& \cellcolor{gray!15}\textbf{0.77}
& 1.70 & 4.46 & 4.97 & 10.02 \\
AAPL
& 2.83 & 1.44 & 1.12
& \cellcolor{gray!15}\textbf{1.18}
& 2.27 & 2.75 & 5.81 \\
GOOGL
& 6.12 & 3.01 & 2.26 & 1.52
& \cellcolor{gray!15}2.22
& 2.60 & 5.76 \\
META
& 13.87 & 8.25 & 6.22 & 5.47
& $\mathbf{2.18}$
& \cellcolor{gray!15}$\mathbf{1.82}$
& 2.65 \\
NVDA
& 19.66 & 14.65 & 12.81 & 11.11 & 4.79 & 4.43
& \cellcolor{gray!15}$\mathbf{1.80}$ \\
\bottomrule
\end{tabular}%
}
\end{minipage}
\hfill
\begin{minipage}[t]{0.49\textwidth}
\centering
\textbf{(b) HyperIV}\\[3pt]
\resizebox{\linewidth}{!}{%
\begin{tabular}{@{}lrrrrrrr@{}}
\toprule
Train $\backslash$ Test
& SPY & QQQ & IWM & AAPL & GOOGL & META & NVDA \\
\midrule
SPY
& \cellcolor{gray!15}\textbf{0.57}
& 0.77 & 1.00 & 1.97 & 6.24 & 6.86 & 14.81 \\
QQQ
& 1.40
& \cellcolor{gray!15}\textbf{0.74}
& \textbf{0.76}
& 1.55 & 3.69 & 4.41 & 9.14 \\
IWM
& 1.65 & 0.85
& \cellcolor{gray!15}0.78
& 1.57 & 4.35 & 4.99 & 10.29 \\
AAPL
& 2.49 & 1.49 & 1.11
& \cellcolor{gray!15}\textbf{1.18}
& 2.31 & 2.87 & 5.60 \\
GOOGL
& 4.36 & 2.66 & 2.11 & 1.43
& \cellcolor{gray!15}\textbf{2.07}
& 2.47 & 5.18 \\
META
& 11.48 & 7.24 & 6.30 & 4.95 & 2.16
& \cellcolor{gray!15}\textbf{1.79}
& 2.74 \\
NVDA
& 16.53 & 12.11 & 10.57 & 9.33 & 4.24 & 3.79
& \cellcolor{gray!15}\textbf{1.80} \\
\bottomrule
\end{tabular}%
}
\end{minipage}

\end{table*}

The directed matrices in Tables~\ref{tab:cross_asset} reveal that asset similarity dominates transfer
quality. For \addmart{}, QQQ$\rightarrow$IWM and IWM$\rightarrow$QQQ attain
overall MAEs of $1.30$ and $1.33$, respectively, close to their in-domain
errors. Transfers between broad-market ETFs and NVDA are much harder:
SPY$\rightarrow$NVDA reaches a reliable MAE of $15.61$, while the reverse
NVDA$\rightarrow$SPY direction reaches $19.66$. Transfer is therefore highly
asymmetric, and reversing the source and target can substantially change the
error. The matrices also show that the best source need not be the target
itself; for example, META transfers slightly better than GOOGL itself when
\addmart{} is evaluated on GOOGL.

Finally, GNO illustrates why both means and medians are necessary in
cross-asset evaluation. Its median off-diagonal MAE remains comparable to the
other neural baselines, but 13 of its 210 seed--direction cells exceed an MAE
of $1$, eight exceed $10$, and the resulting mean becomes several orders of
magnitude larger. These failures indicate unstable zero-shot extrapolation
rather than uniformly poor reconstruction. Overall, the experiment shows that
\addmart{} transfers reliably between related markets but is not
asset-invariant; large shifts in volatility level and smile asymmetry remain a
substantial source of error.

\subsection{Analysis on LEAPS and VIX}
\label{appendix:limitations}

We further examine two regimes that depart substantially from the
liquid equity-option surfaces used in the main experiments: long-dated
LEAPS and options on VIX. These experiments characterize the scope of
the additive martingale assumption rather than extend the main
benchmark universe.

\begin{figure}[h]
    \centering
    \includegraphics[width=0.99\linewidth]{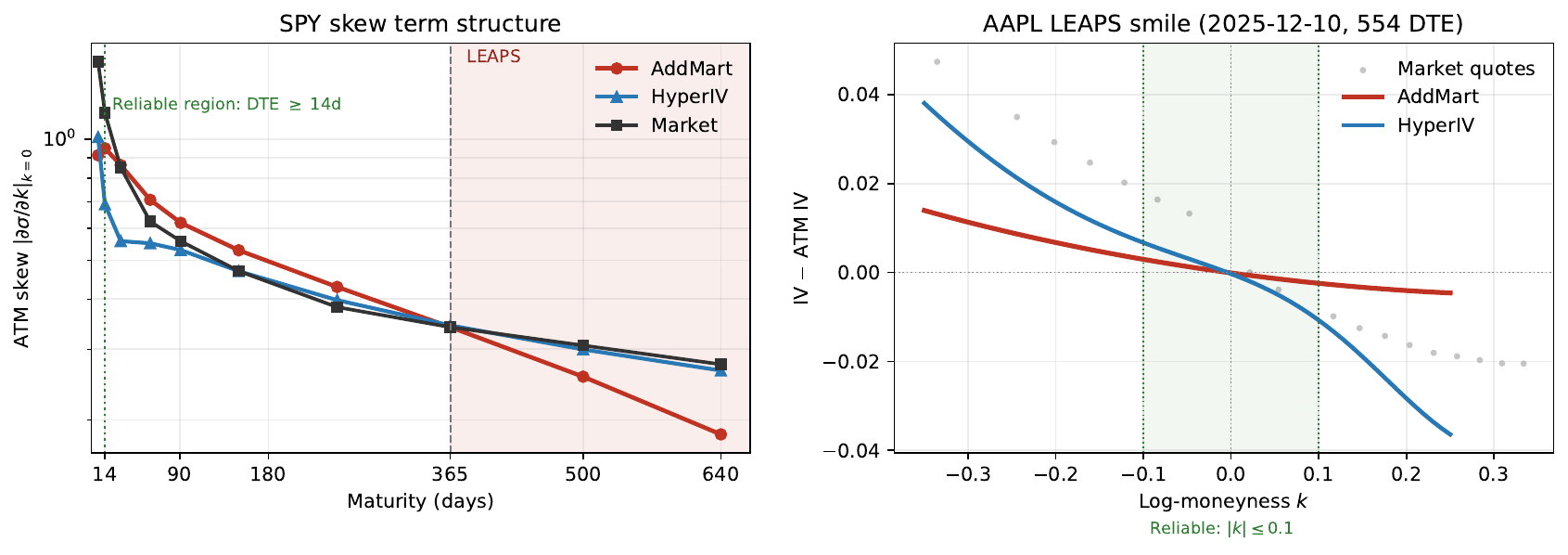}
    \caption{Long-horizon behavior on LEAPS. Left: average ATM-skew
    term structure on the SPY test set, with the reliable-region cutoff
    at 14 days and the LEAPS region beginning at 365 days. Right: a
    representative AAPL LEAPS slice observed on December 10, 2025
    (554 DTE; 18 market quotes). The shaded interval denotes the
    reliable moneyness region $|k|\le0.1$. AddMart increasingly
    underestimates the persistent long-dated skew, whereas HyperIV
    retains more of the observed skew but extrapolates more aggressively
    outside the reliable region.}
    \label{fig:fail_leaps}
\end{figure}

\textbf{Long-dated LEAPS.}
Figure~\ref{fig:fail_leaps} localizes the long-horizon limitation to the
evolution of skew across maturity. On SPY, \addmart{} follows the market
ATM-skew term structure reasonably well through approximately one year,
but its predicted skew subsequently decays faster than observed. The
same behavior is visible in the 554-day AAPL slice, where the predicted
smile is substantially flatter than the market quotes, including within
the reliable region. This is consistent with the repeated convolution
of conditionally independent increments progressively smoothing the
terminal distribution. At sufficiently long horizons, the construction
therefore has difficulty retaining persistent asymmetry and tail shape.
HyperIV is more flexible at these maturities and recovers more of the
observed skew, although its unconstrained extrapolation becomes unstable
away from the quoted region.

\begin{figure}[h]
    \centering
    \includegraphics[width=0.99\linewidth]{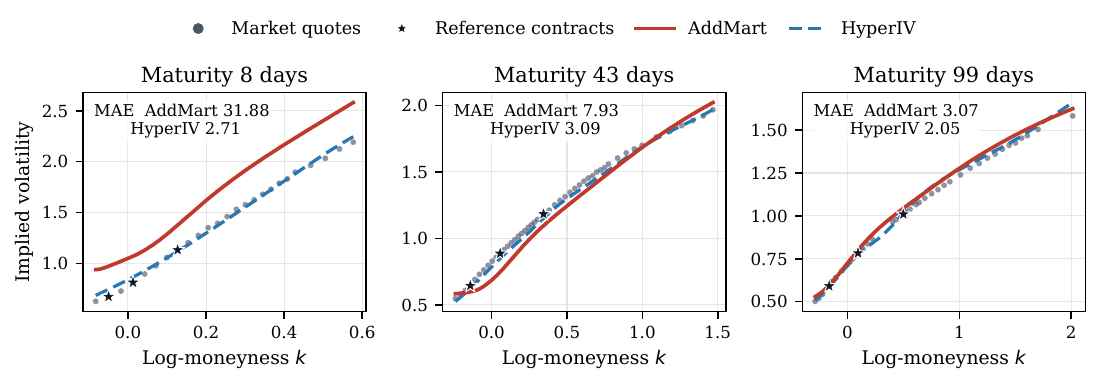}
    \caption{Representative failure case on a VIX daily test surface
    observed on September 9, 2025. The surface is selected by the median
    per-surface MAE gap between AddMart and HyperIV rather than by an
    extreme error. AddMart substantially overestimates the
    short-maturity smile, while the discrepancy decreases with
    maturity.}
    \label{fig:fail_vix}
\end{figure}

\textbf{VIX options.}
VIX produces a distinct short-horizon failure mode.
Figure~\ref{fig:fail_vix} shows that most of the accuracy gap is
concentrated at the shortest maturity and decreases rapidly along the
term structure. VIX options inherit strong mean reversion, jumps,
volatility-of-volatility, and level-dependent skew. Moreover, different
maturities are linked to different points of the VIX-futures term
structure. Representing these marginals through a shared sequence of
additive increments consequently imposes a stronger cross-maturity
restriction than it does for equity options. The consistency of this
pattern across seeds suggests a representation limitation rather than
training instability. Despite the larger IV error, \addmart{} retains
zero measured calendar and butterfly violations on the VIX test set;
the failure concerns market fit rather than measure validity.

\textbf{Practical impact.}
These limitations have limited effect on the primary use case studied
in this paper. Most liquid equity and ETF option activity is
concentrated in shorter maturities and near the central moneyness
region, where \addmart{} remains accurate and the hard no-arbitrage
construction is most valuable. LEAPS quotes are comparatively sparse
and have wider bid--ask spreads, while VIX options constitute a
different underlying process from the equity and ETF surfaces in the
main benchmark. Consequently, neither stress regime materially drives
the aggregate reconstruction results. The limitation is nevertheless
relevant for applications devoted specifically to long-horizon tail
hedging or short-dated VIX trading, for which the current additive
increment parameterization should not be used as a stand-alone pricing
model.

\subsection{Detailed Comparison}

\begin{figure*}[t]
    \centering
    \includegraphics[width=0.95\textwidth,height=0.76\textheight,keepaspectratio]
    {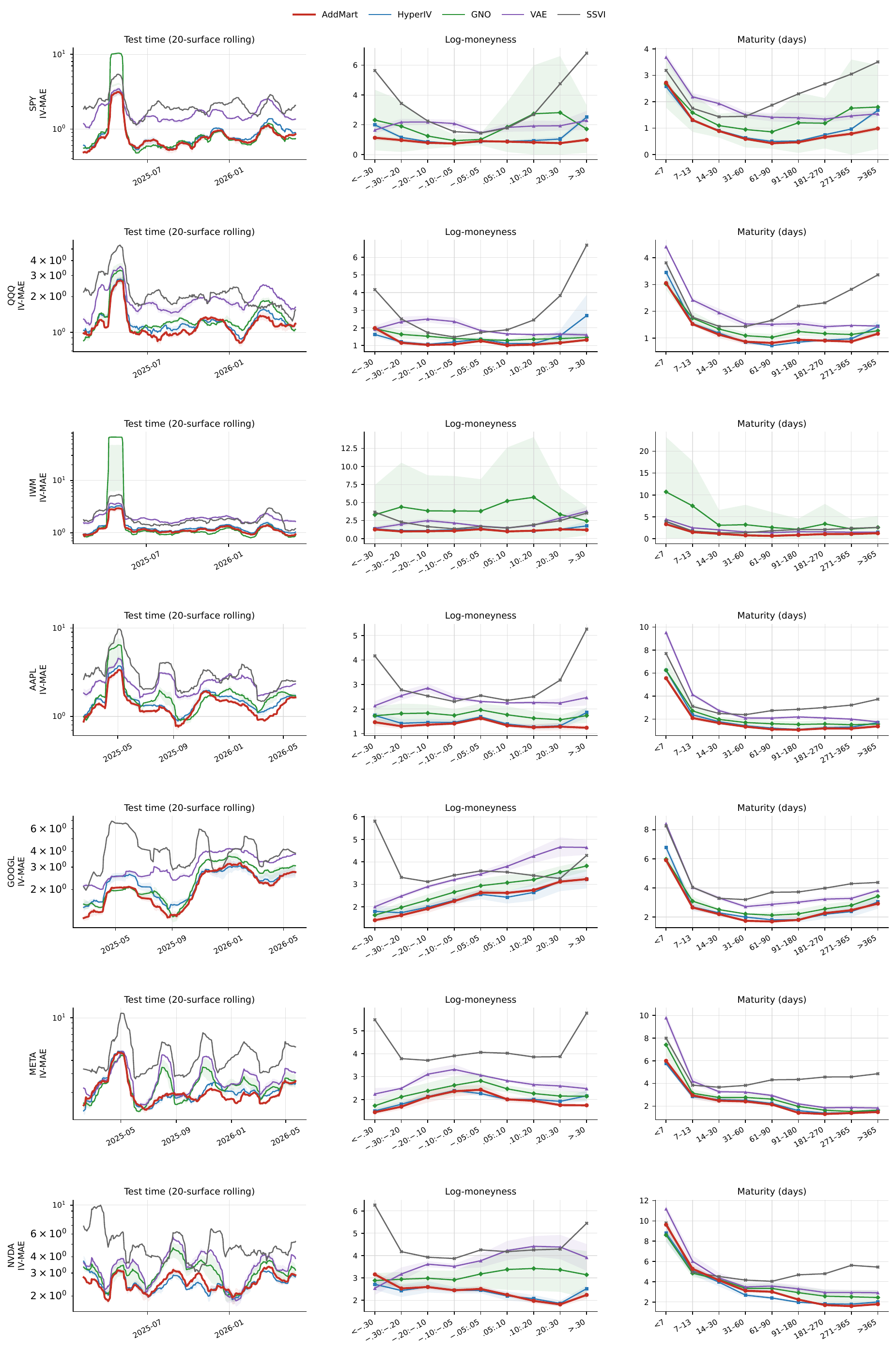}
    \caption{
    Daily-data error profiles for SPY, QQQ, IWM, AAPL, GOOGL,
    META, and NVDA. Rows correspond to underlying assets.
    The columns report 20-surface rolling IV-MAE over the test
    period, IV-MAE by log-moneyness, and IV-MAE by maturity,
    respectively. IV-MAE is reported in $\times 10^{-2}$.
    For amortized methods, lines and shaded regions denote the
    mean and standard deviation over five seeds; SSVI is fitted
    independently to each surface.
    }
    \label{fig:daily-error-profiles}
\end{figure*}

Figures~\ref{fig:daily-error-profiles} compares reconstruction errors along three
complementary dimensions: test time, log-moneyness, and maturity. At the daily
scale, \addmart{} achieves the lowest or near-lowest error across most assets,
market regimes, and contract buckets. The temporal profiles show that several
error spikes occur simultaneously across methods, indicating periods of
greater market-wide surface complexity. GNO exhibits the clearest
method-specific instability on IWM, while SSVI deteriorates substantially in
the deep wings and near the maturity boundaries. HyperIV remains competitive
on several assets and performs slightly better in parts of the NVDA surface.
At the minute scale, \addmart{} is particularly strong on SPY and IWM and
remains competitive on QQQ. Across both sampling frequencies, the shortest
maturities and sparsely populated wings are consistently the most difficult
regions.

\begin{figure*}[p]
    \centering
    \includegraphics[
        width=\textwidth,
        height=0.90\textheight,
        keepaspectratio
    ]{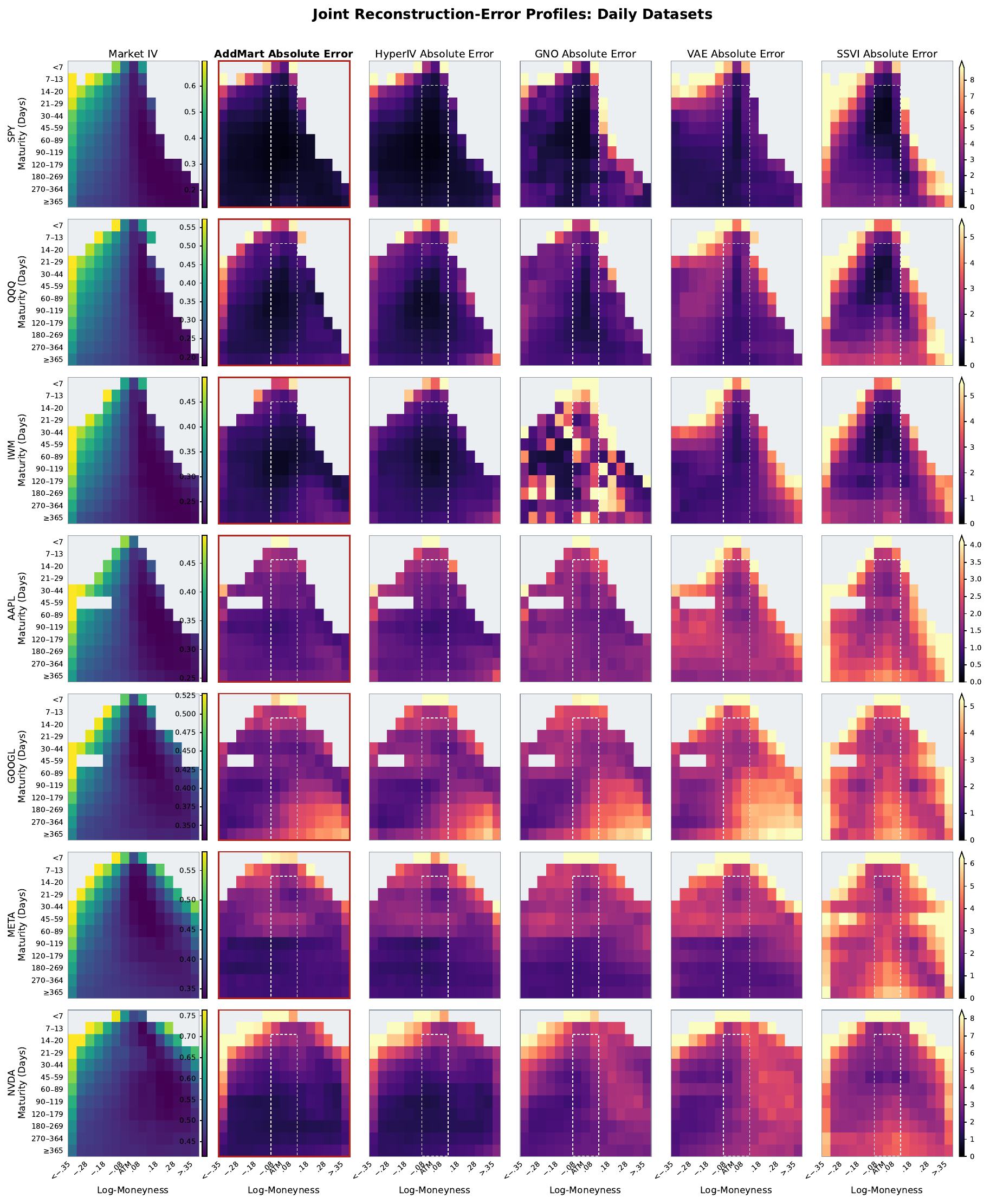}
    \caption{Joint reconstruction-error profiles across the seven daily
    datasets. The first column shows market IV; the remaining columns show
    IV-MAE ($\times10^{-2}$) aggregated by log-moneyness and maturity.
    Error color scales are shared across methods within each asset row.
    The dashed box denotes the reliable region, and gray cells contain
    insufficient observations for reliable aggregation.}
    \label{fig:mae_heatmaps_daily}
\end{figure*}

\begin{figure*}[p]
    \centering
    \includegraphics[
        width=\textwidth,
        height=0.90\textheight,
        keepaspectratio
    ]{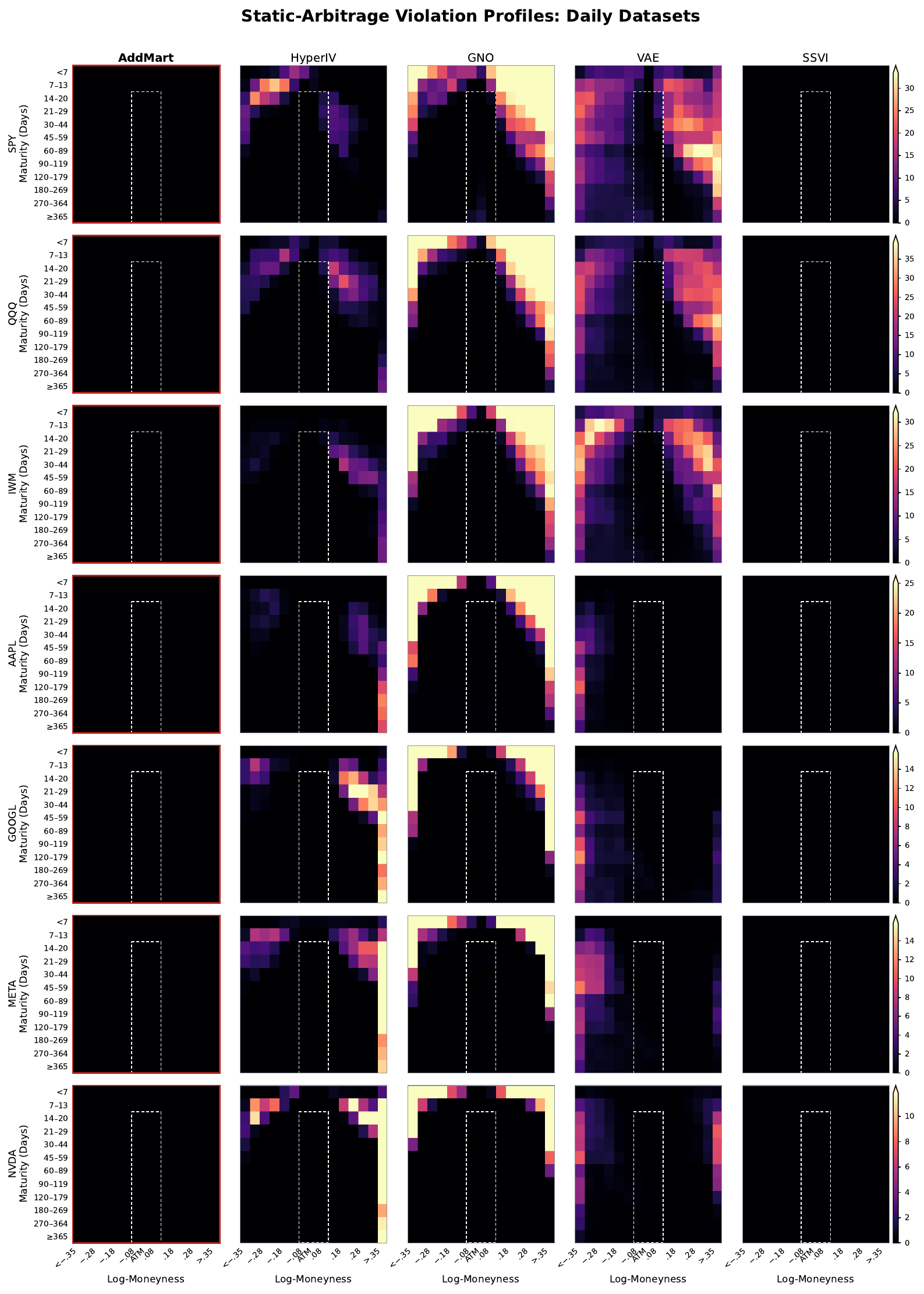}
    \caption{Static-arbitrage violation profiles across the seven daily
    datasets. Each cell reports the pooled calendar and butterfly pair
    violation rate (\%) on a dense price grid, using a tolerance of
    $10^{-7}$. Color scales are shared across methods within each asset
    row; the dashed box denotes the reliable region.}
    \label{fig:arb_heatmaps_daily}
\end{figure*}

\begin{figure*}[p]
    \centering
    \includegraphics[
        width=\textwidth,
        height=0.90\textheight,
        keepaspectratio
    ]{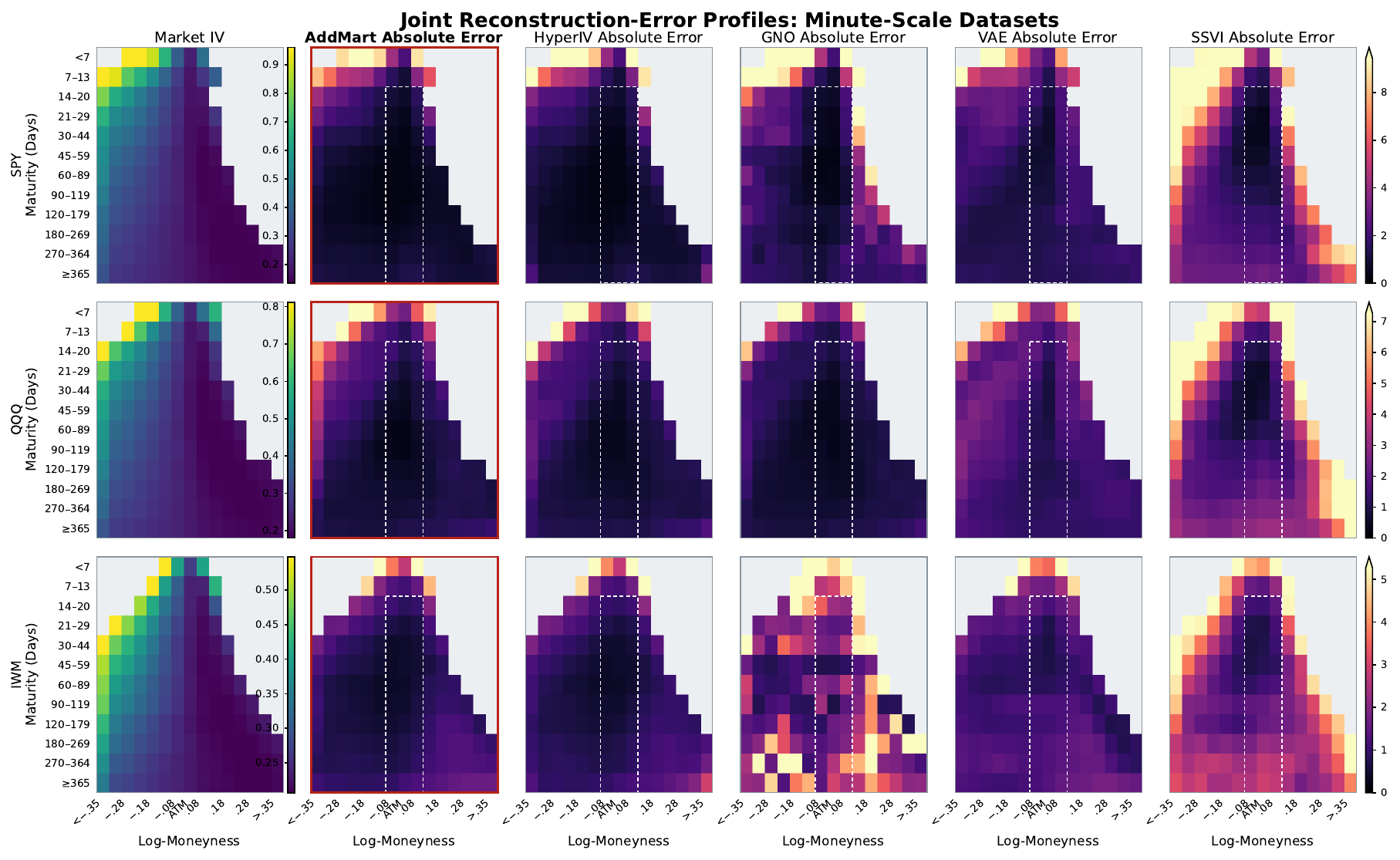}
    \caption{Joint reconstruction-error profiles on the SPY, QQQ, and IWM
    minute-scale datasets. Each heatmap reports IV-MAE
    ($\times10^{-2}$) by log-moneyness and maturity; notation and masking
    follow Figure~\ref{fig:mae_heatmaps_daily}.}
    \label{fig:mae_heatmaps_minute}
\end{figure*}

\begin{figure*}[p]
    \centering
    \includegraphics[
        width=\textwidth,
        height=0.90\textheight,
        keepaspectratio
    ]{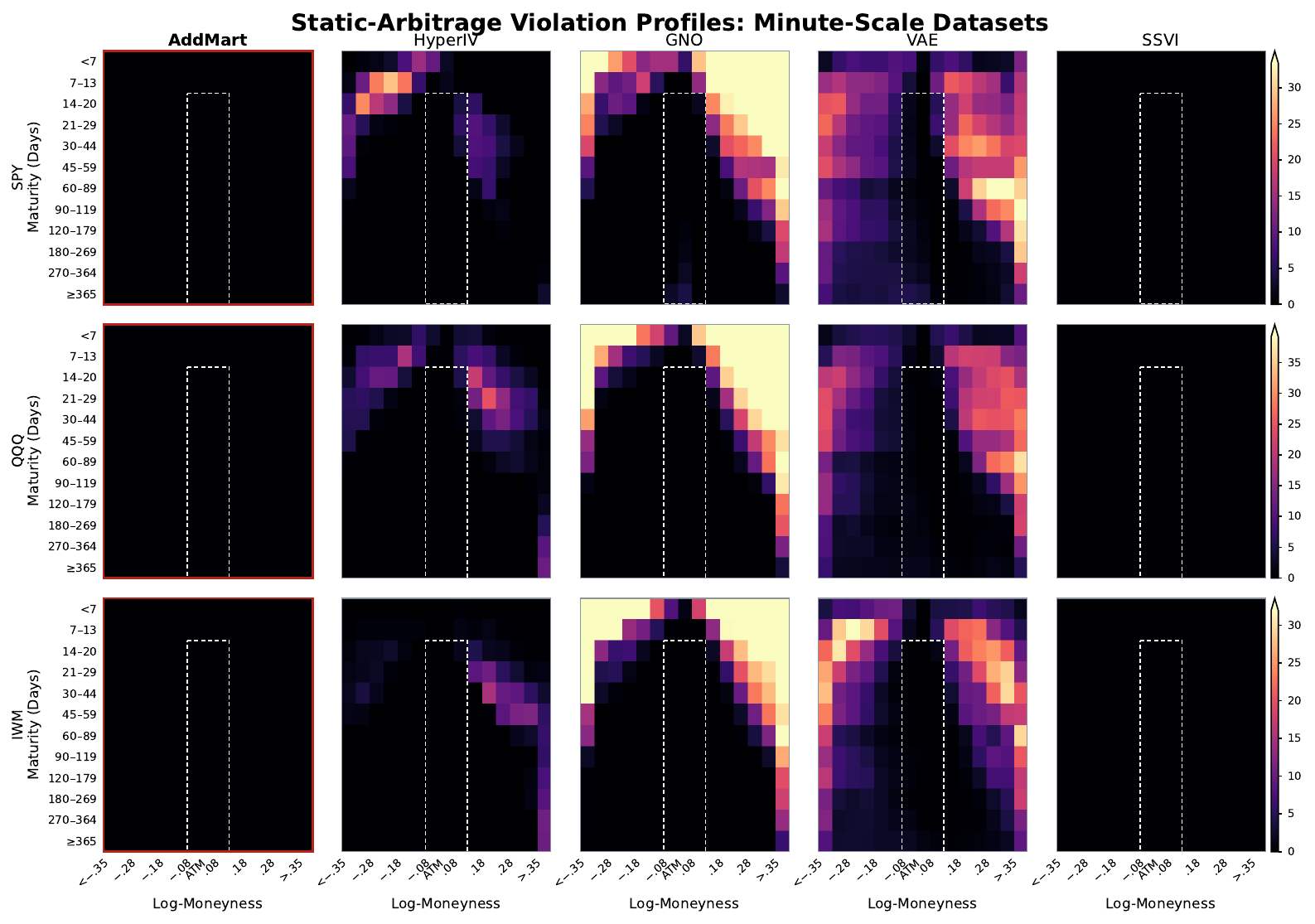}
    \caption{Static-arbitrage violation profiles on the SPY, QQQ, and IWM
    minute-scale datasets. Evaluation and notation follow
    Figure~\ref{fig:arb_heatmaps_daily}.}
    \label{fig:arb_heatmaps_minute}
\end{figure*}

Figures~\ref{fig:mae_heatmaps_daily} and
\ref{fig:mae_heatmaps_minute} further localize reconstruction error jointly
over log-moneyness and maturity. At the daily scale, \addmart{} maintains low
error throughout the reliable region across all seven assets, showing that its
aggregate performance is not driven by a small collection of favorable
contracts. Errors increase mainly at very short maturities and in the deep
wings, where quotes are sparse and implied-volatility inversion is less stable.
The competing methods exhibit more distinct spatial failure patterns:
HyperIV deteriorates near some short-end boundaries, GNO produces pronounced
wing errors---most visibly on IWM---and VAE has broader regions of elevated
error. SSVI is accurate near the liquid interior but extrapolates poorly toward
sparse wings and longer maturities. The minute-scale comparison is tighter,
although \addmart{} continues to degrade smoothly outside the reliable region.
Figures~\ref{fig:arb_heatmaps_daily} and
\ref{fig:arb_heatmaps_minute} show that reconstruction accuracy and structural
validity are separate properties. Each cell reports the pooled violation rate
over calendar and butterfly price-grid pairs. The exact \addmart{} decoder has
zero violations throughout the evaluated domain, including short maturities
and deep wings where reconstruction is more difficult. HyperIV violations are
concentrated mainly at short maturities and in the wings, GNO exhibits
extensive violations under extrapolation, and VAE violations extend closer to
the liquid interior. The constrained per-surface SSVI fit is also
arbitrage-free in this audit, but its error maps reveal the accuracy cost of
its restrictive parametric form.

\begin{figure}[t]
  \centering
  \includegraphics[width=0.95\linewidth]{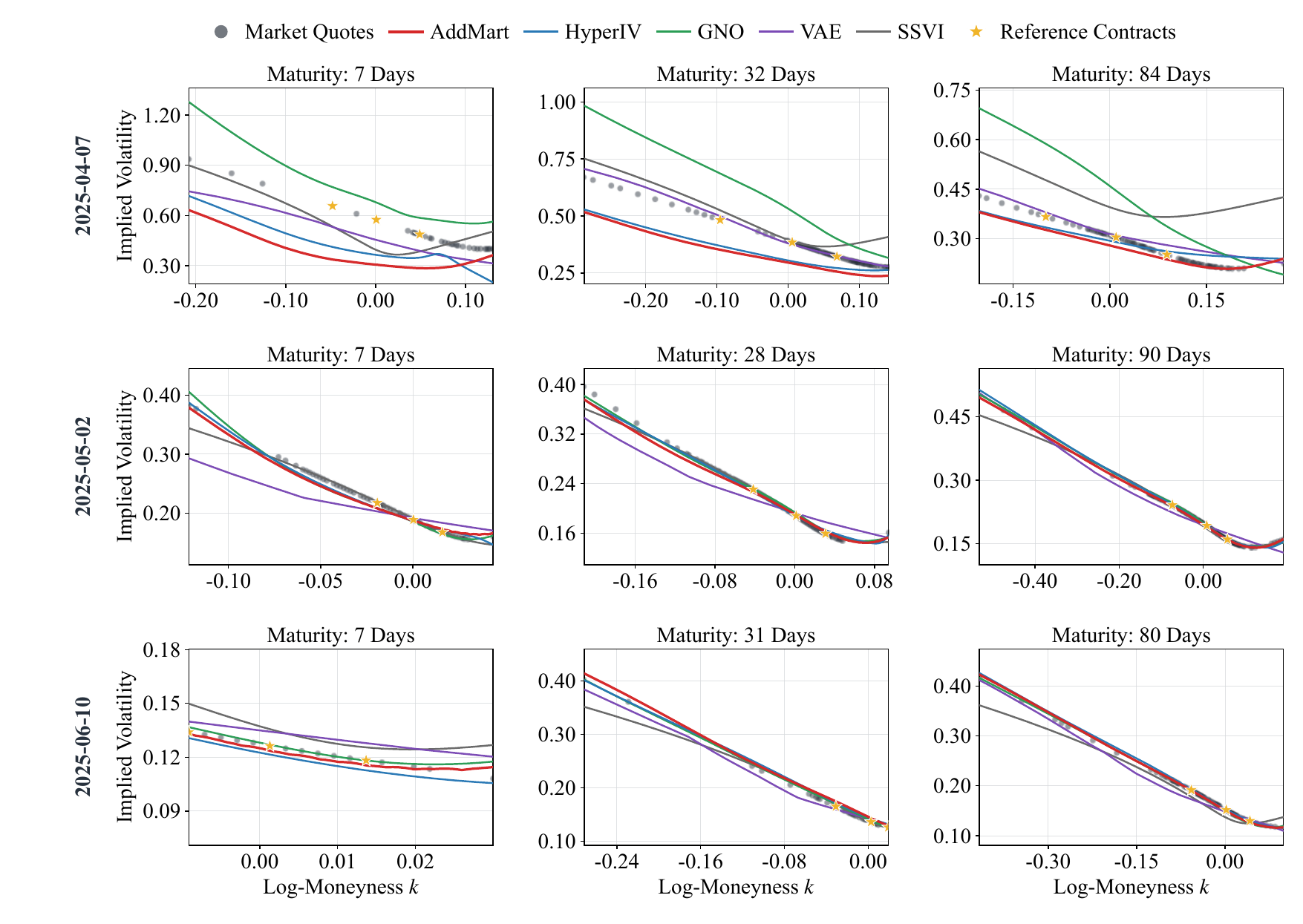}
  \vspace{-1em}
  \caption{SPY implied-volatility slices on 7 April, 2 May, and 10 June 2025. Each row corresponds to one trading day and each column to one reference maturity. Gray dots denote market quotes and gold stars denote the nine reference contracts used by AddMart, HyperIV, GNO, VAE, and SSVI.}
  \label{fig:spy-method-comparison}
\end{figure}

\begin{figure}[t]
  \centering
  \includegraphics[width=0.95\linewidth]{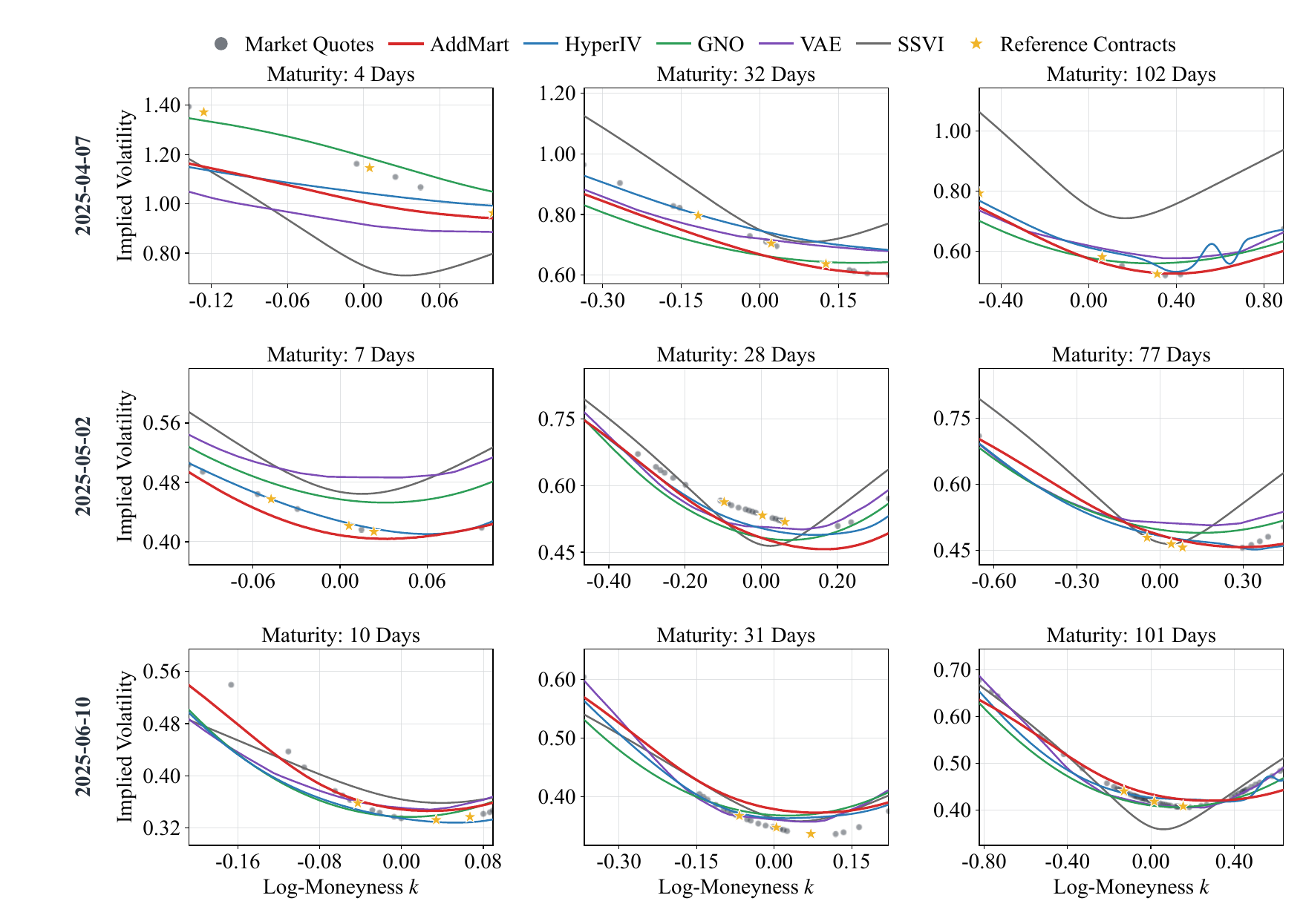}
\vspace{-1em}
  \caption{NVDA implied-volatility slices on 7 April, 2 May, and 10 June 2025. Each row corresponds to one trading day and each column to one reference maturity. Gray dots denote market quotes and gold stars denote the nine reference contracts used by AddMart, HyperIV, GNO, VAE, and SSVI.}
  \label{fig:nvda-method-comparison}
\end{figure}

The representative surfaces in
Figures~\ref{fig:spy-method-comparison} and
\ref{fig:nvda-method-comparison} illustrate these aggregate patterns at the
individual-contract level. The SPY surface on April 7 is especially
challenging at the shortest maturity, where the methods disagree substantially
in the left wing. Their predictions become much closer around the reference
quotes and on the later test dates. On May 2 and June 10, \addmart{} closely
tracks the dense non-reference quotes across the medium and long maturities.
NVDA presents a more difficult cross-sectional fit, and no method dominates
every slice. GNO and SSVI can depart materially from the observed wing shape,
while HyperIV develops an oscillatory long-maturity wing on April 7.
\addmart{} produces smoother extrapolations and competitive fits, although
HyperIV retains a small aggregate advantage on NVDA. Together, the temporal
profiles, joint heatmaps, and representative slices show that \addmart{}
combines stable reconstruction accuracy across market regimes with exact
static-arbitrage control over the full surface.

% \begin{figure*}[h]
%     \centering
%     \includegraphics[width=\textwidth]
%     {figures/minute_error_profiles_overview.pdf}
%     \caption{
%     Minute-data error profiles for SPY, QQQ, and IWM.
%     Rows correspond to underlying assets. The columns report
%     40-surface rolling IV-MAE over the test period, IV-MAE by
%     log-moneyness, and IV-MAE by maturity, respectively.
%     IV-MAE is reported in $\times 10^{-2}$. For amortized
%     methods, lines and shaded regions denote the mean and
%     standard deviation over five seeds; SSVI is fitted
%     independently to each surface.
%     }
%     \label{fig:minute-error-profiles}
% \end{figure*}

\clearpage

\clearpage

\clearpage
\vspace{-2em}

\vspace{-3em}

